\documentclass[a4paper,11pt]{article}
\pdfoutput=1 \usepackage{jheppub_2} 
\usepackage{booktabs}
\usepackage[T1]{fontenc} 
\usepackage[utf8]{inputenc}
\usepackage{color,subfigure}
\usepackage{graphicx, multirow,soul,url,amsmath,amsfonts,amssymb,mathrsfs,amsfonts}
\usepackage[all]{hypcap}
\usepackage{url}
\usepackage{bbm}
\usepackage{cancel}
\usepackage{subfigure}
\usepackage{slashed}
\usepackage[dvipsnames]{xcolor}
\usepackage{multicol, blindtext}
\usepackage[section]{placeins}
\usepackage[version=3]{mhchem}
\usepackage{caption}
\usepackage{lipsum}
\usepackage{hyperref}
\usepackage{float}
\usepackage{mattens}
\usepackage{mathtools}
\usepackage{footnote}
\usepackage{amsmath,amssymb,amsthm,amsxtra,overpic,bbm,bm,epsfig}
\usepackage{color,subfigure}
\usepackage[splitrule]{footmisc}
\usepackage{bbold}
\usepackage{amsmath}
\usepackage{amsbsy}
\usepackage{lipsum}
\newcommand{\overbar}[1]{\mkern 1.1mu\overline{\mkern-1.1mu#1\mkern-1.1mu}\mkern 1.1mu}
\newcommand{\equaref}[1]{Eq.~(\ref{#1})}

\newcommand{\figref}[1]{Fig.~\ref{#1}}

\newcommand{\secref}[1]{Section~\ref{#1}}
\newcommand{\appref}[1]{Appendix~\ref{#1}}
\newcommand{\tabref}[1]{Table~\ref{#1}}

\newcommand{\bq}{\begin{eqnarray}}
\newcommand{\nq}{\end{eqnarray}}

\DeclareMathOperator{\Tr}{Tr}

\title{\bf Leptogenesis from Low Energy $CP$ Violation}

\author[a]{K. Moffat,}
\author[a]{S. Pascoli,}
\affiliation[a]{Institute for Particle Physics Phenomenology, Department of
Physics, Durham University, South Road, Durham DH1 3LE, United Kingdom.}
\author[b,c]{S.T. Petcov}
\affiliation[b]{SISSA/INFN, Via Bonomea 265, I-34136 Trieste, Italy.}
\affiliation[c]{Kavli IPMU (WPI), University of Tokyo, 5-1-5 Kashiwanoha, 277-8583 Kashiwa, Japan.}	
\author[d]{and J. Turner}
\affiliation[d]{Theoretical Physics Department, Fermi National Accelerator Laboratory, P.O. Box 500, Batavia, IL 60510, USA.}

\emailAdd{kristian.p.moffat@durham.ac.uk}
\emailAdd{silvia.pascoli@durham.ac.uk}
\emailAdd{petcov@sissa.it}
\emailAdd{jturner@fnal.gov}

\abstract{We revisit the possibility of producing the 
observed baryon asymmetry of the Universe via thermal leptogenesis,
where $CP$ violation comes exclusively from the low-energy phases of 
the neutrino mixing matrix. We  demonstrate the viability of 
thermal flavoured leptogenesis across seven orders of magnitude 
$\left(10^{6}<T \text{ (GeV)}< 10^{13}\right)$, using modern numerical machinery, where the lower bound can be reached only if flavour effects are taken into account and its value depends on the allowed degree of cancellation between the tree-level and radiative contributions to the light neutrino masses. At very high scales $\left( T \gg 10^{12} \text{ GeV} \right)$, we clarify that thermal leptogenesis is sensitive to the low-energy phases, in contradiction with what is usually claimed in the literature. In particular we demonstrate that Majorana-phase leptogenesis is in general viable while Dirac-phase leptogenesis requires some level of fine-tuning.}

\preprint{
\begin{flushleft} 
IPPP/18/79\\
SISSA  38/2018/FISI\\
IPMU18-0151\\
FERMILAB-PUB-18-382-T
\end{flushleft} 
}


\begin{document}

\thispagestyle{empty}
\def\thefootnote{\fnsymbol{footnote}}
\setcounter{footnote}{1}

\setcounter{page}{0}
\maketitle
\vspace{-1cm}
\flushbottom

\def\thefootnote{\arabic{footnote}}
\setcounter{footnote}{0}

\newpage
\section{Introduction}\label{sec:introduction}
The Standard Model (SM) cannot explain the observed baryon asymmetry 
of the Universe (BAU) in spite of qualitatively satisfying the 
Sakharov conditions.\footnote{We recall that these conditions require $C$-/$CP$ violation, baryon number violation 
and a departure from thermal equilibrium in the production 
of the asymmetry~\cite{Sakharov:1967dj}.} Similarly, physics beyond the SM is often invoked to  explain 
the existence of the non-zero neutrino masses~\cite{Fukuda:1998mi}. 
Leptogenesis is a mechanism by which some lepton-number-violating theories, which may also explain the origin of neutrino masses, produce a lepton asymmetry which is subsequently converted into a baryon asymmetry through the non-perturbative $(B+L)$-violating but $(B-L)$-conserving sphaleron processes of the SM~\cite{Kuzmin:1985mm,Khlebnikov:1988sr}.
A minimal implementation of leptogenesis occurs 
in the type~I seesaw framework in which a number of heavy Majorana 
neutrinos are added to the SM~\cite{Minkowski:1977sc,Yanagida:1979as,GellMann:1980vs,Mohapatra:1979ia}. The decay of these heavy Majorana 
neutrinos in leptons and Higgs bosons
is both lepton-number- and $CP$-violating and occurs out of thermal equilibrium, 
thereby satisfying the Sakharov conditions and potentially producing 
the observed baryon asymmetry~\cite{Fukugita:1986hr}
(see also, e.g., \cite{Buchmuller:1996pa,Buchmuller:2004nz,Buchmuller:2005eh,Pilaftsis:1997jf} 
and articles quoted therein). $CP$ violation is fundamental to the creation of the 
matter-antimatter asymmetry. In thermal leptogenesis, the decays of 
the heavy Majorana
neutrinos are $CP$-asymmetric and this results from 
both $CP$-violating low-scale measurable phases and high-scale 
immeasurable ones. In the original conception of leptogenesis, 
the flavour-dependent interactions due to charged lepton Yukawa couplings between the leptons and 
the early Universe plasma were disregarded. If leptogenesis occurs 
at high scales, where the temperature $T \gg 10^{12}$ GeV, then 
this approximation is ordinarily justified and a basis 
may be chosen in which essentially only one flavour of 
lepton ever appears in the theoretical description. Consequently,  
it was expected that the low-energy $CP$-violating phases 
contained in the neutrino mixing matrix play no physical 
role in the production of the lepton and therefore baryon asymmetry 
\cite{Buchmuller:2004nz,Buchmuller:2005eh}. 

 Under certain ad hoc assumptions the high-scale $CP$-violating 
phases can be related to the $CP$-violating phases in 
the Pontecorvo-Maki Nakagawa-Sakata (PMNS) matrix 
 (see, e.g., \cite{PDG2018}) 
which then participate in the production of the lepton asymmetry ---  a possibility that was investigated in \cite{Branco:2001pq,Branco:2002xf,Endoh:2002wm,Frampton:2002qc,Shimizu:2017vwi}. If  leptogenesis occurs at 
temperatures somewhat below $10^{12}$ GeV $\left( 10^{9} \text{ GeV} \right)$, the Yukawa 
interactions of the tau charged lepton (of the muon) come into 
thermal equilibrium, causing decoherence between this and the 
remaining flavour components of the charged lepton state ~\cite{Nardi:2005hs,Abada:2006ea,Abada:2006fw,Barbieri:1999ma,Nielsen:2002pc} such that two 
(three) lepton flavour states must be separately considered and 
the $CP$-violating phases of the PMNS matrix have physical significance. 
Historically, the possibility that the $CP$ violation in leptogenesis 
may be strictly due to Dirac and/or Majorana phases of the PMNS matrix 
was first apparent in this regime~\cite{Pascoli:2006ie,Pascoli:2006ci,Blanchet:2006be,Branco:2006ce,Anisimov:2007mw,Molinaro:2008rg,Molinaro:2008cw,Dolan:2018qpy}
(for a review see, e.g., \cite{Hagedorn:2017wjy}).

There have been other works which have investigated the impact of low 
energy phases on the BAU. Indeed, $CP$ conservation at the high-scale and 
$CP$ violation at the low-scale in the context of leptogenesis can 
be theoretically motivated by 
minimal flavour violation \cite{Branco:2006hz,Merlo:2018rin}, 
flavour symmetries \cite{Nishi:2018vlz,Meroni:2012ze,Mohapatra:2015gwa} 
or a generalised $CP$ symmetry \cite{Hagedorn:2016lva,Chen:2016ptr,Li:2017zmk}.
Beyond the type~I seesaw mechanism, there have been other 
 studies
which connect the Dirac phase, $\delta$, with the BAU using
an extended Higgs sector \cite{Kim:2016xyi}. 

The primary aims of this work are twofold:
\begin{itemize}

\item We revisit the possibility of producing the observed BAU where 
the only source of $CP$ violation comes from the phases of the 
PMNS matrix and calculate the lepton asymmetry generated across 
seven orders of magnitude ($10^{6}< T (\text{GeV})< 10^{13}$) using our 
modern numerical machinery which properly incorporates 
radiative corrections to the light neutrino masses.

\item We clarify a misunderstanding regarding the high-scale regime: 
as discussed above, it was initially thought that at very high scales, 
$T\sim10^{12}$ GeV, the low-energy $CP$ phases could not produce 
the observed BAU. This came from the observation that in the 
one-flavoured regime, and barring any relation between the high-scale 
and low-energy $CP$-violating phases, 
the $CP$ asymmetry is vanishing  $\left( \epsilon^{(1)}=0 \right)$ 
for $CP$ violation solely coming from low energy phases.
This led to the conclusion that in this case
no lepton asymmetry could be generated. 
However, the individual flavour contributions to the $CP$ asymmetries are not zero and, even in this scenario, where flavour effects are very weak, the washouts of lepton asymmetries are flavour dependent. We show that these flavour effects may be sufficient to produce the observed baryon asymmetry. We discuss this analytically and demonstrate the viability of leptogenesis in this scenario numerically. We demonstrate with Dirac-phase leptogenesis fine-tuned cancellations in the radiative expansion of the light neutrino masses are required, however with Majorana-phase leptogenesis this is not the case. This implies the $CP$ violation present in the low energy phases can, 
in principle, generate the observed matter antimatter asymmetry 
 between $10^{6}\lesssim T (\text{GeV}) \lesssim 10^{13}$.
\end{itemize}

The remainder of this work is structured as follows: 
in \secref{sec:framework} we present the relevant theoretical 
framework beginning with 
 \secref{sec:numassnadmix} where we briefly review neutrino masses, 
mixing and radiative corrections to the light neutrino mass matrix 
in the type~I seesaw. In \secref{subsec:CCP} we 
present the $C$ and $CP$ properties of the light and heavy Majorana 
neutrinos of the type I seesaw. This will be crucial
in understanding the textures of the neutrino Yukawa matrices 
which contain $CP$-violating low energy phases and
$CP$-conserving high energy phases. Subsequently, 
in \secref{sec:reviewintlepto} we discuss the allowed structures of 
the Yukawa matrix when there is a high-scale $CP$ symmetry. 
The kinetic equations and effects of flavour are discussed 
in \secref{subsec:flav}. In
 \secref{sec:typicalM1} we revisit the link between leptonic 
$CP$ violation at low energies and successful leptogenesis in 
the two-flavour regime (or more precisely when 
$10^9 \leq T \text{ (GeV) } \leq 10^{12})$. We bring a more sophisticated 
set of numerical tools to more convincingly answer the question than 
in earlier work on the subject~\cite{Pascoli:2006ie}. In \secref{sec:lowM1} 
we demonstrate that successful leptogenesis with only low energy 
$CP$ violation is possible at much lower scales than previously 
considered $\left( T \sim 10^6 \text{ GeV} \right)$, if one allows for fine-tuning 
in the light neutrino masses. In \secref{sec:HSLG} we establish 
that even if leptogenesis occurs at very high scales 
($T \gg 10^{12}$ GeV), then it is still possible for successful 
leptogenesis to result from purely low energy $CP$ violation. We find that no fine-tuning is necessary if Majorana phases play a role in the $CP$ violation. However, purely Dirac-phase $CP$ violation is sufficient only in scenarios with a certain degree of fine-tuning.
This latter case is in stark contrast with the conclusions 
of the current literature.

\section{Framework}
\label{sec:framework}
\subsection{Neutrino Masses and Mixing}
\label{sec:numassnadmix}

 Neutrino oscillation experiments  have provided compelling 
evidence that neutrinos have small but non-zero masses and mix 
 (for a review see, e.g., \cite{PDG2018}).
The mass and flavour states of neutrinos are misaligned, 
with this misalignment described by the PMNS matrix $U$:
\begin{equation}
\nu_{\alpha L} = \sum_{i=1}^3 U_{\alpha i} \nu_{i L},
\end{equation}
where $\alpha \in \{e,\mu,\tau\}$ is the flavour of the 
given neutrino flavour field, $\nu_{\alpha L}$, and $\nu_{iL}$ 
is the left-handed component of the $i$th massive neutrino. 
Throughout this work we employ the conventional 
PDG parametrisation~\cite{PDG2018}:
\begin{equation}
U =\begin{pmatrix}
c_{12}c_{13} & s_{12}c_{13} & s_{13} e^{-i \delta} \\
-s_{12}c_{23}-c_{12}s_{23}s_{13} e^{i \delta} & c_{12} c_{23} - s_{12} s_{23} s_{13} e^{i \delta} & s_{23} c_{13}  \\
s_{12}s_{23}-c_{12}c_{23}s_{13} e^{i \delta} & -c_{12}s_{23}-s_{12}c_{23}s_{13} e^{i \delta} & c_{23}c_{13} 
\end{pmatrix}
\begin{pmatrix}
1 & 0 & 0\\
0&e^{i\frac{\alpha_{21}}{2}} & 0\\
0 & 0 &  e^{i\frac{\alpha_{31}}{2}}
\end{pmatrix},
\end{equation}
where $c_{ij} \equiv \cos\theta_{ij}$, $s_{ij} \equiv \sin\theta_{ij}$, $\delta$ 
is the Dirac  phase and $\alpha_{21}$, $\alpha_{31}$ are the 
Majorana phases~\cite{Bilenky:1980cx} with best-fit values of $\theta_{12}$, $\theta_{23}$,  $\theta_{13}$ and $\delta$ given in 
\tabref{tab:table}. Neutrino oscillation experiments also allow to measure with precision the two independent neutrino mass squared differences  $\Delta m^2_{12}$ and 
 $\Delta m^2_{31}$ ($\Delta m^2_{32}$) which are given in \tabref{tab:table}. In order to accommodate them, the three neutrino masses $m_1$, $m_2$ $m_3$ can be arranged into two possible orderings, normal ordering (NO) for $m_1<m_2<m_3$ and inverted ordering (IO) $m_3<m_1<m_2$. The ordering is not yet known although data show some mild preference for normal ordering~\cite{Esteban:2016qun}. 
\begin{table}[t]
\centering
\begin{tabular}{ c  c  c  c  c  c  c }
 \toprule
$\theta_{13}$ & $\theta_{12}$ & $\theta_{23}$ & $\delta$ & $\Delta m_{21}^2$ & $\Delta m_{31}^2$ & $\Delta m_{32}^2$\\
$(^{\circ})$ & $(^{\circ})$ & $(^{\circ})$ & $(^{\circ})$ & ($10^{-5} \text{eV}{}^2$) & ($10^{-3} \text{eV}{}^2$) & ($10^{-3} \text{eV}{}^2$) \\
\specialrule{2.5pt}{1pt}{1pt}
$8.52^{+0.15}_{-0.15}$ & $33.63^{+0.78}_{-0.75}$ & $48.7^{+1.4}_{-6.9}$ & $228^{+51}_{-33}$ & $7.40^{+0.21}_{-0.20}$ & $2.515^{+0.035}_{-0.035}$ & $-2.483^{+0.034}_{-0.035}$ \\
\bottomrule
\end{tabular}\caption{Best fit and 1$\sigma$ ranges from a global fit to neutrino data \cite{Esteban:2016qun}. }\label{tab:table}
\end{table}

A simple means of explaining the smallness of neutrino masses 
is the type I seesaw framework~\cite{Minkowski:1977sc,Yanagida:1979as,GellMann:1980vs,Mohapatra:1979ia} in which heavy Majorana neutrinos are added to the SM particle spectrum. 
We shall work within a realisation of this framework which incorporates 
three heavy Majorana neutrinos, $N_i$ ($i \in \{1,2,3\}$), such that 
after electroweak symmetry breaking, when the Higgs has developed 
a vacuum expectation value (vev)  $v \approx 174 \text{ GeV}$, 
the neutrino mass terms of the Lagrangian are given by
\begin{equation}
\label{Mdefinition}
\begin{aligned}
\mathcal{L}_m & = -\frac{1}{2} \left(\bar{\nu}_L, \bar{N}^c_L \right) 
\left(\begin{array}{cc}
0 & v Y\\
v Y^T & M\\
\end{array}\right)
\left(\begin{array}{c}
\nu_R^c\\
N_R\\
\end{array}\right)
+\text{h.c.},
\end{aligned}
\end{equation}
where $\nu_L^T \equiv ( \nu_{eL}^T, \nu_{\mu L}^T, \nu_{\tau L}^T )$ 
($N_R^T \equiv \left( N_{1R}^T, N_{2 R}^T, N_{3 R}^T \right)$) 
are the light flavour (heavy mass eigenstates) neutrino fields and $Y$ is the 
neutrino Yukawa matrix which couples the heavy Majorana neutrinos $N_{1,2,3}$
to the leptonic and Higgs doublets. 
In \equaref{Mdefinition}
$(\nu_R^c)^T = ((\nu^c_{eR})^T, (\nu^c_{\mu R})^T, (\nu^c_{\tau R})^T))$, 
$(N^c_L)^T = \left( (N^c_{1L})^T, (N^c_{2 L})^T, (N^c_{3 L})^T \right)$,
with  $\nu^c_{lR} = C(\bar{\nu}_{lL})^T$, $l \in \{e,\mu,\tau\}$, 
and  $N^c_{j L} = C (\bar{N}_{jR})^T$ for $j \in \{1,2,3\}$. Finally, 
$C$ denotes charge conjugation matrix.

The mass terms in \equaref{Mdefinition}
are written in the basis where  the charged lepton 
Yukawa couplings are flavour diagonal, which we use throughout the present study.
We shall also work, without loss of generality, 
in a basis in which $M$ is diagonal and positive.

  To first order in the seesaw expansion, 
the tree-level light neutrino mass matrix is given by
\begin{equation}
m^{\text{tree}}_{\nu} = v^2 Y M^{-1} Y^T,
\end{equation}
where we employ the sign conventions of~\cite{Pascoli:2006ie}.
In the generic type I seesaw, there is no symmetry protecting the tree-level neutrino masses from radiative corrections. Moreover, the one-loop radiative corrections may be comparable to the tree-level mass. For this reason, when exploring the parameter space of type I seesaw models, it is relevant to compute the additional contribution to the light masses arising from the one-loop self-energy\footnote{
 In this expression terms of 
order $m^{\text{tree}}_{\nu} \frac{Y^2}{32 \pi^2}$ have been neglected.
This is equivalent to neglecting the tree-level mass when 
computing the one-loop contribution. 
Thus,  the physically irrelevant
divergent pieces and renormalisation-scale dependent 
pieces are not present in the expression we give.}~
\cite{Pilaftsis:1991ug,AristizabalSierra:2011mn,LopezPavon:2012zg} 
(see also, e.g., \cite{Grimus:2002nk}):
\begin{equation}\label{eq:lightmass}
m^{\text{1-loop}}_{\nu} = - Y \frac{M}{32 \pi^2} \left(\frac{\log\left(\frac{M^2}{m_H^2}\right)}{\frac{M^2}{m_H^2}-1} + 3 \frac{\log\left(\frac{M^2}{m_Z^2}\right)}{\frac{M^2}{m_Z^2}-1}\right) Y^T,
\end{equation}
where $m_Z$ and $m_H$ are the Z and Higgs boson masses respectively. 
With these included, the light neutrino mass matrix is given by
\begin{equation}\label{eq:mneutrino}
\begin{aligned}
m^{}_{\nu} & = m^{\text{tree}}_{\nu} + m^{\text{1-loop}}_{\nu} \\ & = v^2 Y f\left(M\right) Y^T,
\\ & = m_D f\left(M\right) m_D^T,
\end{aligned}
\end{equation}
where $m_D \equiv v Y$ and
\begin{equation}\label{eq:fdef}
f(M) \equiv M^{-1} - \frac{M}{32 \pi^2 v^2} \left(\frac{\log\left(\frac{M^2}{m_H^2}\right)}{\frac{M^2}{m_H^2}-1} + 3 \frac{\log\left(\frac{M^2}{m_Z^2}\right)}{\frac{M^2}{m_Z^2}-1}\right).
\end{equation}

From the structure of  \equaref{eq:lightmass},  we observe that the tree 
and one-loop level contributions to the light neutrino masses
are both quadratic in the Yukawa couplings. The only suppression  
of the one-loop derives from the $\mathcal{O}\left(10^{-2}\right)$ 
loop-factor. Furthermore, from \equaref{eq:fdef}
it follows that  
the one-loop contribution generically tends to reduce the 
tree-level mass. This suggests that there exists the possibility 
of a cancellation between the tree and one-loop contributions. In this work 
we do not preclude the possibility that these cancellations may be present. 
The higher-order corrections to the light neutrino masses are generally 
suppressed by additional loop factors and couplings and generally do 
not experience fine-tuned cancellations, instead they contribute in 
the usual way to the perturbative expansion. We define a fine-tuning measure, 
$\mathcal{F}$, to quantify the level of cancellation:
\begin{equation}
\mathcal{F} \equiv \frac{m_{\nu}^{\text{1-loop}}}{m^{\text{tree}}_{\nu} + m^{\text{1-loop}}_{\nu}}.
\end{equation}

The one-loop correct light neutrino mass matrix of 
\equaref{eq:mneutrino}, $m_{\nu}$, may be transformed into a positive 
diagonal form (denoted by a caret) using the Takagi factorisation
\begin{equation}
\hat{m}_{\nu} = U^{\dagger} m_{\nu} U^*,
\end{equation}
such that $\hat{m}_{\nu} = \text{diag}(m_1,m_2,m_3)$, with $m_i$ 
the mass corresponding to $\nu_i$.
By analogy with the method of Casas and Ibarra~\cite{Casas:2001sr}, 
we parametrise the Yukawa matrix to include the relevant radiative 
corrections as~\cite{Lopez-Pavon:2015cga}
\begin{equation}
\label{eq:CIloop}
Y=\frac{1}{v}U\sqrt{\hat{m}_{\nu}}R^T\sqrt{f(M)^{-1}},
\end{equation}
where $R$ is a $3 \times 3$ complex orthogonal matrix. 
Explicitly, we choose to work with the parametrisation\footnote{A phase factor $\xi = \pm 1$ could have been included in the definition of $R$ to allow for both the cases $\det(R)=\pm 1$ however we have chosen to extend the range of the Majorana phases such that the choice of signs of $\det(R)$ have effectively already been incorporated.}:
\begin{equation}
\label{eq:Rparametrisation}
R=\begin{pmatrix}
1 & 0 & 0 \\
0 & c_{1} & s_{1} \\
0 &- s_{1} & c_{1} 
\end{pmatrix}
\begin{pmatrix}
c_{2} & 0 & s_{2} \\
0 & 1 & 0\\
-s_{2} & 0 & c_{2} 
\end{pmatrix}\\
\begin{pmatrix}
c_{3} & s_{3} & 0\\
-s_{3} & c_{3} & 0\\
0 & 0 & 1
\end{pmatrix},
\end{equation}
where $c_{i}=\cos w_{i}$, $s_{i}=\sin w_{i}$ and the complex angles 
are given by $w_{i}=x_{i}+iy_{i}$ ($i \in \{1, 2, 3\}$).

\subsection{$C$ and $CP$ Properties of Majorana Neutrinos}\label{subsec:CCP}

As we focus on the possibility that low-scale $CP$ phases are responsible 
for the BAU, the $C$ and $CP$ properties of neutrinos
will be  crucial in  understanding the structure of the $R$-matrix
 which results in 
the $CP$ conservation of the high-scale phases. 
In the type I seesaw, the light ($\nu_i$) and the heavy ($N_i$) neutrino 
mass states are both Majorana in nature and thus satisfy 
the following conditions:
\begin{equation}
\begin{aligned}
C \overline{\nu}_i^T & = \nu_i,\\
C \overline{N}_i^T & = N_i,
\end{aligned}
\end{equation}
where $C$  denotes the charge conjugation matrix. 

Following~\cite{Pascoli:2006ci}, we express the $CP$-conjugated 
neutrino fields in terms of the $CP$ operator $U_{CP}$ as
\begin{equation}
\begin{aligned}
U^{}_{CP} N_i\left(x\right) U_{CP}^{\dagger} & = i \rho^N_i N_i\left(x'\right), \\
U^{}_{CP} \nu_i\left(x\right) U_{CP}^{\dagger} & = i \rho^{\nu}_i \nu_i\left(x'\right),
\end{aligned}
\end{equation}
 where $x'$ is the parity-transformed coordinate and 
$i\rho^{N}_i = \pm i$ and  $i\rho^{\nu}_i = \pm i$ are  
the $CP$ parities of the respective Majorana fields. 
The conditions for $CP$ invariance impose the 
following restrictions on the elements of 
the matrix of neutrino Yukawa couplings (setting the unphysical phases in 
the $CP$ transformations of the lepton and Higgs doublets 
to unity) is given by,
\begin{equation}
\label{eq:YCPconditions}
Y_{\alpha i}^{*} = Y_{\alpha i} \rho^{N}_i\,,
\end{equation}
and on the elements of the PMNS matrix \cite{Bilenky:1987ty}:
\begin{equation}
\label{eq:UCPconditions}
U^*_{\alpha j} = U_{\alpha j} \rho^{\nu}_j\,,~j \in \{1,2,3\},~
\alpha \in \{e,\mu,\tau\}\,.
\end{equation}
From  the parametrisation of the Yukawa matrix of \equaref{eq:CIloop}, 
this imposes the following conditions on the elements of the
$R$-matrix~\cite{Pascoli:2006ci}:
\begin{equation}
\label{eq:UCPRCPconditions}
 R^*_{ij} = R_{ij} \rho^{N}_i \rho^{\nu}_j\,,~~i,j \in \{1,2,3\}\,.
\end{equation}

The leptogenesis scenarios considered in this work have 
$CP$ violation provided only by the phases of the PMNS matrix. 
 This corresponds to imposing the condition 
of \equaref{eq:UCPRCPconditions} onto the $R$-matrix 
but not the condition \equaref{eq:UCPconditions}
on $U$. In these scenarios the values of the Dirac and 
Majorana phases of the PMNS matrix determine 
the success of leptogenesis.
One should bear in mind, however,
that there are certain intuitively 
unexpected possibilities for $CP$ violation in 
(non-resonant) leptogenesis 
even when the PMNS- and $R$-matrices are
$CP$-conserving, i.e., conditions 
(\equaref{eq:UCPconditions}) and 
(\equaref{eq:UCPRCPconditions})   
are individually fulfilled and 
the elements  $U_{lj}$ and $R_{jk}$
are real or purely imaginary ~\cite{Pascoli:2006ci}.\footnote{This unusual possibility is realised when  
$\rho^{N}_i$ and $\rho^{\nu}_j$ are fixed by conditions
(\equaref{eq:YCPconditions}) and (\equaref{eq:UCPconditions}),
but the product of the so fixed values of 
$\rho^{N}_i$ and $\rho^{\nu}_j$ differs from the 
value of $\rho^{N}_i \rho^{\nu}_j$ in
(\equaref{eq:UCPRCPconditions}) \cite{Pascoli:2006ci}. 
Under these conditions the low energy PMNS matrix $U$ 
and the high-scale $R$-matrix are individually 
$CP$-conserving, but the interplay between the two 
in leptogenesis is $CP$-violating.
}

 $CP$ violation due to the Dirac phase $\delta$ can only be 
practically investigated in neutrino oscillation experiments. 
There has been a slight statistical preference from 
the existing
data for  maximally $CP$-violating $\delta\sim 270^\circ$.
This hint has been obtained 
from the combination of results from long-baseline experiments 
such as T2K \cite{Abe:2011ks} and NO$\nu$A \cite{Ayres:2004js} 
with reactor experiments like Daya-Bay \cite{An:2012eh}, 
RENO \cite{Ahn:2012nd} and Double-Chooz \cite{Ardellier:2006mn}. 
In principle, the difference in oscillation 
probabilities~\cite{CABIBBO1978333,BILENKY1980495,Barger:1980jm},

\begin{equation}
A^{\alpha,\beta}_{CP} \equiv P(\nu_{\alpha} \rightarrow \nu_{\beta}) - P(\overline{\nu}_{\alpha} \rightarrow \overline{\nu}_{\beta}) \quad (\alpha \neq \beta),
\end{equation}
is a measure of $CP$ violation in neutrino oscillations in vacuum and 
can be measured experimentally. For vacuum oscillations in the 
three-neutrino case we have \cite{Krastev:1988yu}
\begin{equation}
A^{e,\mu}_{CP} = 4 J^{}_{CP} F^{\text{vac}}_{\text{osc}},
\end{equation}
\begin{equation}
F^{\text{vac}}_{\text{osc}} \equiv \sin \left( \frac{\Delta m^2_{21}}{2E} L  \right)+\sin \left( \frac{\Delta m^2_{32}}{2E} L  \right)+\sin \left( \frac{\Delta m^2_{13}}{2E} L  \right),
\end{equation}
\begin{equation}
J_{CP} \equiv \Im \left[ U_{e1} U_{\mu 2} U_{e2}^* U_{\mu 1}^*  \right].
\end{equation}
 $J_{CP}$ is the analogue of the Jarlskog invariant
for the lepton sector, which gives a parametrisation-independent 
measure of $CP$ violation in neutrino oscillations,
$L$ is the distance travelled by the neutrinos,
$E$ the neutrino energy and $\Delta m^2_{ij} \equiv m^2_i - m^2_j$. 
In the case of $CP$-invariance we have $\delta = 0,\pi$ and therefore $J_{CP} = 0$. 
By measuring, for example, $A^{e, \mu}_{CP}$, one can determine 
$J_{CP}$
which has the following 
expression
in the standard parametrisation of the PMNS matrix:
\begin{equation}
J_{CP} = \frac{1}{4} \sin 2\theta_{12} \sin 2\theta_{23} \cos^2 \theta_{13} \sin \theta_{13} \sin \delta\,.
\end{equation}
The best-fit value and $1\sigma$ uncertainty of  $J_{CP}$ 
reported in ~\cite{Esteban:2016qun} are
\begin{equation}
J^{\text{max}}_{CP} = 0.0329 \pm 0.0007 \, (\pm 1\sigma).
\end{equation}
 In the longer term,  the next generation of neutrino oscillation 
experiments such as DUNE \cite{DUNE} and T2HK \cite{T2HK}, will be able 
to measure the Dirac $CP$-violating phase $\delta$ with greater precision 
and determine whether $CP$-symmetry is indeed violated 
in the lepton sector.

 Information on $CP$-violating Majorana phases can, in principle, 
be obtained in neutrinoless double beta decay experiments 
~\cite{Bilenky:2001rz,Pascoli:2001by,Pascoli:2005zb}
(see, however, also \cite{Barger:2002vy}).
These experiments  are the most sensitive probes of the possible 
Majorana nature of massive neutrinos. They can also provide information 
on the neutrino mass ordering  
\cite{Pascoli:2002xq} (see also \cite{Pascoli:2005zb}).
 The rate of neutrinoless double beta decay 
is given by (see, e.g.,~\cite{Blennow:2010th})
\begin{equation}
\frac{\Gamma_{0\nu\beta\beta}}{\log 2} = \frac{G_{01}}{m^2_e} \lvert \mathcal{A} \rvert^2,
\end{equation}
where $G_{01}$ is a kinematic factor and $\mathcal{A}$ denotes 
the amplitude which has the following form
\begin{equation}
\label{eq:0vbb}
\mathcal{A} \propto \sum_{i=1}^3 m_i U_{ei}^2 \mathcal{M}^{0\nu\beta\beta}(m_i) + \sum_{i=1}^3 M_i V_{ei}^2 \mathcal{M}^{0\nu\beta\beta}(M_i).
\end{equation}
The amplitude is dependent on the nuclear matrix elements 
$\mathcal{M}^{0\nu\beta\beta}$ for which 
$\mathcal{M}^{0\nu\beta\beta}(m_i) \approx \mathcal{M}^{0\nu\beta\beta}(0) \gg \mathcal{M}^{0\nu\beta\beta}(M_i)$ if $M_i \gg 10^{3} \text{ MeV}$ 
(see, e.g., ~\cite{Blennow:2010th,Vergados:2016hso}), 
which shall always be the case in this work. 
The mixing elements $V_{ei}$ for the heavy states are 
$\mathcal{O}\left(m_D/M\right)$ and thus the second term of 
\equaref{eq:0vbb} is 
$\mathcal{O}(m_D^2/M) \mathcal{M}^{0\nu\beta\beta}(M_i) 
\sim \mathcal{O}(m_i) \mathcal{M}^{0\nu\beta\beta}(M_i)$. 
As $U_{ei} \sim \mathcal{O}(1)$, the second term is negligible in comparison 
with the first and we find~\cite{DOI1981323} 
(see e.g., \cite{Bilenky:1987ty}):
\begin{equation}
A \propto \langle m_{\nu} \rangle \equiv  m_1 U_{e1}^2 + m_2 \lvert U_{e2} \rvert ^2 e^{i\alpha_{21}} + m_3 \lvert U_{e3} \rvert^2 e^{i(\alpha_{31} - 2\delta)},
\end{equation}
where $\langle m_{\nu} \rangle$ is the neutrinoless double beta decay 
effective Majorana mass in the case of 3-neutrino mixing. 
In the case of $CP$-invariance we have 
$\alpha_{21} = k\pi$, $\alpha_{31}=q\pi$, $k,q=0,1,2,\hdots$ \cite{Wolfenstein:1981rk,Bilenky:1984fg,Kayser:1984ge}.\footnote{
Thus, in order for a value of $\alpha_{21(31)}$ to be 
$CP$-violating both $\sin \frac{\alpha_{21(31)}}{2}$ and $\cos \frac{\alpha_{21(31)}}{2}$ 
at this value should be different from zero.
}
The most stringent upper  bound on 
$\lvert \langle m_{\nu} \rangle \rvert$ 
was reported by the KamLAND-Zen collaboration~\cite{KamLAND-Zen:2016pfg} 
searching for  neutrinoless double beta decay of  $^{136}$Xe:
\begin{equation}
\lvert \langle m_{\nu} \rangle \rvert < (0.061\text{ -- } 0.165) \text{ eV},
\end{equation}
 where the uncertainty in the knowledge of the 
nuclear matrix element of  $^{136}$Xe decay have been accounted for.
In terms of the half-lives for neutrinoless double beta decay the best lower limits are: for germanium-76,  tellurium-130,  and  xenon-136: $T^{0 \nu}_{1/2} > 8.0 \times 10^{25}$~yr (reported  by  the  GERDA-II collaboration), $T^{0 \nu}_{1/2} > 1.5 \times 10^{25}$~yr (from  the combined results of the  Cuoricino, CUORE-0, and CUORE experiments),  and $T^{0 \nu}_{1/2} > 1.07 \times 10^{26}$~yr  (from  the  KamLAND-Zen collaboration), with all limits given at the 90\% CL. Most importantly, a large number of 
experiments of a new generation aim at
sensitivities to 
$\lvert \langle m_{\nu} \rangle \rvert \sim (0.01\div 0.05)$ eV (see, e.g., 
\cite{Vergados:2016hso,DellOro:2016tmg}): 
CUORE ($^{130}$Te), SNO+ ($^{130}\text{Te}$),
GERDA ($^{76}$Ge), MAJORANA ($^{76}$Ge), LEGEND ($^{76}$Ge),
SuperNEMO ($^{82}$Se, $^{150}$Nd), 
 KamLAND-Zen ($^{136}$Xe), 
EXO  and nEXO ($^{136}$Xe), 
PANDAX-III  ($^{136}$Xe),
NEXT ($^{136}$Xe), 
AMoRE ($^{100}$Mo), 
MOON ($^{100}$Mo),  
CANDLES ($^{48}$Ca), 
XMASS  ($^{136}$Xe),
DCBA  ($^{82}$Se, $^{150}$Nd),
ZICOS  ($^{96}$Zr), etc.
The GERDA-II and KamLAND-Zen 
experiments have already provided  
the best lower limits on the double beta decay
half-lives of $^{76}$Ge and $^{136}$Xe. 
The experiments listed above
aim to probe the ranges of predictions of  
$\lvert \langle m_{\nu} \rangle \rvert$
corresponding to neutrino mass spectra  
of quasi-degenerate type and with inverted 
ordering (see, e.g., \cite{PDG2018}).

The primary focus of this work is to answer the question: 
\emph{at what scales can low-energy $CP$-violating phases produce 
the observed BAU?} We shall show that the scale of successful 
leptogenesis in the case of interest may indeed vary across many orders 
of magnitude from $10^6-10^{13}$ GeV.  The observation of low-scale  
leptonic Dirac $CP$ violation, in combination with the positive 
determination of the Majorana nature of the massive neutrinos, 
would make more plausible, but will not be a proof of, 
the existence of high-scale thermal leptogenesis. 
These remarkable discoveries would indicate, 
in particular, that thermal leptogenesis \emph{could} produce the BAU 
with the requisite $CP$ violation provided by the 
Dirac $CP$-violating phase in the neutrino mixing matrix.

\subsection{$CP$-Conserving $R$-Matrix  
and the Structure of the Light Neutrino Mass Matrix}\label{sec:reviewintlepto} 

If the orthogonal matrix $R$ is allowed to have large elements, then the scale of leptogenesis may be lowered to $M_1 \sim 10^{6}$ GeV~\cite{Blanchet:2008pw,Antusch:2009gn,Moffat:2018wke}. In such scenarios, care must be taken with the radiative corrections to the light neutrino masses which may grow large (and non-negligible) with the elements of the $R$-matrix. One can either impose a near-lepton-number-symmetry to prevent this (see~\cite{Antusch:2009gn}), or more generically, incorporate the one-loop contribution to the light neutrino masses (in the manner we have discussed) and remain agnostic about fine-tuned cancellations between the tree-level and one-loop contributions. We proceed with this approach following the attitude taken in~\cite{Moffat:2018wke}, in which the figure  $M_1 \sim 10^{6}$ GeV was first demonstrated.
\begin{equation}
\label{eq:RStructure}
R \approx
\left(
\begin{array}{ccc}
 R_{11} & R_{12} & R_{13} \\
 \pm i R_{22} & R_{22} & R_{23} \\
 -R_{22} & \pm i R_{22} & \pm i R_{23} \\
\end{array}
\right),
\end{equation}
$|R_{22}| \gg |R_{1i}|, |R_{23}|$ for $i \in \{1,2,3\}$. 
The cancellation of large tree-level and large one-loop light 
neutrino mass matrices occurs as a result of relations between 
the magnitudes and phases of the $R$-matrix elements 
which lead to the following structure for the Dirac mass matrix:
\begin{equation}
\label{eq:ftDiracmass}
m_D \sqrt{f} =\left(
\begin{array}{ccc}
 \Delta, & u, &  \pm i u \\
\end{array}
\right),
\end{equation}
with $\Delta = U \left( \sqrt{m_1} R_{11}, \sqrt{m_2} R_{12}, 
\sqrt{m_3} R_{13} \right)^T$ and 
$u = U \left( \pm i \sqrt{m_1} R_{22}, \sqrt{m_2} R_{22}, \sqrt{m_3} R_{23} \right)^T$, such that 
$\lvert \Delta_i \rvert \ll \lvert u_j \rvert$,  $i, j \in \{1,2,3\}$.
We may rewrite the tree and one-loop masses in terms of this 
relatively simple matrix $m_D \sqrt{f}$, such that
\begin{equation}
m^{\text{tree}} = 
\left(m_D \sqrt{f} \right) M^{-1} f^{-1} \left(m_D \sqrt{f} \right)^T,
\end{equation}
where  the commutativity of the diagonal matrices 
$M$ and $f$ has been exploited and
\begin{equation}
m^{\text{1-loop}} 
= \left(m_D \sqrt{f} \right) 
\left(f-M^{-1}\right) f^{-1} \left(m_D \sqrt{f} \right)^T.
\end{equation}
This ensures that the sum of the tree-level and one-loop masses is
\begin{equation}
\begin{aligned}
m_{\nu} & = m_D \sqrt{f} \left(m_D \sqrt{f}\right)^T \\
& = \Delta \Delta^T.
\end{aligned}
\end{equation}
Due to the relative smallness of the elements of 
$\Delta$, the matrix $m_{\nu}$ may be considerably smaller than $m^{\text{tree}}$. Immediately, we have
\begin{equation}
m^{\text{tree}} = - m^{\text{1-loop}} + \mathcal{O}(\Delta^2),
\end{equation}
which is an explicit expression of the fine-tuned cancellation.

As the $R$-matrix structure of \equaref{eq:RStructure} is required 
for successful leptogenesis at intermediate scales, we are 
tasked with the problem of finding the $R$-matrices which assume 
this form and obey the $CP$-invariance conditions of 
\equaref{eq:UCPRCPconditions}. We intend to translate the conditions 
in \equaref{eq:RStructure} and \equaref{eq:UCPRCPconditions} 
into constraints on $x_i$ and $y_i$. However, we know 
\textit{a priori} from the work of~\cite{Moffat:2018wke} that one
must have 
$y_2 \sim 0^\circ$ and $y_1 \gtrsim 180^\circ$, $y_3 \gtrsim 180^\circ$ 
to produce the relative magnitudes of the elements of $R$ in 
\equaref{eq:RStructure}, crucial to the successful production 
of the observed baryon asymmetry.

We begin with the elements
\begin{equation}
R_{22} = \cos w_1 \cos w_3 -\sin w_1 \sin w_2 \sin w_3,
\end{equation}
and
\begin{equation}
R_{31} = -\cos w_1 \cos w_3\sin w_2+\sin w_1 \sin w_3,
\end{equation}
which result from the expansion of the $R$-matrix parametrised 
as in \equaref{eq:Rparametrisation}. The condition of 
\equaref{eq:RStructure} that $R_{22} \approx -R_{31}$ implies that 
$\sin w_2 \approx 1$, which in turn imposes 
$\sin x_2  \approx 1$ and $y_2 \approx 0^\circ$. In order to 
simplify future expressions, we promote the condition on $y_2$ 
to the exact equality $y_2=0^\circ$. 
With conditions on $x_2$ and $y_2$ determined, we now examine
\begin{equation}
R_{13} = \cos x_2 \left( \cos x_1 \cosh y_1-i \sin x_1 \sinh y_1 \right).
\end{equation}
According to the condition \equaref{eq:UCPRCPconditions}, 
$R_{13}$ (like all the elements of $R$) must be purely real or 
imaginary and thus we should choose one of, 
$\cos x_1 = 0$ or $\sin x_1 = 0$. We exclude the possibility 
of $y_1 = 0$ for the reason given above. 
Likewise, consider
\begin{equation}
R_{11} = \cos x_2 \left(\cos x_3 \cosh y_3-i \sin x_3 \sinh y_3 \right),
\end{equation}
and select $\cos x_3=0$ or $\sin x_3 = 0$ by the same reasoning.

In summary, we have the following set of constraints
\begin{equation}
\label{eq:RCPIoptions}
\begin{aligned}
      \cos x_2  \approx 0 & \text{ and } y_2 = 0,\\
      |\cos x_1| & = 0 \text{ or } 1, \\
      |\cos x_3| & = 0 \text{ or } 1, 
\end{aligned}
\end{equation}
which lead to an $R$-matrix of purely real and imaginary 
components and are therefore good candidates for $CP$-invariant 
$R$-matrices. We shall make use of these conditions in considerations 
where enhancement of the $R$-matrix is necessary for successful leptogenesis.

\subsection{The Effects of Flavour and Scale in Leptogenesis}
\label{subsec:flav}

In addition to explaining the smallness of neutrino masses, the 
type I seesaw provides a framework under which the 
matter-antimatter asymmetry of the Universe is explicable. 
The heavy Majorana neutrinos $N_i$ may undergo out-of-equilibrium, 
$C$-/$CP$- and lepton-number-violating decays in the early Universe. 
The resulting leptonic matter-antimatter asymmetry is then partially 
converted into a baryonic asymmetry by SM sphaleron processes 
which violate $B+L$ 
 but conserve  $B-L$.
The baryon asymmetry, which quantifies the excess of matter 
over antimatter in the Universe, is defined by
\begin{equation}
\eta_B \equiv \frac{n_B-n_{\overbar{B}}}{n_\gamma},
\end{equation}
where $n_B$, $n_{\overline{B}}$ and $n_\gamma$ are the number 
densities of baryons, antibaryons and photons respectively.
This quantity has been measured using two independent methods. 
There is the measurement of the baryon-to-photon ratio 
from Big-Bang nucleosynthesis (BBN), a process which occurs 
when the temperature of the Universe drops below 
$T \lesssim 1 \text{ MeV}$~\cite{Patrignani:2016xqp}:
\[
{\eta_{B}}_{\text{BBN}}  = \left(5.80-6.60\right)\times 10^{-10}.
\]
In complement, there is the determination of $\eta_B$ from 
Cosmic Microwave Background radiation (CMB) 
data~\cite{Ade:2015xua} for which the relevant cosmological 
period is that of recombination, for which $T \lesssim 1 \text{ eV}$:
\[
{\eta_{B}} _{\text{CMB}} = \left(6.02-6.18\right)\times 10^{-10}.
\]
Throughout our numerical study we apply the latter, more precisely 
measured value. 

In the simplest scenario, thermal leptogenesis 
describes the time evolution of a lepton asymmetry as a result of 
the $CP$-violating decays of the heavy Majorana neutrinos. 
In these processes, the lepton and anti-lepton states are
\begin{equation}
| i \rangle \equiv \sum_{\alpha} C_{i \alpha} | {\alpha} \rangle, \quad | \overline{i} \rangle \equiv \sum_{\alpha} \overline{C}_{i \alpha} | \overline{\alpha} \rangle, 
\end{equation}
where, at tree-level, the projection coefficients are expressed as
\begin{equation}
C_{i \alpha} = \overline{C}_{i \alpha} = \frac{Y_{\alpha i}}{\sqrt{\left(Y^{\dagger}Y\right)_{ii}}},
\end{equation}
for $i \in \{1,2,3\}$ and $\alpha \in \{e, \mu, \tau\}$. 
The simplest scenario for leptogenesis is the 
\textit{single flavour} regime, in which, the leptons resulting 
from the decay of each  heavy Majorana neutrino $N_i$ are always 
found in the coherent superposition of flavours described by 
the corresponding $| i \rangle$. This condition is valid if 
flavour-dependent interactions mediated by the SM charged 
lepton Yukawa couplings are negligible. This is usually a sufficiently 
good approximation for temperatures $T \gg 10^{12}$ GeV, when the 
charged lepton Yukawa interactions proceed at a slower rate 
than the  expansion of the Universe. However, in this work, 
we shall demonstrate that such an approximation fails in 
certain regions of the model parameter space.

The single flavoured Boltzmann equations for thermal leptogenesis 
provide a semi-classical description of the time evolution of 
the heavy neutrino densities, $n_{N_i}$ ($i \in \{1,2,3\}$), 
and the lepton asymmetry, $n_{B-L}$.\footnote{All number densities are normalised to a volume containing 
a single heavy Majorana neutrino in ultra-relativistic thermal equilibrium.}
Introducing the parameter $z \equiv M_1/T$, which increases monotonically 
with time, these kinetic equations are written
\begin{equation}
\label{eq:BE1F}
\begin{aligned}
\frac{dn_{N_{i}}}{dz}=&-D_{i}(n^{}_{N_{i}}-n^\text{eq}_{N_{i}}),\\
 \frac{dn_{B-L}}{dz} =&\sum_{i=1}^{3} \left( \epsilon^{(i)} D_{i}(n^{}_{N_{i}}-n^\text{eq}_{N_{i}})-W_{i}n_{B-L} \right).
\end{aligned}
\end{equation}
In general, the decay parameter $D_i$, describing the decay of 
$N_i$ is defined in terms of the heavy neutrino decay rate 
$\Gamma_i \equiv \Gamma_i \left(N_i \rightarrow \phi^{\dagger} l_i \right)$ 
(with $\phi$ and $l_i$ the Higgs and lepton doublets), 
the $CP$-conjugate rate, $\overline{\Gamma}_i$, 
and Hubble rate, $H$~\cite{Buchmuller:2004nz}:
\begin{equation}
D_i \equiv \frac{\Gamma_i+\overline{\Gamma}_i}{Hz}.
\end{equation}
Likewise, the washout factor is defined in terms of the heavy 
neutrino inverse decay rate $\Gamma^{\text{ID}}_i$ and the $CP$-conjugate 
inverse decay rates $\overline{\Gamma}^{\text{ID}}_i$
\begin{equation}
W_i \equiv \frac{1}{2}
\frac{\Gamma_i^{\text{ID}}+\overline{\Gamma}_i^{\text{ID}}}{Hz}.
\end{equation}
Finally, the $CP$-asymmetry parameter $\epsilon^{(i)}$ is defined as\footnote{Note that the Yukawas enter only in the combination 
$Y^{\dagger}Y$ and hence, from the \equaref{eq:CIloop}, there is no 
dependence on the PMNS matrix, $U$. Thus, in the one-flavour case, 
there can be no contribution to the $CP$-asymmetry from the Dirac 
and Majorana phases.}
\begin{equation}
\begin{aligned}
\epsilon^{(i)} & \equiv -\frac{\Gamma_i-\overline{\Gamma}_i}{\Gamma_i+\overline{\Gamma}_i} = -\frac{3}{16 \pi \left(Y^{\dagger}Y\right)_{ii}} \sum_{j \neq i} \Im \left[ \left(Y^{\dagger}Y\right)^2_{ij} \right] \frac{\xi \left(x_j/x_i\right)}{\sqrt{x_j/x_i}},
\end{aligned}
\end{equation}
with
\begin{equation}
x_{i} \equiv M^2_{i}/M^2_{1}, \quad \xi\left(x\right) \equiv \frac{2}{3}x\left[ \left(1+x\right)\log\left( \frac{1+x}{x}  \right) -\frac{2-x}{1-x}   \right].
\end{equation}
As discussed, the light neutrino masses may have an 
accidental cancellation in the tree-level mass that makes 
it comparable to the one-loop mass. However, there is no reason 
to expect an accidental cancellation like this to occur in the $CP$-asymmetry. 
This is because the cancellation in the light neutrino masses was due 
to the presence of terms of the form $Y M^{-1} Y^T$ while in 
the $CP$-asymmetries, the Yukawa matrix is multiplied by their 
conjugates in combinations of $Y^{\dagger} Y$ and therefore a 
similar cancellation cannot occur. As such, the higher-order corrections 
to the $CP$-asymmetry should not be any more significant in our 
case than they usually are.

\equaref{eq:BE1F} describes the time evolution of lepton number density asymmetry, $n_{B-L}$, from an initial value 
(usually it is assumed that there is a vanishing initial abundance) 
to a final value, $n_{B-L}\left(z_{\text{final}}\right)$. The final leptonic 
asymmetry is then partially converted into a baryonic asymmetry 
through sphaleron processes. This is expressed quantitatively 
by the relation $\eta_B \approx a/f \times n_{B-L} \approx 10^{-2} n_{B-L}$ 
\cite{Buchmuller:2004nz}, where $a=28/79$ describes the partial 
conversion of the $B-L$ asymmetry into a baryon asymmetry by 
sphaleron processes, and 
$f \equiv n^{\text{rec}}_{\gamma}/n^{*}_{\gamma}=2387/86$ accounts 
for the dilution of the asymmetry due the change of 
photon densities ($n_{\gamma}$) between leptogenesis 
($n_{\gamma}=n^{*}_{\gamma}$) and recombination ($n_{\gamma}=n^{\text{rec}}_{\gamma}$).

\begin{figure}[t]
\centering
\includegraphics[width=0.6\textwidth]{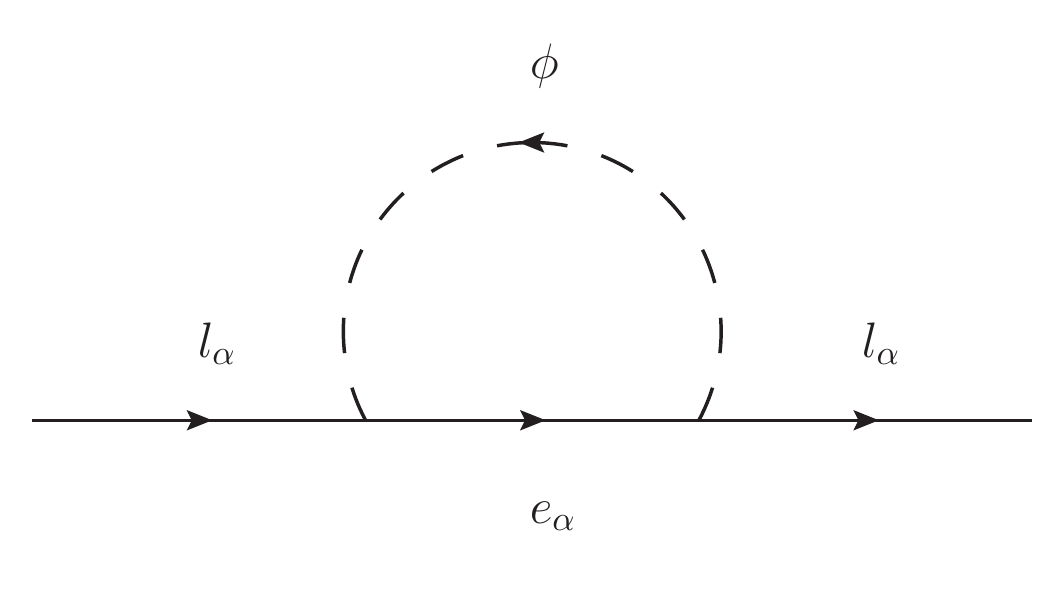}
\caption{The one-loop contribution to the thermal-width for left-handed leptonic doublet of flavour $\alpha$ ($l_{\alpha}$) through the right-handed singlet of the same flavour ($e_{\alpha}$).}
\label{fig:thermalwidth}
\end{figure}

If the era of leptogenesis is lowered below $T \sim 10^{12}$ GeV, 
the interactions of the tau charged lepton, mediated by its 
SM Yukawa coupling, come into thermal equilibrium. The effect is that 
the $\tau$-component of each $|i\rangle$ experiences relatively 
rapid interactions with the early Universe plasma. The left-handed 
$\tau$ component is rapidly converted to a right-handed $\tau$ via 
scattering with the Higgs. Similarly, the reverse process repopulates 
the $\tau$ asymmetry density at the same rate. This rate is determined 
by the imaginary part of the thermal self-energy of $\tau$, 
$\Im(\Lambda_{\tau})$, which by the optical theorem determines 
the mean free path of the $\tau$ state (see \figref{fig:thermalwidth}).
When this process is sufficiently rapid, the coherent superposition 
of flavours is destroyed and the $\tau$ component can no longer 
contribute at the level of amplitudes to the decay and inverse decay 
processes with $e$ and $\mu$ (which form a single coherent flavour 
state which we shall refer to as $\tau^{\perp}$ such that 
$\langle \tau | \tau^{\perp} \rangle = 0$). Instead $\tau$ undergoes 
decay and inverse decay as a separate decoherent state. 
Correspondingly, the kinetic equations must separately describe 
the time evolution of $n_{\tau \tau}$ and $n_{\tau^{\perp} \tau^{\perp}}$ and 
the total baryon asymmetry is simply the sum: 
$n_{B-L} = n_{\tau \tau} + n_{\tau^{\perp} \tau^{\perp}}$. 
This is the \textit{two-flavoured} regime and the  
Boltzmann equations are given by
\begin{equation}
\begin{aligned}
\label{eq:BE2F}
\frac{dn_{N_{i}}}{dz}&=-D_{i}(n^{}_{N_{i}}-n^\text{eq}_{N_{i}}),\\
\frac{dn_{\tau \tau}}{dz}& =\sum_i\left(\epsilon^{(i)}_{\tau \tau} D_{i}(n^{}_{N_{i}}-n^\text{eq}_{N_{i}})-p_{i \tau}W_{i}n_{\tau \tau} \right),\\
 \frac{dn_{\tau^{\perp} \tau^{\perp}}}{dz} & =\sum_i \left( \epsilon^{(i)}_{\tau^{\perp} \tau^{\perp}} D_{i}(n^{}_{N_{i}}-n^\text{eq}_{N_{i}})-p_{i \tau^{\perp}}W_{i} n_{\tau^{\perp} \tau^{\perp}}\right),
\end{aligned}
\end{equation}
where $p_{i \alpha} \equiv |C_{i \alpha}|^2$, $\overline{p}_{i \alpha} 
\equiv |\overline{C}_{i \alpha}|^2$ are the projection probabilities 
expected from the decoherence - the classical measurement of 
$\tau$ by the early Universe plasma. Furthermore, the $CP$-asymmetries are
\begin{equation}
\begin{aligned}
\epsilon^{(i)}_{\alpha \alpha} & = -\frac{p_{i \alpha} \Gamma_i-\overline{p}_{i \alpha}\overline{\Gamma}_i}{\Gamma_i+\overline{\Gamma}_i},
\end{aligned}
\end{equation}
for $\alpha = \tau^{\perp}, \tau$.
Analogously, there exists the possibility that the 
out-of-equilibrium decays of the heavy Majorana neutrinos occur at temperatures
where the SM muon Yukawa interactions has thermalised 
$T \sim 10^9$ GeV and the \textit{three-flavoured} Boltzmann equations 
are the relevant kinetic equations. This possibility was explored in~\cite{Blanchet:2008pw} and it was shown that thermal leptogenesis can be lowered to $T \sim 10^8$ GeV.

 It has been shown~\cite{Barbieri:1999ma, Abada:2006fw,DeSimone:2006nrs,Blanchet:2006ch,Blanchet:2011xq} that the density matrix equations produce a 
more physically accurate description of leptogenesis where 
the density matrix may be expressed as
\begin{equation}
n \equiv \sum_{\alpha, \beta} n_{\alpha \beta} |\alpha \rangle \langle \beta|,
\end{equation}
where $|\alpha \rangle$ are states of definite lepton flavour, 
defining the flavour basis. Using this description, the 
diagonal elements, $n_{\alpha \alpha}$, are the differences of 
the normalised densities of $\alpha$ and $\overline{\alpha}$ 
particles such that $n_{B-L} = \Tr n$. The off-diagonals describe 
the degree of coherence between the flavour states. The advantage of 
a density matrix description is that decoherence effects are easily 
incorporated and as such, the dynamical process by which 
different flavour states decohere is readily incorporated into 
the equations. This allows for a single set of equations with solutions 
which transition between flavour-regimes as appropriate. 
Furthermore, the equations should remain accurate even in the 
regions of transition where the interactions leading 
to decoherence are not infinitely fast.

Explicitly, the \textit{density matrix equations} of leptogenesis are~\cite{Barbieri:1999ma,Abada:2006fw,DeSimone:2006nrs,Blanchet:2006ch,Blanchet:2011xq}
\begin{equation}
\label{eq:full3}
\begin{aligned}
\frac{dn_{N_{i}}}{dz}=&-D_{i}(n_{N^{}_{i}}-n^\text{eq}_{N_{i}})\\
 \frac{dn_{\alpha\beta}}{dz} =& \sum_i \left(\epsilon^{(i)}_{\alpha\beta}D_{i}(n^{}_{N_{i}}-n^\text{eq}_{N_{i}})-\frac{1}{2}W_{i}\left\{P^{0(i)},n\right\}_{\alpha\beta}\right) \\
-&\frac{\Im(\Lambda_{\tau})}{Hz}\left[\begin{pmatrix}1&0&0\\ 0&0&0 \\
0&0&0 \end{pmatrix},\left[\begin{pmatrix}1&0&0\\ 0&0&0 \\
0&0&0 \end{pmatrix},n\right]\right]_{\alpha\beta}
-\frac{\Im(\Lambda_{\mu})}{Hz}\left[\begin{pmatrix}0&0&0\\ 0&1&0 \\
0&0&0 \end{pmatrix},\left[\begin{pmatrix}0&0&0\\ 0&1&0 \\
0&0&0 \end{pmatrix},n\right]\right]_{\alpha\beta},
\end{aligned}
\end{equation}
where the projection matrices are
\begin{equation}
P^{0(i)}_{\alpha \beta} \equiv C_{i \alpha} C_{i \beta}^*,
\end{equation}
which generalise the notion of the projection probability and 
appear in the double commutator structure. The double-commutator 
structures in \equaref{eq:full3} give rise to an exponentially 
damping term proportional to $\Im(\Lambda_{\alpha})/Hz$ for the 
equations describing the off-diagonal elements of $n$. 
In the flavour basis, if these terms are dominant,  the density matrix 
is driven towards a diagonal form. The $CP$-asymmetry parameters 
are~\cite{Covi:1996wh,Blanchet:2011xq,Abada:2006ea,DeSimone:2006nrs,
Abada:2006fw,Biondini:2017rpb,Biondini:2015gyw,Biondini:2016arl} 
\begin{equation}
\begin{aligned}
\epsilon^{(i)}_{\alpha\beta}&=\frac{3}{32\pi\left(Y^{\dagger} Y\right)_{ii}}
\sum_{j\neq i}\Bigg\{ i[Y_{\alpha i}Y^{*}_{\beta j}(Y^{\dagger}Y)_{ji}
- Y^{*}_{\beta i}Y_{\alpha j}(Y^{\dagger}Y)_{ij}] f_1\left(\frac{x_{j}}{x_{i}}\right) \\
&+i[Y_{\alpha i}Y^{*}_{\beta j}(Y^{\dagger}Y)_{ij}-Y^{*}_{\beta i}Y_{\alpha j}(Y^{\dagger}Y)_{ji}] f_2\left(\frac{x_{j}}{x_{i}}\right) \Bigg\},
\label{eq:CPoff}
 \end{aligned}
\end{equation}
where
\begin{equation}
\begin{aligned}
f_1\left(\frac{x_{j}}{x_{i}}\right) \equiv \frac{\xi\left(\frac{x_{j}}{x_{i}}\right)}{\sqrt{\frac{x_{j}}{x_{i}}}}, \quad
 f_2\left(\frac{x_{j}}{x_{i}}\right) \equiv \frac{2}{3\left(\frac{x_{j}}{x_{i}}-1\right)}.
\end{aligned}
\end{equation}
The diagonal components of the $\epsilon^{(i)}$ matrix simplify 
to the following form
\begin{equation}
\begin{aligned}
\epsilon^{(i)}_{\alpha\alpha}  =\frac{3}{16\pi\left(Y^\dagger Y\right)_{ii}}\sum_{j\neq i} & \Bigg\{\Im\left[ {Y_{\alpha i}}^{*}Y_{\alpha j}(Y^{\dagger}Y)_{ij} \right] f_1\left(\frac{x_{j}}{x_{i}}\right)+\Im\left[  {Y_{\alpha i}}^{*}Y_{\alpha j}(Y^{\dagger}Y)_{ji} \right]f_2\left(\frac{x_{j}}{x_{i}}\right)
 \Bigg\}.
  \end{aligned}\label{eq:CPa2}
\end{equation}

\section{Leptogenesis in the regime
$\mathbf{10^9} < \mathbf{M_1} \text{\textbf{ (GeV)}} < \mathbf{10^{12}}$}
\label{sec:typicalM1}
\begin{figure}[t]
\centering
\includegraphics[width=1.0\textwidth]{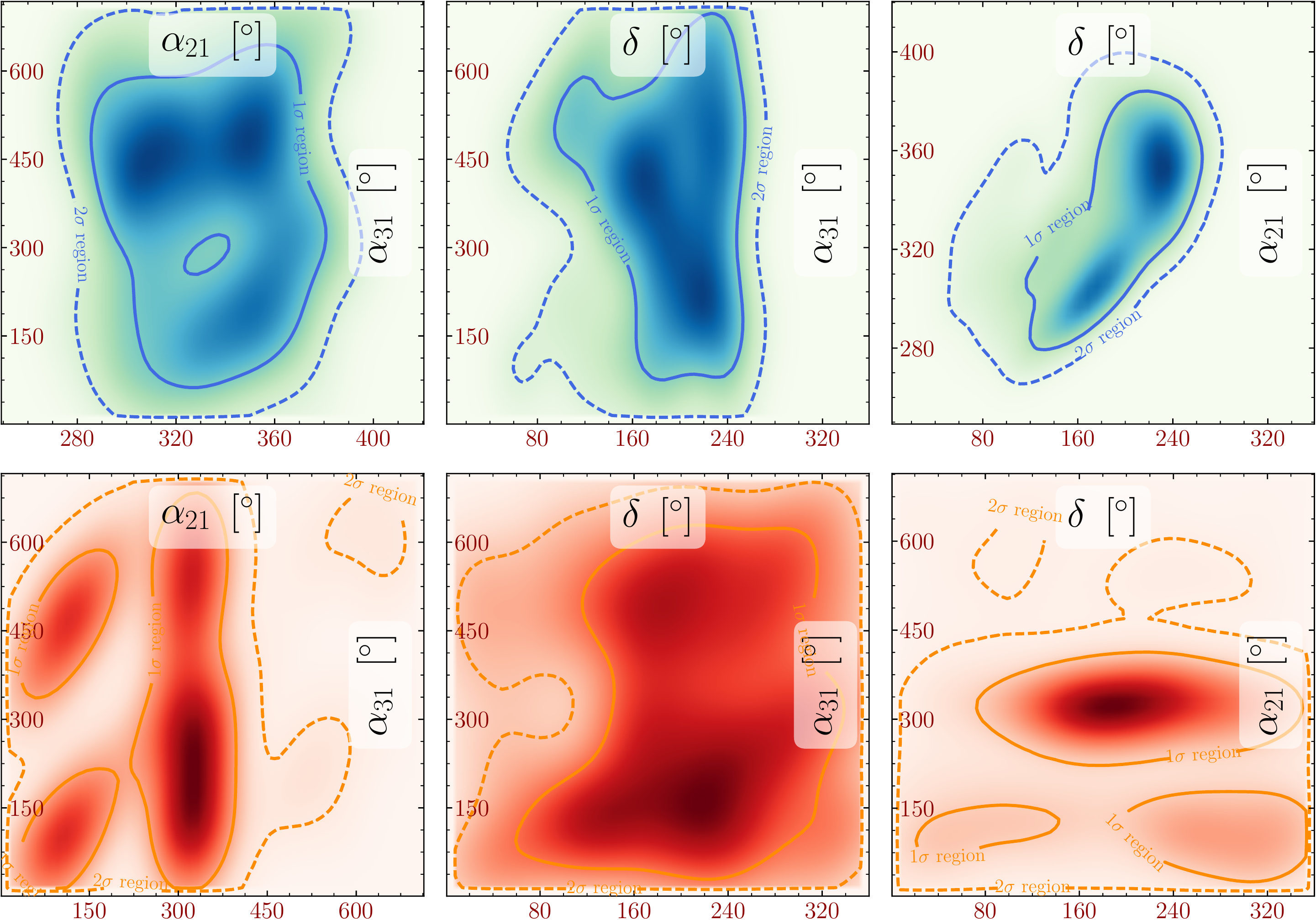}
\caption{The two-dimensional projections for leptogenesis with 
$M_1 = 10^{10}$ GeV and $CP$ violation 
provided only by the phases of the PMNS matrix. 
The NO case is coloured blue/green and the IO one 
is orange/red. The contours correspond 
to $68\%$ and $95\%$ confidence levels.
This plot was created using {\sc SuperPlot}~\cite{Fowlie:2016hew}.}
\label{fig:genericscale1}
\end{figure}

In this section, we explore the possibility that successful leptogenesis  
derives solely from the $CP$-violating PMNS phases and the mass scale 
is between  $10^9 \leq M_1\text{ (GeV)} \leq 10^{12}$, which generally 
corresponds to the two-flavour regime. Historically, the link between 
low-energy $CP$ violation and the baryon asymmetry was first established 
in this regime and thus our main purpose in this section is to revisit 
the scenario with more robust numerical methods than have previously 
been applied. We shall perform a comprehensive exploration of the 
parameter space for a model with three heavy Majorana neutrinos 
in both the normal ordered and inverted ordered scenarios. 
We shall then investigate a subset of scenarios in which only 
the Dirac or only the Majorana phases are varied.

\subsection{Results of Parameter Exploration}\label{sec:ParamGenericScale}

In this particular explorations of the parameter space, we fix $M_1$ and 
vary $M_2$ and $M_3$ such that $M_3> 3 M_2 > 9 M_1$, ensuring that 
resonant regimes are avoided~
\cite{Pilaftsis:2003gt,Blanchet:2008pw,Covi:1996wh,Covi:1996fm,Pilaftsis:1997jf,Buchmuller:1997yu}. We choose to set $M_1=10^{10}$ GeV, this being 
typical of the mass window under 
consideration. 

We fix, $x_1 = 90^\circ$ and $x_3 = 180^\circ$ and 
$y_2 =  0^\circ$ such that there is a complete leptonic $CP$-symmetry 
when $\delta = 0^\circ$, $\alpha_{21} = 180^\circ$ and $\alpha_{31} = 0^\circ$.\footnote{This choice of parameters for the low-energy phases is made such that the $CP$-symmetry holds for the Yukawa matrix when the $R$-matrix is taken in to account. It would not suffice to choose, eg., $\delta = \alpha_{21} = \alpha_{31} = 0^\circ$.}
With the specified parameters fixed or constrained as stated, 
we explore the parameter space using a flat prior and log-likelihood 
function evaluated at a point 
$\mathbf{p}=\left(\delta,\alpha_{21},\alpha_{31},m_{1,3},M_2,M_3\right)$ 
(varying $m_1$ or $m_3$ for normal or inverted ordering respectively) by
\begin{equation}
\log L  = - \frac{1}{2} 
\left( \frac{\eta^2_B(\mathbf{p})-\eta_{B_{CMB}}^2}{\Delta \eta_{B_{CMB}}^2} \right),
\end{equation}
to define regions of $1 \sigma$ and $2 \sigma$ agreement with the 
observed value of the asymmetry. In addition we impose a bound on 
the sum of neutrino masses of $1 \text{ eV}$ which is consistent with 
the tritium beta-decay experiments~
\cite{Weinheimer:2003fj,Kraus:2004zw,Lobashev:2001uu} 
 but 
more conservative than recent constraints from Planck~\cite{Ade:2015xua}. 
In the numerical work of this section we allow only for the two lightest 
heavy Majorana neutrinos to decay (an excellent approximation) and we neglect 
lepton number-changing scattering processes, spectator effects 
\cite{Buchmuller:2001sr,Nardi:2005hs}, thermal corrections 
\cite{Kiessig:2010pr,Giudice:2003jh} and the inclusion of quantum 
statistical factors  
\cite{DeSimone:2007gkc,Beneke:2010dz,Anisimov:2010dk,Beneke:2010wd} 
which typically introduces an $\mathcal{O}\left(10\%\right)$ 
error~\cite{Blanchet:2006be,Frossard:2013bra,Nardi:2007jp,Garbrecht:2014kda}. 
To solve the density matrix equations we use the {\sc Python} 
interface~\cite{odeintw} to the {\sc LSODA} algorithm~\cite{odepack} 
that is available in {\sc Scientific Python}~\cite{scipy}. 
Due to the high-dimensionality of the parameter space we found the use of 
{\sc Multinest}~\cite{Feroz:2008xx,Feroz:2007kg,2013arXiv1306.2144F} 
(more explicitly,
{\sc pyMultiNest}~\cite{pymultinest}, a wrapper around {\sc Multinest} written
in {\sc Python}) particularly useful. In Fig. \ref{fig:2DIntermediatePlots} 
we display the two-dimensional posterior probability plots for 
the $CP$-violating PMNS phases in the normal ordered and 
inverted ordered cases.

The results of this parameter search are shown in the form of 
two-dimensional projections in \figref{fig:genericscale1}. 
For points in these regions of parameter space for 
which $\eta_B = \eta_{B_{CMB}}$, the fine-tuning is 
$\mathcal{F} \approx 0.23$ which corresponds only to a very 
slight enhancement of the $R$-matrix. 
The values of lightest neutrino mass for NO (IO) neutrino 
mass spectrum corresponding to this case are $m_{1(3)} = 0.0215$ eV. 
For the best-fit values of the fitted parameters in the NO (IO) case 
we find: $\delta = 133.76^\circ~(139.8^\circ)$, 
$\alpha_{21} = 315.5^\circ~(165.3^\circ)$, 
$\alpha_{31} = 551.0^\circ~(565.5^\circ)$,
$M_2 = 4.90~(4.97)\times 10^{11}$ GeV, 
$M_3 = 2.19\times 10^{12}$ GeV, 
$x_2 = 113.4^\circ~(13.9^\circ)$.
For the case of an NO light neutrino mass spectrum,
we find that the observed baryon 
asymmetry may be obtained to within $1 \sigma$ ($2 \sigma$) with 
$\delta$ between $[95,265]^\circ$ ($[52,282]^\circ$). For IO, 
the $1 \sigma$ ($2 \sigma$) range is $[60, 338]^\circ$ ($[8,360]^\circ$). 
Both of these scenarios comfortably incorporate the measured bounds 
on $\delta$ (\tabref{tab:table}).
In what follows, we provide some 
explanation of these 
 results and plots
by introducing an analytical approximation 
which we use to study the scenarios where only the Dirac or only the 
Majorana phases provide $CP$ violation.

\subsection{Dependence of $\eta_B$ on the Dirac and Majorana Phases}
\begin{table}[t!]
\centering
\begin{tabular}{ c c  c c  c  c c  c  c c  }
 \toprule
$\delta$ & $\alpha_{21}$ & $\alpha_{31}$ &$M_1 $ & $M_2 $ & $M_3 $ 
& $x_1$ & $x_2$ & $x_3$ & $y_2$   \\
(${}^{\circ}$) & (${}^{\circ}$) & (${}^{\circ}$)
& ($\text{GeV})$ & $(\text{GeV})$ & $(\text{GeV})$ 
& (${}^{\circ}$) & (${}^{\circ}$) & (${}^{\circ}$) &
  (${}^{\circ}$) \\
\specialrule{2.5pt}{1pt}{1pt}
$228$ & $447$ & $570$&
$2.82 \times 10^{10}$ & $1.00 \times 10^{13}$ & $3.16 \times 10^{13}$
&$90$ & $18$ & $180$
 & $0$ \\
\bottomrule
\end{tabular}
\caption{A benchmark point for leptogenesis with $M_1=2.82 \times 10^{10}$ GeV, 
with normal ordering. Here, we have $m_{1}=0.02$ eV and 
$y_1 = y_3 = - 33^\circ$, corresponding to $\mathcal{F} = 0.27$. 
This point produces $\eta_B = 6.1 \times 10^{-10}$.}
\label{tab:BFGenericScale}
\end{table}

In the scenario  $10^9 < M_1 \text{(GeV)} < 10^{12}$, it is appropriate 
to apply the two-flavour Boltzmann equations (\equaref{eq:BE2F}). 
These equations have the following analytical solution~\cite{Fong:2013wr}
\begin{equation}
\label{eq:2FSol}
\begin{aligned}
n_{B - L} \approx \frac{\pi^2}{6 z_d K_1} n^{\text{eq}}_{N_1} (z_0) \left( \epsilon^{(1)}_{\tau \tau} \frac{1}{P^{0(1)}_{\tau \tau}} 
+ \epsilon^{(1)}_{\tau^{\perp} \tau^{\perp}}  
\frac{1}{P^{0(1)}_{\tau^{\perp} \tau^{\perp}}} \right),
\end{aligned}
\end{equation}
where it is assumed that the dominant contribution to the 
final asymmetry is from the lightest of the heavy Majorana 
neutrinos and that leptogenesis occurs in the strong washout regime. 
We denote the $z$ for which the washout becomes less than one 
as $z_d$, $W(z_d) < 1$, $K_1 \equiv \Gamma_1 / H(M_1)$ 
is the decay parameter for $N_1$, and the $z$ for which leptogenesis 
is initiated as $z_0$.
As we are interested in those scenarios in 
which $CP$ violation derives only from the phases of the PMNS matrix, 
we have the supplementary condition $\Tr \epsilon^{(1)} = 0$ 
(or $\epsilon^{(1)}_{\tau \tau} = - \epsilon^{(1)}_{\tau^\perp \tau^\perp}$) 
which we may use to simplify the solution to
\begin{equation}
\begin{aligned}
\label{eq:2FSol}
n_{B - L} = \frac{\pi^2}{6 z_d K_1} n^{\text{eq}}_{N_1} (z_0) \epsilon^{(1)}_{\tau \tau} \Delta F,
\end{aligned}
\end{equation}
with
\begin{equation}
\begin{aligned}
\Delta F & \equiv \frac{1}{P^{0(1)}_{\tau \tau}} 
- \frac{1}{P^{0(1)}_{\tau^{\perp} \tau^{\perp}}}
 = \frac{1}{P^{0(1)}_{\tau \tau}} - \frac{1}{1-P^{0(1)}_{\tau \tau}}.
\end{aligned}
\end{equation}
\begin{figure}[t!]
\centering
\includegraphics[width=1.0\textwidth]{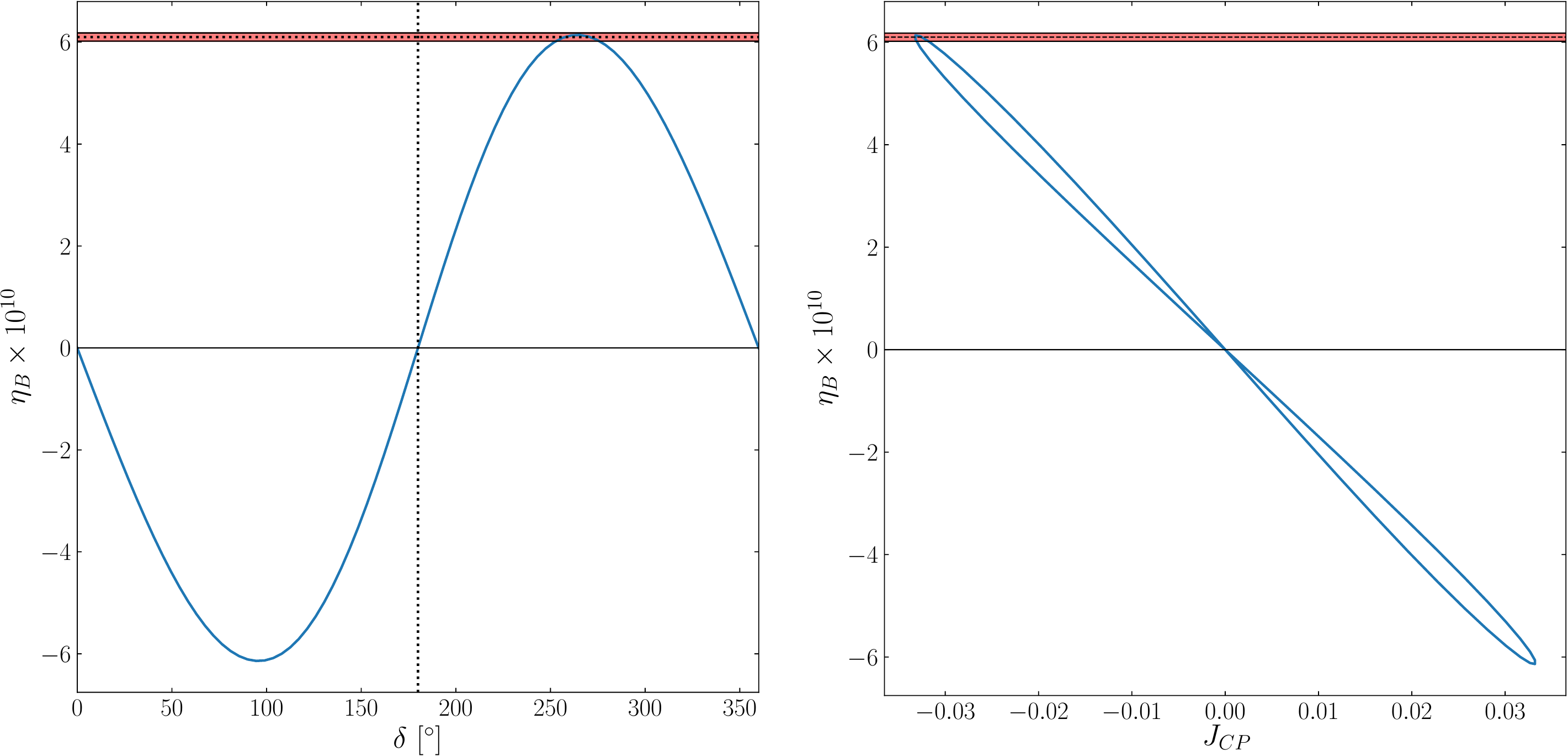}
\caption{The baryon asymmetry with 
$M_1 = 5.13 \times 10^{10} \text{ GeV}$ and $CP$ violation 
provided solely by $\delta$. The Majorana phases are fixed 
at $\alpha_{21} = 180^{\circ}$ and $\alpha_{31} = 0^{\circ}$. 
The red band indicates the $1 \sigma$ observed values 
for $\eta_{B_{CMB}}$ with the best-fit value indicated by the 
horizontal black dotted line. 
Left: The final baryon asymmetry as a function of $\delta$ with 
exact $CP$-invariance when $\delta = 0^{\circ}$ and $180^{\circ}$ 
(vertical black dotted line). 
Right: A parametric plot of $\eta_B$ against $J_{CP}$ as $\delta$ is varied. See the text for further details.}
\label{fig:genericscale2}
\end{figure}
\begin{figure}[t!]
\centering
\includegraphics[width=0.5\textwidth]{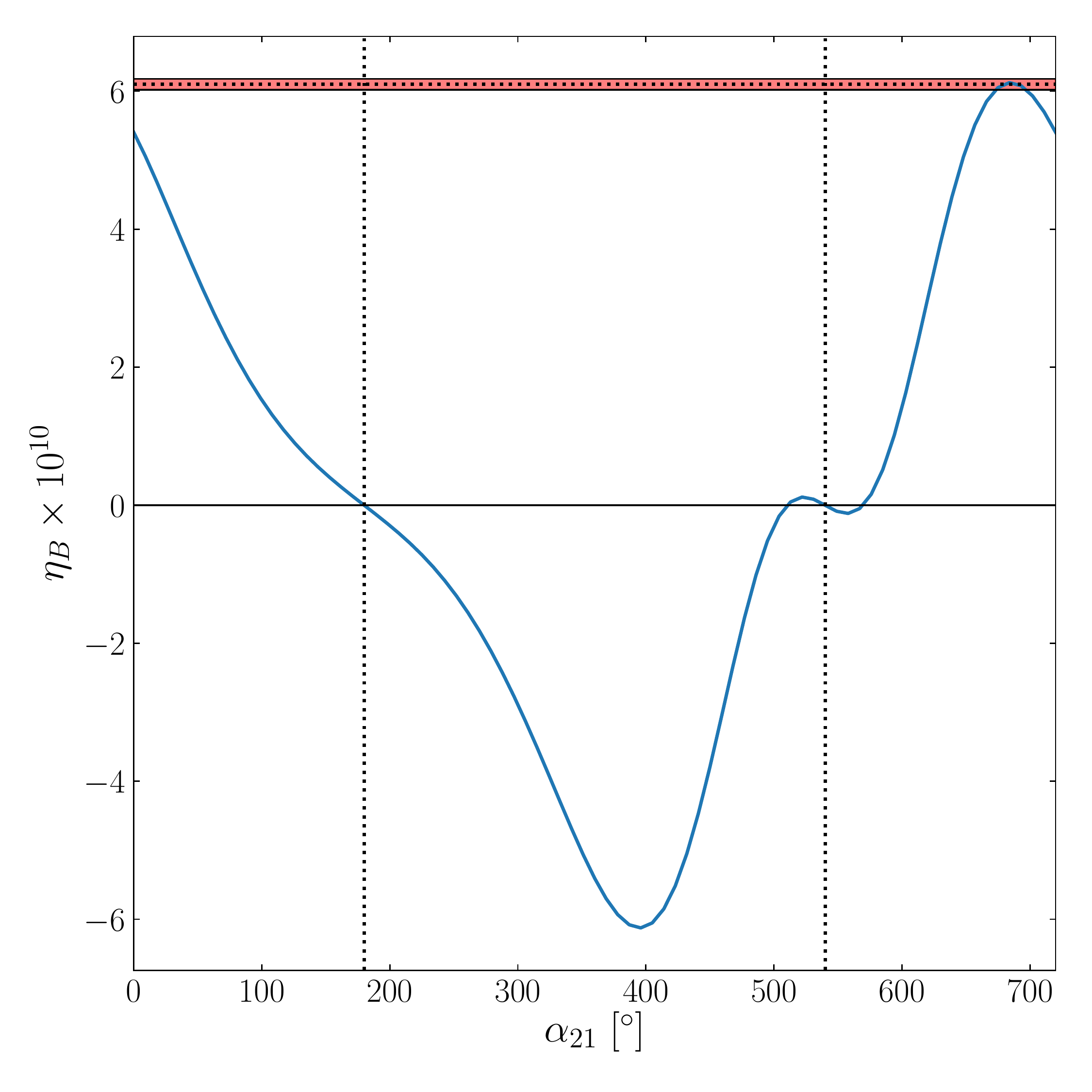}
\caption{The baryon asymmetry with $M_1 = 3.05 \times 10^{10}$ GeV and 
$CP$ violation provided solely by $\alpha_{21}$ (corresponding 
to $\delta = \alpha_{31} = 0^\circ$). The red band indicates 
the $1 \sigma$ observed values for $\eta_B$ with the best-fit 
value indicated by the horizontal black dotted lines. 
Here we show the baryon asymmetry against $\alpha_{21}$ with exact 
$CP$-invariance at $\alpha_{21} = 180^{\circ}$ and $540^{\circ}$ 
(vertical black dotted lines).}
\label{fig:genericscale3}
\end{figure}
\begin{figure}[t!]
\centering
\includegraphics[width=0.5\textwidth]{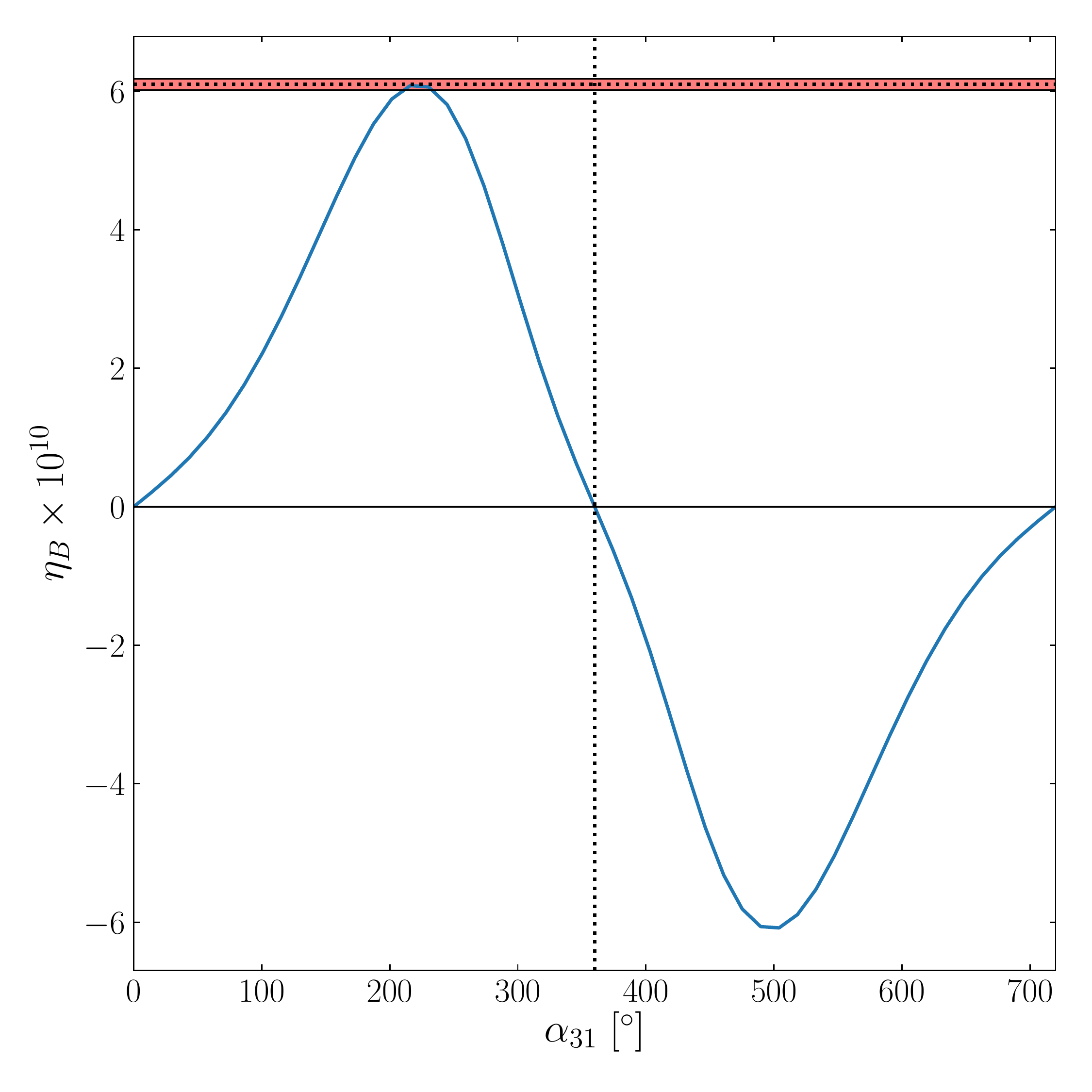}
\caption{The baryon asymmetry with $M_1 = 5.13\times 10^{10}$ GeV and  
$CP$ violation provided solely by $\alpha_{31}$ (corresponding to 
$\delta = 0^{\circ}$, $\alpha_{21} = 180^{\circ}$). The red band indicates 
the $1 \sigma$ observed values for $\eta_B$ with the best-fit value 
indicated by the horizontal black dotted lines. 
Here we show the baryon asymmetry as a function of $\alpha_{31}$, 
exact $CP$-invariance exists for $\alpha_{31} = 0^{\circ}$ 
and $360^{\circ}$ (vertical black dotted lines).}
\label{fig:genericscale4}
\end{figure}

At the benchmark point for normal ordering
defined in Table \ref{tab:BFGenericScale}, which we will use in the further analyses in the present section,
we have:
\begin{equation}
\begin{aligned}
\label{eq:YukawaTypical}
Y_{\tau 1} & = 1.37\times 10^{-3}-1.67\times 10^{-4} e^{i \delta},\\
Y_{\tau 1} & = 6.64\times 10^{-4}-8.74\times 10^{-4} e^{i \frac{\alpha_{21}+\pi}{2}},\\
Y_{\tau 1} & = 4.71\times 10^{-4}+1.07\times 10^{-3} e^{\frac{i \alpha_{31}}{2}},
\end{aligned}
\end{equation}
for $CP$ violation from $\delta$, $\alpha_{21}$ and $\alpha_{31}$ 
respectively. For the case in which $\delta$ provides the 
$CP$ violation in \equaref{eq:YukawaTypical}, this phase gives a subdominant contribution to $\lvert Y_{\tau 1} \rvert$. 
As can be shown,  $P^{0(1)}_{\tau \tau}$ is similarly 
weakly dependent on the phases.
Thus, the phase dependence of the solutions of \equaref{eq:2FSol} does 
not come predominantly from the flavour factor $\Delta F$ but from 
the $CP$-asymmetry $\epsilon^{(1)}_{\tau \tau}$. However, in the case 
 of $\alpha_{21}$ providing
the $CP$ violation, the two terms of 
\equaref{eq:YukawaTypical} are similar in magnitude and we 
may get a strong enhancement in $\Delta F$. The final case where 
$\alpha_{31}$ provides the $CP$ violation is intermediate 
and should experience a slight phase-dependent enhancement from $\Delta F$.
\subsubsection{Dirac Phase $CP$ Violation}

In this subsection, we consider deviations from the benchmark point 
of \tabref{tab:BFGenericScale} where we allow $\delta$ to vary but 
fix $\alpha_{21} = 180^{\circ}$ and $\alpha_{31} = 0^\circ$. Given the pattern 
of $R$-matrix angles, this ensures that any $CP$ violation 
comes  solely
from $\delta$. In this case, the $\tau \tau$-component 
of the $CP$-asymmetry is given by
\begin{equation}
\epsilon^{(1)}_{\tau \tau} = \left(0.515 -3.94 c_{13} 
\right) s_{13} \times 10^{-8} \sin \delta = -0.501 \times 10^{-8} \sin \delta.
\end{equation}
Thus, given the approximate phase-independence of $\Delta F$, 
we obtain a sinusoidal dependence of $\eta_B$ on $\delta$, 
with $\eta_B = 0$ when $\delta = 0^{\circ}$ or $180^{\circ}$. 
Keeping all other parameters fixed, we find that 
for $M_1 =2.82\times 10^{10}$ GeV
no value of $\delta$ can produce the observed baryon 
asymmetry of the Universe, 
the maximum value of $\eta_B$ as a function of $\delta$ 
is $4.07 \times 10^{-11}$. We might scale the heavy Majorana neutrino 
masses by a constant value, as when the two-flavour approximation 
of \equaref{eq:2FSol} is valid, the factor $\epsilon^{(1)}_{\tau \tau}$ 
scales in proportion with this constant and thus so does $\eta_B$. 
In doing so, we find that the final asymmetry rises until 
$M_1 = 7.08 \times 10^{11}$ GeV, where $\eta_B$ takes maximum value 
$4.01 \times 10^{-10}$. After this, the simple scaling fails as one 
begins to enter the transition to what is usually the single-flavour regime. 

Performing a detailed numerical parameter 
exploration we find that  purely Dirac phase $CP$ violation 
leads to successful leptogenesis for  
$M_1 = 5.13 \times 10^{10}$ GeV, $M_2 = 2.19 \times 10^{12}$ GeV and
$M_3 = 1.01 \times 10^{13}$ GeV. This is illustrated 
in \figref{fig:genericscale2} in which the plotted 
$\eta_B$ comes from solving the full density matrix equations. 
In this case, we have:
\begin{equation}
Y_{\tau 1}  = 1.11 \times 10^{-2}-2.40\times 10^{-4} e^{i \delta}.
\end{equation}
Given the different order of magnitude of the two terms in 
the expression for $Y_{\tau 1}$, the baryon asymmetry  
should exhibit dependence on $\delta$ only from $\epsilon^{(1)}_{\tau \tau}$ 
and not from $\Delta F$. Our theoretical expectations 
are borne out by the approximate sinusoidal dependence
of $\eta_B$ on $\delta$ seen in \figref{fig:genericscale2}.

\subsubsection{$CP$ Violation from the Majorana Phase $\alpha_{21}$}

Here, we set $\delta = \alpha_{31} = 0^\circ$ but 
allow $CP$ violation from $\alpha_{21}$. Setting all other parameters 
to their benchmark values we find
\begin{equation}
\epsilon^{(1)}_{\tau \tau} = 3.14 \times 10^{-7} \cos \frac{\alpha_{21}}{2}.
\end{equation}
It follows from this expression for $\epsilon^{(1)}_{\tau \tau}$ that at the $CP$-conserving values for $\alpha_{21}=0^\circ, 360^\circ$ we have $\epsilon^{(1)}_{\tau \tau} \neq 0$ (see also \figref{fig:genericscale3}). This corresponds to the case of $CP$-conserving $R$-matrix, $CP$-conserving PMNS matrix, but $CP$-violating interplay between the $R$ and PMNS matrix elements in leptogenesis \cite{Pascoli:2006ie}.
In a similar way to the previous subsection, we find that no 
value of $\alpha_{21}$ can achieve successful leptogenesis using 
this combination of phases and the benchmark values from 
\tabref{tab:BFGenericScale}. Thus, we find it necessary to scale all 
of the heavy Majorana neutrino masses by a common factor such that 
$M_1 = 3.05 \times 10^{10} \text{ GeV}$, may allow for  successful 
leptogenesis. With this scaling we obtain the results plotted in 
\figref{fig:genericscale3}. The deviation from pure (co)sinusoidal 
behaviour is explained by the $\alpha_{21}$-dependence of 
$\Delta F$. For $\alpha_{21} < 360^\circ$, $\Delta F$ varies relatively 
slowly exhibiting a global minimum at $\alpha_{21} = 180^\circ$,
resulting in a slightly modified sinusoidal dependence through this point 
in $\eta_B$. A strong peak exists for $\Delta F$ around 
$\alpha_{21} = 540^\circ$, which results in the peak of 
$\eta_B$ occurring before $720^\circ$, as would be expected from 
the dependence of $\epsilon^{(1)}_{\tau \tau}$. The small sign-changing 
fluctuation around the zero at $\alpha_{21} = 540^\circ$ is a feature that 
does not appear in the solution of two-flavour Boltzmann equations 
and thus cannot be explained in terms of the analytic solution 
\equaref{eq:2FSol}. However, the extra zeros of 
$\eta_B$ that are 
 seen in \figref{fig:genericscale3} are  
due only to accidental cancellations and do not correspond 
to cases of $CP$-symmetry
(unlike those at $\alpha_{21} = 180^\circ$ and $\alpha_{21} = 540^\circ$).

 \subsubsection{$CP$ Violation from the Majorana Phase $\alpha_{31}$}
 We set $\delta = 0^\circ$ and $\alpha_{21} = 180^\circ$ such 
that $CP$ violation is provided by $\alpha_{31}$. 
Using the 
benchmark values for the other parameters from 
\tabref{tab:BFGenericScale} we find:
\begin{equation}
\epsilon^{(1)}_{\tau \tau} = 2.11 \times 10^{-7} \sin \frac{\alpha_{31}}{2}.
\label{eq:genericscaleepsalpha31}
\end{equation}
Again we find that without scaling the heavy Majorana neutrino masses, 
no value of $\alpha_{31}$ corresponds to successful leptogenesis. 
At $M_1 = 5.13 \times 10^{10} \text{ GeV}$ we obtain the first point for 
which the observed baryon asymmetry is created and 
 this is plotted in \figref{fig:genericscale4}.
We see
that analytical expectation of a sinusoidal 
dependence of the baryon asymmetry 
($\eta_B \propto \epsilon^{(1)}_{\tau \tau} \propto \sin(\alpha_{31}/2)$) 
from \equaref{eq:genericscaleepsalpha31} is present. 
$\Delta F$ exhibits a broad peak around $\alpha_{31} = 360^\circ$ 
which results in the slight shift to the centre of the 
otherwise sinusoidal peaks.

\subsection{The Case of $N_3$ Decoupled}
\begin{figure}[t!]
  \includegraphics[width=1.0\textwidth]{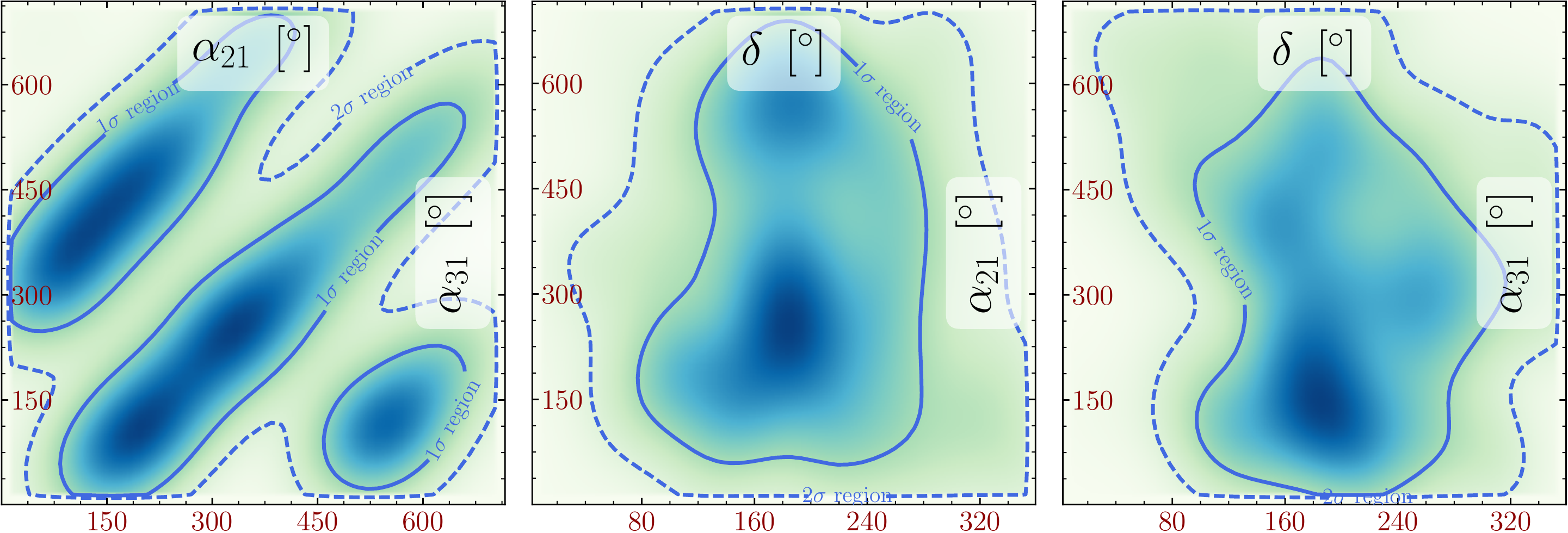}
\caption{The two-dimensional projections for leptogenesis with 
 $M_1 = 10^{11}$ GeV,  $M_2 = 10^{12}$ GeV
and $N_3$ decoupled, with $CP$ violation provided only by 
the phases of the PMNS matrix. Here it is assumed that the 
light neutrino mass spectrum has normal ordering. 
Contours correspond to $68\%$ and $95\%$ confidence levels. 
This plot was created using {\sc SuperPlot}~\cite{Fowlie:2016hew}.
}
\label{fig:2DDecoupled}
\end{figure}

 In this section, we review the case that 
the heaviest Majorana neutrino, $N_3$, physically decouples.
We restrict ourselves to normal ordered light neutrino masses.
The resultant scenario with two relevant heavy Majorana neutrinos is 
the simplest (minimal) type I framework compatible with all neutrino data. 
 In this scenario only two of the 
light neutrinos have non-zero masses since $m_1 =  0$. For normal ordering,
the $R$-matrix may be parametrised 
as \cite{Ibarra:2003up,Antusch:2011nz,Blanchet:2007qs}
\begin{equation}
R = 
\begin{pmatrix}
0 & \cos \theta& \sin \theta \\
0 & - \sin \theta & \cos\theta \\
1& 0 & 0
\end{pmatrix}.
\label{eq:RN3decoup} 
\end{equation}
The resulting neutrino Yukawa matrix thus has $Y_{\alpha 3} = 0$, 
consistent with the premise that $N_3$ has decoupled.
We choose to take $\theta$ in \equaref{eq:RN3decoup} 
to be real in order 
to have the condition of \equaref{eq:UCPRCPconditions} satisfied. 
We assume further that at least one of the three 
phases in the PMNS matrix has a $CP$-violating value.

As with the previous sections, we have performed an exhaustive 
exploration of the parameter space where again we are primarily 
concerned with the situation in which $CP$ violation is provided 
only by the PMNS phases. We choose to fix $M_1 = 10^{11}$ GeV and 
$M_2 = 10^{12}$ GeV such that the parameter space to explore is described 
by ${\bf{p}} = (\delta, \alpha_{21}, \alpha_{31}, \theta)$. 
In \figref{fig:2DDecoupled}, we present the two-dimensional posterior 
projection for the case of normal ordering. Here, it is seen that 
with $M_1 = 10^{11}$ GeV, for normal ordering, successful leptogenesis 
may produce a baryon asymmetry with $1\sigma$ ($2\sigma$) agreement 
with the observed value for 
$\delta \in [95,315]^\circ$, ($\delta \in [25,360]^\circ$). 

\begin{figure}[t!]
  \includegraphics[width=1.0\textwidth]{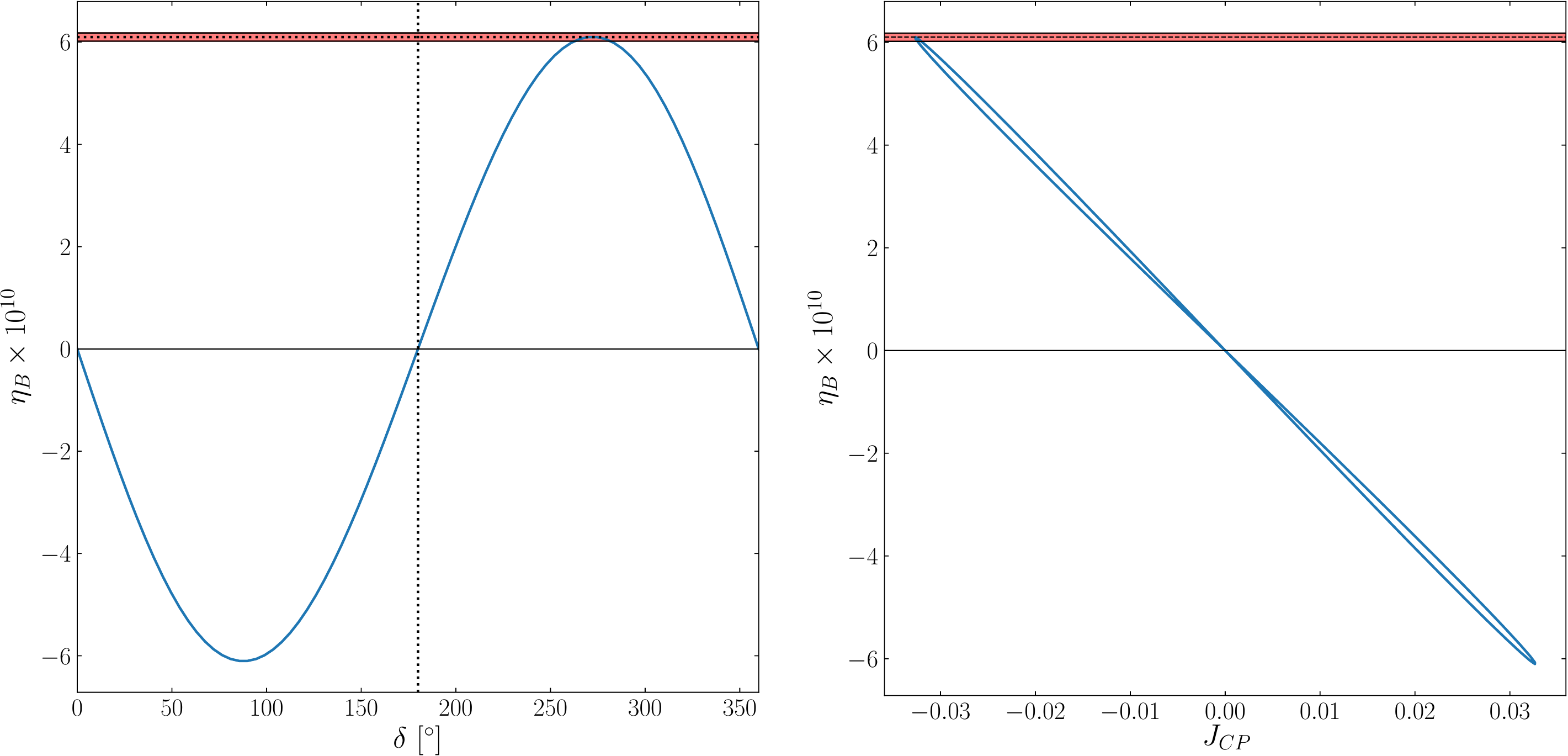}
\caption{The baryon asymmetry from leptogenesis with
$M_1 = 10^{11}$ GeV and $N_3$ decoupled, where $CP$ violation 
is provided only by the Dirac 
phase $\delta$. The red bands indicate the values in $1\sigma$ 
agreement with the observed value $\eta_{B_{CMB}}$. 
Left: A plot of $\eta_B$ against $\delta$, showing successful leptogenesis 
near the maximal $CP$-violating value $\delta = 270^\circ$. 
Right: The corresponding parametric plot of $\eta_B$ with 
$J_{CP}$ as $\delta$ is varied. See the text for further details.}
\label{fig:DecoupledPlot2}
\end{figure}

In \tabref{tab:BM1011}, we provide a benchmark point for normal 
ordered leptogenesis, with purely low-energy $CP$ violation and 
$N_3$ decoupled. 
 At this point, 
the observed BAU is produced 
with a corresponding fine-tuning of $\mathcal{F} = 0.23$.
In \figref{fig:DecoupledPlot2}, we illustrate a similar scenario, 
in which the $CP$ violation is provided only by $\delta$ ($\alpha_{21}= 180^\circ$, $\alpha_{31}=0^\circ$), 
and where the observed baryon asymmetry is produced near 
$\delta = 270^\circ$.
We conclude that, even for the minimal type I seesaw 
scenario with two heavy Majorana neutrinos 
exhibiting hierarchical mass spectrum,
it is possible  to generate the observed 
value of the baryon asymmetry
with the requisite $CP$ violation 
provided exclusively  by the Dirac phase $\delta$,
and/or  by the Majorana phase $\alpha_{21}$ or $\alpha_{31}$.

Furthermore, we note that, in performing a 
similar investigation for the inverted ordering scenario, 
we find no point in the parameter space which 
corresponds to successful leptogenesis with $N_3$ 
decoupled in this mass window\footnote{
 For IO light neutrino mass spectrum 
the decoupling of $N_3$ implies $R_{13}=0$. 
In this case $m_3=0$ as well.
}
with real $R$-matrix. If, however, e.g., $R_{11}R_{12} = \pm i |R_{11}R_{12}|$ 
($R_{13}=0$ in the case of interest), 
we can have successful leptogenesis 
with the $CP$ violation provided by the Dirac and/or 
Majorana phases in PMNS matrix also 
for the IO spectrum. These conclusions are in agreement with  
the results of \cite{Pascoli:2006ie} 
wherein one may find a detailed discussion 
of the cases considered in the present subsection.
\begin{table}[t!]
\centering
\begin{tabular}{ c c  c c  c  c c  c  c c  }
 \toprule
$\delta$ & $\alpha_{21}$ & $\alpha_{31}$
&$M_1 $ & $M_2 $ & $\theta$   \\
(${}^{\circ}$) & (${}^{\circ}$) & (${}^{\circ}$)
& ($\text{GeV})$ & $(\text{GeV})$ & (${}^{\circ}$) \\
\specialrule{2.5pt}{1pt}{1pt}
$228$ & $516$ & $100$&
$10^{11}$ & $10^{12}$ & $25.05$ \\
\bottomrule
\end{tabular}
\caption{A benchmark point for leptogenesis with $M_1=10^{11}$ GeV and $N_3$ decoupled with a normal ordered light mass spectrum.}\label{tab:BM1011}
\end{table}

Finally, in \cite{Pascoli:2006ie} the following necessary condition 
for successful leptogenesis in the case of NO spectrum 
with the requisite $CP$ violation provided exclusively by
the Dirac phase $\delta$ was obtained: 
\begin{equation}
|\sin\theta_{13}\, \sin \delta|  \gtrsim 0.09\,.
\label{eq:LFth13d}
\end{equation}
We recall that this condition was derived  
by using values of the the $CP$-conserving $R$-matrix 
elements maximising the lepton asymmetry 
and assuming that the transition from two-flavour 
to one-flavour regime starts at $T\cong 5\times 10^{11}$ GeV, 
i.e., that at $M_1 \lesssim 5\times 10^{11}$ GeV the two-flavour 
regime is fully effective.

\section{Leptogenesis in the regime {$\mathbf{M_1 < 10^9}$} GeV}
\label{sec:lowM1}

Successful thermal leptogenesis at intermediate scales may be 
accomplished through the combination of flavour effects and 
fine-tuned Yukawa matrices with 
$\mathcal{F} \gtrsim\mathcal{O}(10)$~\cite{Lavignac:2002gf,Moffat:2018wke}. 
In Section~\ref{sec:reviewintlepto}, we first review these 
fine-tuned scenarios and then proceed to determine the subset 
among them in which the $R$-matrix is $CP$-conserving while 
the PMNS matrix contains $CP$-violating phases. 
In Section~\ref{sec:resultsintlepto} we present and analyse 
the results of a comprehensive search of the model parameter 
space for regions with successful leptogenesis compatible 
with these subsets where we have numerically solved the density matrix 
equations, for two-decaying heavy Majorana neutrinos, exactly. 
Following this,  we consider in detail the scenarios in 
which $CP$ violation is due solely to the Dirac phase in 
Section~\ref{sec:onlydelta}, or due only to the Majorana phases 
in Section~\ref{sec:onlyalpha21} and Section~\ref{sec:onlyalpha31}.
In \appref{appendix:FurtherResults}, we display results for $M_1 = 10^9$ GeV, where 
$\mathcal{O}(10)$  fine-tuning is also required.

We present an analytic approximation of the baryon asymmetry 
to find that the detailed dependence of the baryon asymmetry 
on the low energy phases may be roughly explained by the 
features of $Y_{\tau 1}$ and $Y_{\mu 1}$. We reiterate that 
we apply these approximation simply to illustrate the qualitative 
behaviour of the solutions but we  numerically solve the  
density matrix to produce all plots in this paper.

\subsection{Results of Parameter Exploration}
\label{sec:resultsintlepto}
\begin{figure}[t!]
  \includegraphics[width=1.0\textwidth]{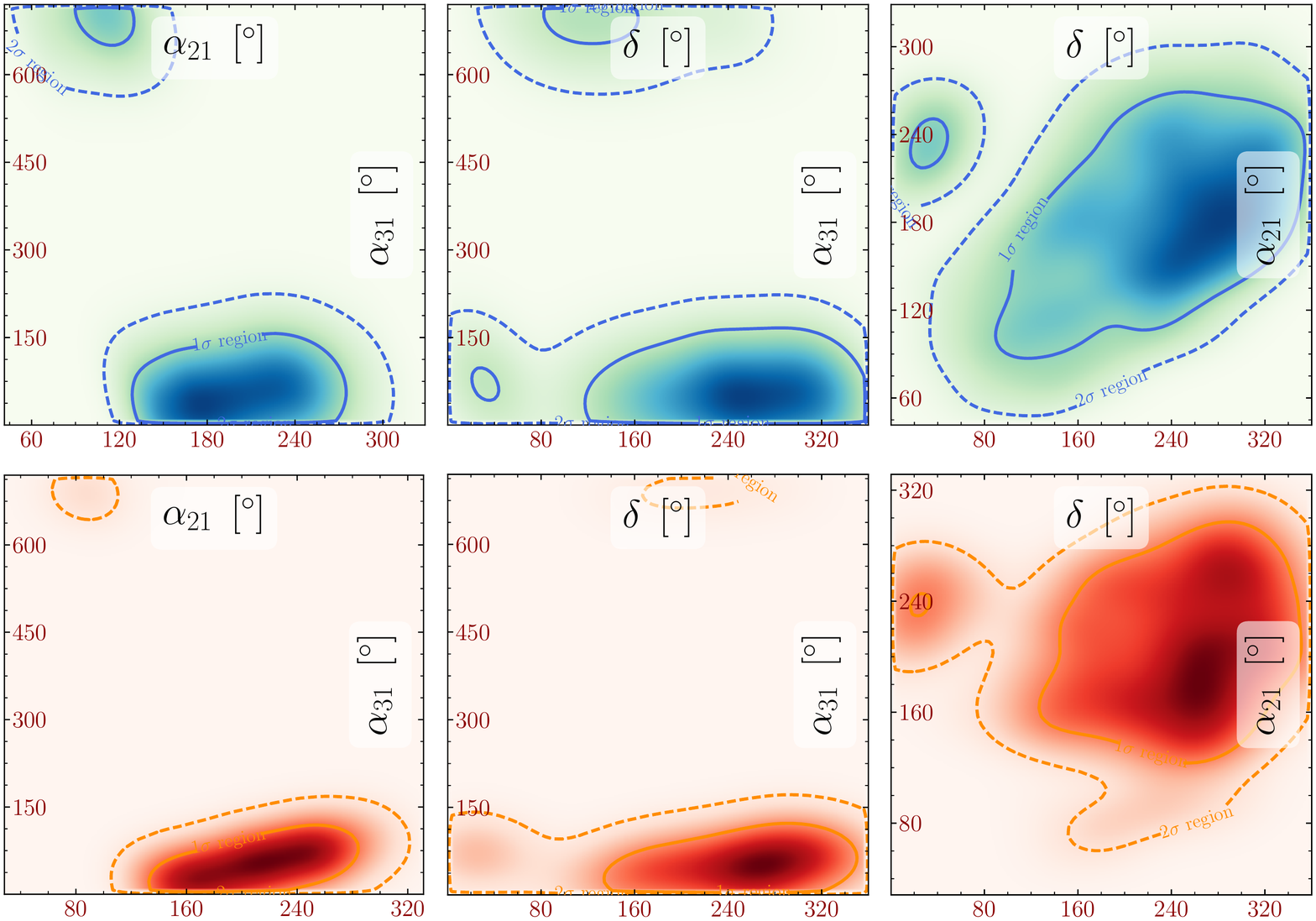}
\caption{The two-dimensional projections for intermediate scale leptogenesis 
with $M_1 = 3.16 \times 10^{6}$ GeV  for $x_1=0$, $y_2=0$,
$x_3=180^\circ$, $y_1=y_2=180^\circ$,
 with $CP$ violation provided only 
by the phases of the PMNS matrix. The normal ordered case is coloured 
blue/green and inverted ordering orange/red and contours correspond 
to $68\%$ and $95\%$ confidence levels. 
This plot was created using {\sc SuperPlot}~\cite{Fowlie:2016hew}.}
\label{fig:2DIntermediatePlots}
\end{figure}
\begin{figure}[t!]
  \includegraphics[width=1.0\textwidth]{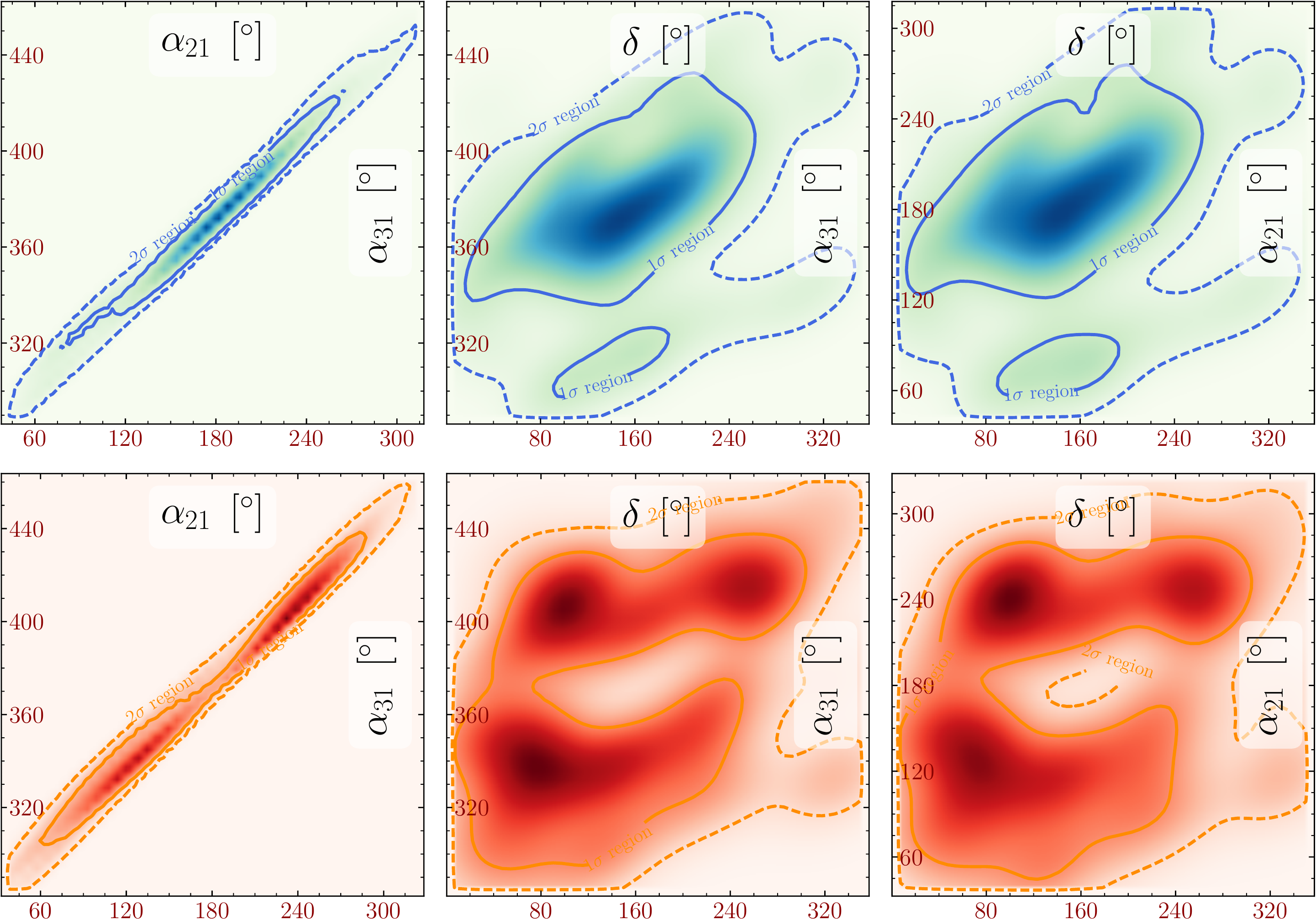}
\caption{The two-dimensional projections for intermediate scale 
leptogenesis with $M_1 = 1.29 \times 10^{8}$ GeV 
for $x_1=0$, $y_2=0$, $x_3=180^\circ$, $y_1=y_2=180^\circ$,
with $CP$ violation 
provided only by the phases of the PMNS matrix. The normal ordered 
case is coloured blue/green and inverted ordering orange/red and 
contours correspond to $68\%$ and $95\%$ confidence levels. 
This plot was created using {\sc SuperPlot}~\cite{Fowlie:2016hew}.}
\label{fig:2DPlot108}
\end{figure}

The options of \equaref{eq:RCPIoptions} are satisfied by 
sixteen distinct $R$-matrices which may be divided into 
four classes according to the corresponding parity 
vectors $\rho^{\nu}$, $\rho^N$ (see \appref{sec:RMatrices} 
for definitions and further details). All such matrices are 
identical except for the placement of factors 
$\pm 1$ or $\pm i$. The phenomenological implications of each 
will be qualitatively similar except for the precise positions 
in parameter space that certain features occur. As we are primarily 
concerned with demonstrating the viability of leptogenesis with 
the $\mathcal{O}(100)$ fine-tuned Yukawa matrices 
(of the type in \equaref{eq:ftDiracmass}) then we shall focus 
our numerical efforts on just one possible $R$-matrix of the set 
of sixteen. Namely, we choose a scenario corresponding 
to $\cos x_1 = 0$, $\cos x_3 =-1$ such 
that $\rho^{\nu} = \pm(+1, -1, +1)^T$, $\rho^N = \pm(+1, +1, -1)^T$.

 For the numerical analysis, we follow the same procedure outlined 
in \secref{sec:ParamGenericScale} with one additional constraint. 
At values of $\mathcal{F} \gtrsim 1000$, higher-order corrections 
to the light neutrino mass become important. For this reason we 
fix $y_1 = y_3 = 180^\circ$ and thereby avoid these problematic 
regions of the model parameter space. 

 In the parameter searches of this section, we consider two cases, 
in one we fix $M_1 \sim 10^6$ GeV (\figref{fig:2DIntermediatePlots}) and in the other we 
fix $M_1 \sim 10^8$ GeV (\figref{fig:2DPlot108}). We note that $M_{2} =  3.5 M_{1}$ and $M_{3} =  3.5 M_{2}$ and $m_1 = 0.21$ eV.
\footnote{In \appref{appendix:FurtherResults}, we demonstrate that one may lower $m_1$ as far as $0.05$ eV and still have successful leptogenesis in a albeit rather constrained parameter space.} This allows for a comparison of the effects 
of two different degrees of fine-tuning, with the former corresponding 
usually to $\mathcal{F} \sim 500$. This is close to the 
maximum fine-tuning (and correspondingly, the smallest 
non-resonant leptogenesis scale) for which second-order 
radiative corrections to the mass can be ignored~\cite{Moffat:2018wke}.

For the scenario in which $M_1 = 3.16 \times 10^6$ GeV, as anticipated, there is a large fine-tuning of $\mathcal{F} = 745$. In the normal ordered case, we find that the observed baryon asymmetry may be obtained to within $1 \sigma$ ($2 \sigma$) with $\delta$ between $[84, 360]^\circ$ ($[0,360]^\circ$). For inverted ordering, the $1 \sigma$ ($2 \sigma$) range is $[134, 350]^\circ$ ($[0,360]^\circ$). With $M_1 = 1.29 \times 10^8$ GeV, the fine-tuning is considerably less, at $\mathcal{F} = 12$.
In the normal ordered case, we find that the observed baryon asymmetry may be obtained to within $1 \sigma$ ($2 \sigma$) with $\delta$ between $[16, 263]^\circ$ ($[0,360]^\circ$). For inverted ordering, the $1 \sigma$ ($2 \sigma$) range is $[0,305]^\circ$ ($[0,360]^\circ$).  As in the previous section, we may explain these plots in detail by introducing an analytical approximation and the considering the simpler scenarios in which only the Dirac or only the Majorana phases provide $CP$ violation. For brevity, we choose to perform this analysis only for $M_1 \sim 10^8$ GeV in the normal ordered scenario.

\begin{table}[t!]
     \centering
\begin{tabular}{ c  c  c c  c  c c  c  c c }
 \toprule
$\delta$ & $\alpha_{21}$ & $\alpha_{31}$
&$M_1 $ & $M_2 $ & $M_3 $ 
& $x_1$ & $x_2$ & $x_3$ & $y_2$ \\
(${}^{\circ}$) & (${}^{\circ}$) & (${}^{\circ}$)
& ($\text{GeV})$ & $(\text{GeV})$ & $(\text{GeV})$ 
& (${}^{\circ}$) & (${}^{\circ}$) & (${}^{\circ}$) & (${}^{\circ}$)  \\
\specialrule{2.5pt}{1pt}{1pt}
$228$ & $189$ & $327.6$&
$7.00 \times 10^{8}$ & $1.55 \times 10^{10}$ & $3.80 \times 10^{10}$
&$90$ & $110$ & $180$
& $0$ \\
\bottomrule
\end{tabular}
\caption{A benchmark point for intermediate scale leptogenesis with 
 quasi-degenerate (QD) spectrum 
of the light neutrino masses.
In addition to the parameters listed we have 
$m_{1}=0.215$ eV and $y_1 = y_3 = -140^{\circ}$ and 
corresponding fine-tuning $\mathcal{F} \approx 30$.
}
\label{tab:BFIntermediateNO}
\end{table}

\subsection{Dependence of $\eta_B$ on Dirac and Majorana Phases}
\begin{figure}[t]
\centering
\includegraphics[width=1.0\textwidth]{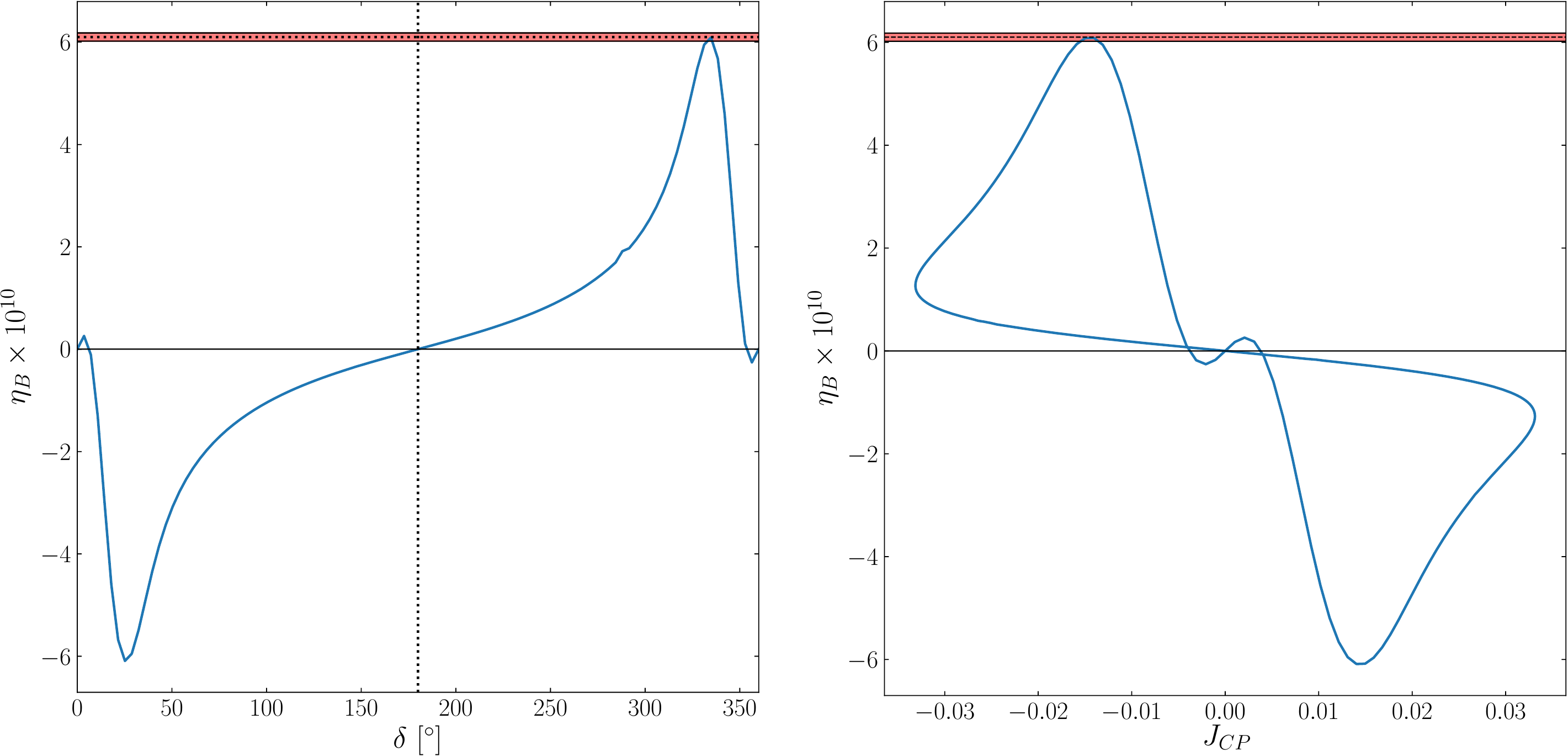}
\caption{Intermediate scale leptogenesis 
($M_1 = 7.00\times 10^8 \text{ GeV}$), with $CP$ violation provided 
solely by $\delta$, with $\alpha_{21} = 180^{\circ}$ and 
$\alpha_{31} = 0^{\circ}$. The red band indicates the $1 \sigma$ observed 
values for $\eta_{B_{CMB}}$ with the best-fit value indicated by 
the horizontal black dotted line. 
Left: The final baryon asymmetry as a function of $\delta$ with exact 
$CP$-invariance when $\delta = 0^{\circ}$ and $180^{\circ}$ 
(vertical black dotted line). 
Right: A parametric plot of $\eta_B$ against $J_{CP}$ as 
$\delta$ is varied at intermediate scales 
($M_1 = 7.00\times 10^8 \text{ GeV}$). See the text for further details.}
\label{fig:delta_etaB_plot_example}
\end{figure}

In this section, we use the benchmark point given in 
\tabref{tab:BFIntermediateNO}, in order to analytically study 
leptogenesis from low-energy $CP$ violation in the case that the 
lightest heavy Majorana neutrino has mass $M_1 < 10^{9}$ GeV, 
such that a relatively high degree of fine-tuning in the light 
neutrino masses is required. We choose this benchmark point as it 
allows us to accurately neglect the contributions from decays 
of the other heavy Majorana neutrinos and thus simplify the analysis.

With $M_1 < 10^9$ GeV, leptogenesis occurs in the three-flavour 
regime for which the three-flavoured Boltzmann equations 
are a good approximation to the density matrix equations 
and have approximate analytical solution~\cite{Fong:2013wr}:
\begin{equation}\label{eq:analyticalsolution1}
n_{B-L} = \frac{\pi^2}{6 z_d K_1} n^{\text{eq}}_{N_{1}} \left(z_0\right)  \left(\frac{\epsilon^{(1)}_{\tau \tau}}{P^{0(1)}_{\tau \tau}} + \frac{\epsilon^{(1)}_{\mu \mu}}{P^{0(1)}_{\mu \mu}}+\frac{\epsilon^{(1)}_{e e}}{P^{0(1)}_{e e}}\right),
\end{equation}
where we take into account the decays of only the lightest 
heavy Majorana neutrino. As we are most interested in the scenarios 
in which $CP$ violation is due to PMNS phases only, 
i.e.\ $\Tr \epsilon^{(1)} = 0$, then we can re-express 
\equaref{eq:analyticalsolution1} as
\begin{equation}\label{eq:analyticsolBE}
n_{B-L} = \frac{\pi^2}{6 z_d K_1} n^{\text{eq}}_{N_{1}} \left(z_0\right)  \left( \epsilon^{(1)}_{\tau \tau} \Delta F_{\tau e}  + \epsilon^{(1)}_{\mu \mu} \Delta F_{\mu e} \right),
\end{equation}
where the asymmetry depends on the low-energy phases 
via $\epsilon^{(1)}_{\tau \tau}$ and $\epsilon^{(1)}_{\mu \mu}$ 
and from the difference of the inverse flavour projections:
\begin{equation}
\Delta F_{\tau e} \equiv \frac{1}{P^{0(1)}_{\tau \tau}} 
- \frac{1}{P^{0(1)}_{e e}} \text{,} 
\quad \Delta F_{\mu e} \equiv \frac{1}{P^{0(1)}_{\mu \mu}}-\frac{1}{P^{0(1)}_{e e}}.
\end{equation}
However, for the case of \tabref{tab:BFIntermediateNO}, the two and 
three-flavour regime Boltzmann equations, to a high degree of accuracy 
give the same value of $\eta_B$. Given the comparative simplicity of 
the two-flavour solution \equaref{eq:2FSol}, we choose to use this 
for the practical purpose of simplifying the analysis.

For the benchmark parameter values listed in 
\tabref{tab:BFIntermediateNO}, we may find analytical approximations 
for the $CP$-asymmetries $\epsilon^{(1)}_{\alpha \alpha}$. Under the relatively 
good approximation, that $m_1=m_2$
\footnote{The approximation  $m_1=m_2$ is sufficiently 
precise as long as 
$m^2_1 \gg 0.5\Delta  m^2_{21} 
\cong 3.7\times 10^{-5}$ eV$^2$.}, 
the asymmetry is given by
\begin{equation}
\begin{aligned}
\epsilon^{(1)}_{\alpha \alpha} & = \frac{3}{16 \pi \left(Y^{\dagger} Y\right)_{11}} \frac{M_1^2}{v^4} m_1 \sqrt{m_1 m_3} (e^{y_3} \sin 2 x_2) \lvert R_{21} \rvert^2 \\ & \left( \frac{2}{3} \left( \frac{M_1}{M_2} + \frac{M_1}{M_3}\right) \Im \left[e^{- i x_3} \left( U_{\alpha 1} + i U_{\alpha 2} \right) U^{*}_{\alpha 3} \right]  - \frac{5}{9} \frac{M_1^2}{M_2^2} \Im \left[ e^{- 2 i (x_1 + x_3)} e^{- i x_3} (U_{\alpha 1} + i U_{\alpha 2} ) U^{*}_{\alpha 3} \right] \right).
\end{aligned}
\end{equation}
Selecting $x_1 = (2 k_1 +1) \pi / 2$ and $x_3 = k_3 \pi$ for $k_1$, 
$k_3 \in  \mathbb{Z}$, such that $\cos x_1 = 0$ and 
$\vert \cos x_3 \rvert = 1$ and $\cos x_3 = (-1)^{k_3}$ is satisfied, 
we find the $CP$-asymmetry $\epsilon^{(1)}_{\alpha \alpha}$ to be
\begin{equation}
\begin{aligned}
\epsilon^{(1)}_{\alpha \alpha} & = \frac{3}{16 \pi \left(Y^{\dagger} Y\right)_{11}} 
\frac{M_1^2}{v^4} m_1^{\frac{3}{2}} m_3^{\frac{1}{2}} (e^{y_3} \sin 2 x_2) 
(-1)^{k_3} \lvert R_{21} \rvert^2 \\ 
& \times \left( \frac{2}{3} \left( \frac{M_1}{M_2} 
+ \frac{M_1}{M_3}\right) + \frac{5}{9} \frac{M_1^2}{M_2^2} \right) 
\Im \left[ \left( U_{\alpha 1} + i U_{\alpha 2} \right) U^{*}_{\alpha 3} \right],
\end{aligned}
\end{equation}
where, at our benchmark point, the coefficient of 
$\Im\left[ U^{*}_{\alpha 3} \left( U_{\alpha 1} + i U_{\alpha 2} \right) \right]$ 
has magnitude approximately equal to $3.7 \times 10^{-6}$. This form is 
particularly useful in order to isolate the dependence of the 
$CP$-asymmetry on the PMNS phases in the factor 
$\Im\left[ U^{*}_{\alpha 3} \left( U_{\alpha 1} + i U_{\alpha 2} \right) \right]$.

\subsubsection{Dirac Phase $CP$ Violation}
\label{sec:onlydelta}
\begin{figure*}[t]
\centering
\includegraphics[width=1.0\textwidth]{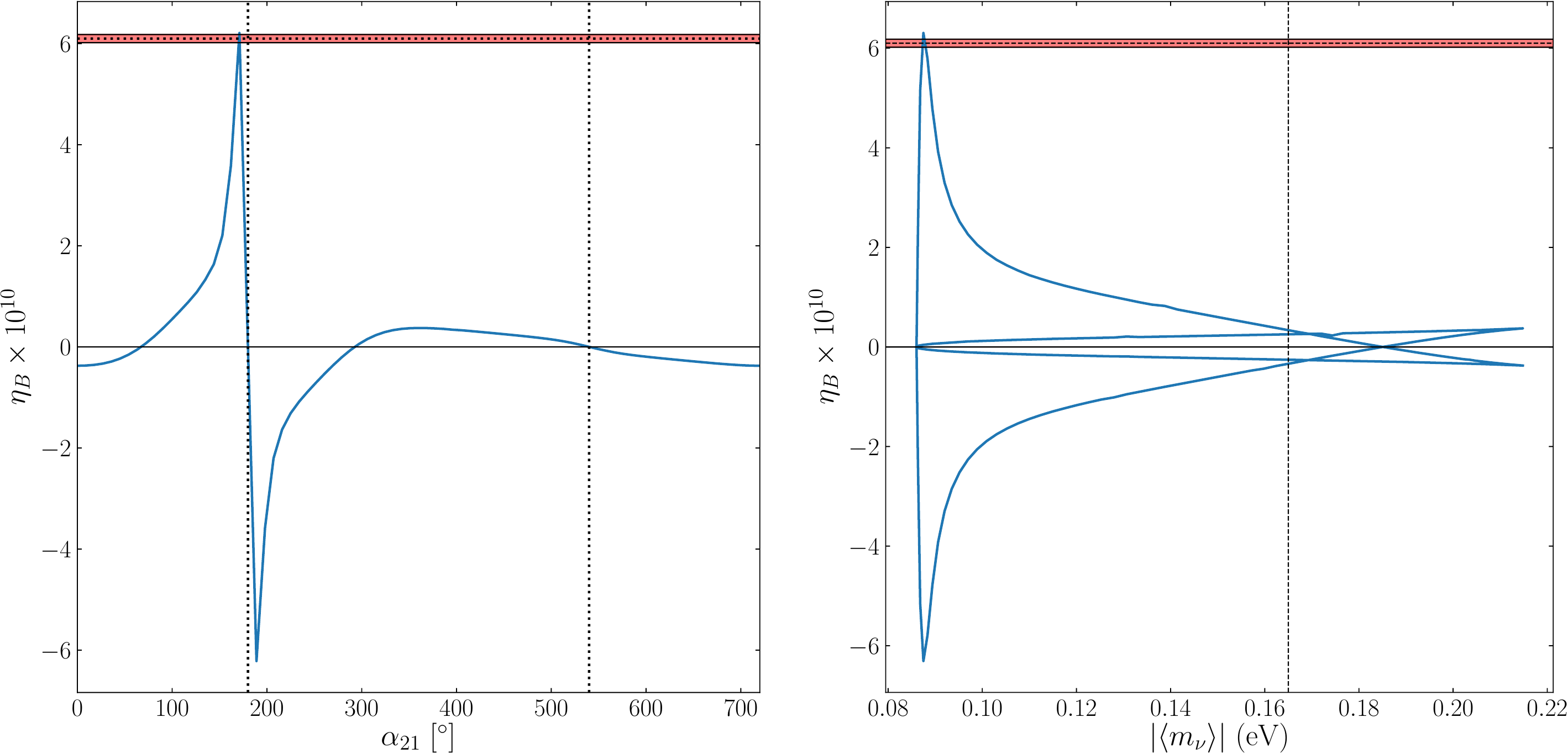}
\caption{Intermediate scale leptogenesis ($M_1 = 7.00 \times 10^8$ GeV) 
with $CP$ violation provided solely by $\alpha_{21}$ and with 
$\delta = \alpha_{31} = 0$. The red band indicates the $1 \sigma$ observed 
values for $\eta_B$ with the best-fit value indicated by the horizontal 
black dotted lines. Left: The baryon asymmetry against 
$\alpha_{21}$ with exact $CP$-invariance at 
$\alpha_{21} = 180^{\circ}$ and $540^{\circ}$ 
(vertical black dotted lines). Right: A parametric plot of $\eta_B$ against 
the effective neutrino mass $\lvert \langle m_{\nu} \rangle \rvert$ 
as $\alpha_{21}$ is varied with the vertical dashed black line denoting
the upper value of the KamLAND-Zen bound 
$0.165 \text{ eV}$~\cite{KamLAND-Zen:2016pfg}. Successful leptogenesis is achieved for $\lvert \langle m_{\nu} \rangle \rvert = 0.0877$ eV.
See the text for further details.
}
\label{fig:alpha21_etaB_plot_example}
\end{figure*}

 We consider the possibility that the Majorana phases are $CP$-conserving: 
$\alpha_{21} = 180^{\circ}$, $\alpha_{31} = 0^{\circ}$  
(given the $R$-matrix under consideration). 
The sole source of $CP$ violation 
is $\delta$ and there is exact $CP$-invariance 
 if $\delta = 0^{\circ},180^\circ$.
The corresponding $\eta_B$ is plotted in  
\figref{fig:delta_etaB_plot_example} alongside a 
parametric plot of 
$\eta_B$ against $J_{CP}$ with parameter $\delta$.\footnote{All plots involving $\eta_B$ in this work have been obtain 
by solving the full density matrix equations, allowing for the lightest 
pair of heavy Majorana neutrinos to decay and possibly 
(if indicated) include scattering effects.
}

From the $CP$-asymmetry, one expects to find $\eta_B$ proportional to
\begin{equation}\label{eq:intdeltacp}
\begin{aligned}
\Im\left[ U^{*}_{\tau 3} \left( U_{\tau 1} + i U_{\tau 2} \right) \right] 
& = s_{13} c_{13} c_{23}^2 (s_{12}  -  c_{12}) \sin \delta 
\approx -0.0178 \sin \delta,
\end{aligned}
\end{equation}
and thus sinusoidal in $\delta$. However, the phase-dependent 
 efficiency (flavour-factor)
$\Delta F$ exhibits a sharp maximum around the 
region $\delta = 0^\circ$ or $\delta = 360^\circ$. This modifies the 
sinusoidal dependence from the $CP$-asymmetries such that the 
extrema
of $\eta_B$ are shifted towards the extreme values of $\delta$, 
as in seen in
\figref{fig:delta_etaB_plot_example}. 
The small fluctuations around $\delta = 0^\circ$, $\delta = 360^\circ$ 
are not captured by the Boltzmann equations (neither two-flavoured nor 
three-flavoured) and are only present when solving the full density matrix 
equations which take account of the finite size of the lepton thermal widths. 
The result is the addition of some accidental zeros in the variation 
$\eta_B$ which do not correspond to  
 $CP$-conserving values of $\delta$.

\subsubsection{$CP$ Violation from the Majorana Phase $\alpha_{21}$}
\label{sec:onlyalpha21}

Alternatively, consider the case of $CP$ violation from $\alpha_{21}$, 
where $\delta = 0^{\circ}$, $\alpha_{31} = 0^{\circ}$ and all other 
parameters are set to their benchmark values of \tabref{tab:BFIntermediateNO}. 
The variation of $\eta_B$ with $\alpha_{21}$ in this scenario is plotted 
on the left of \figref{fig:alpha21_etaB_plot_example}. On the right of 
the same figure, we parametrically plot $\eta_B$ against 
$\vert \langle m_{\nu} \rangle \rvert$ with parameter $\alpha_{21}$.
The baryon asymmetry $\eta_B$ vanishes 
at the $CP$-conserving values of $\alpha_{21} = 180^{\circ}$ and  $540^{\circ}$.
However, as is seen in \figref{fig:alpha21_etaB_plot_example},
$\eta_B\neq 0$ at the $CP$-conserving values of 
$\alpha_{21} = 0^{\circ}$, $360^{\circ}$ and  $720^{\circ}$ 
since at these values the interplay between 
the $CP$-conserving $R$ and  PMNS matrices leads to 
$CP$ violation in leptogenesis \cite{Pascoli:2006ie}.

 The  efficiency
function $\Delta F$, when plotted as a function of $\alpha_{21}$, 
exhibits a very strong narrow peak at $\alpha_{21} = 180^{\circ}$ and a 
much less pronounced peak at $\alpha_{21} = 540^{\circ}$. As a consequence, 
the corresponding $\eta_B$ is modified from the simple cosine curve 
expected from the dependence of $\epsilon^{(1)}_{\tau \tau}$ and 
$\epsilon^{(1)}_{\mu \mu}$ on $\alpha_{21}$, which arises 
in the factors:
\begin{equation}
\begin{aligned}
\Im\left[ U^{*}_{\tau 3} \left( U_{\tau 1} + i U_{\tau 2} \right) \right] 
& = - c_{13} c_{23} (c_{12} s_{23} +  s_{12} c_{23} s_{13}) 
\cos \frac{\alpha_{21}}{2} \approx -0.444 \cos \frac{\alpha_{21}}{2},
\end{aligned}
\end{equation}
Thus, there is a sharp transition around $\alpha_{21} = 180^{\circ}$. 
We can conclude then that the strong peak in $\Delta F$ is what has 
allowed the observed baryon asymmetry of the Universe to be reproduced. 
This peak originates in an accidental cancellation of terms in the 
function $P^{0(1)}_{\tau \tau}$. There is no \textit{a priori} reason to 
expect such a cancellation and it should be understood as a feature of 
the fine-tuned solutions that are being here studied. Thus, we see that 
the flavour-effects introduce a pair of accidental zeros of $\eta_B$, 
one in the range $[180,540]^\circ$ and the other in $[0,180]^\circ$.

\subsubsection{$CP$ Violation from the Majorana Phase $\alpha_{31}$}
\label{sec:onlyalpha31}
\begin{figure}[t]
\centering
\includegraphics[width=1.0\textwidth]{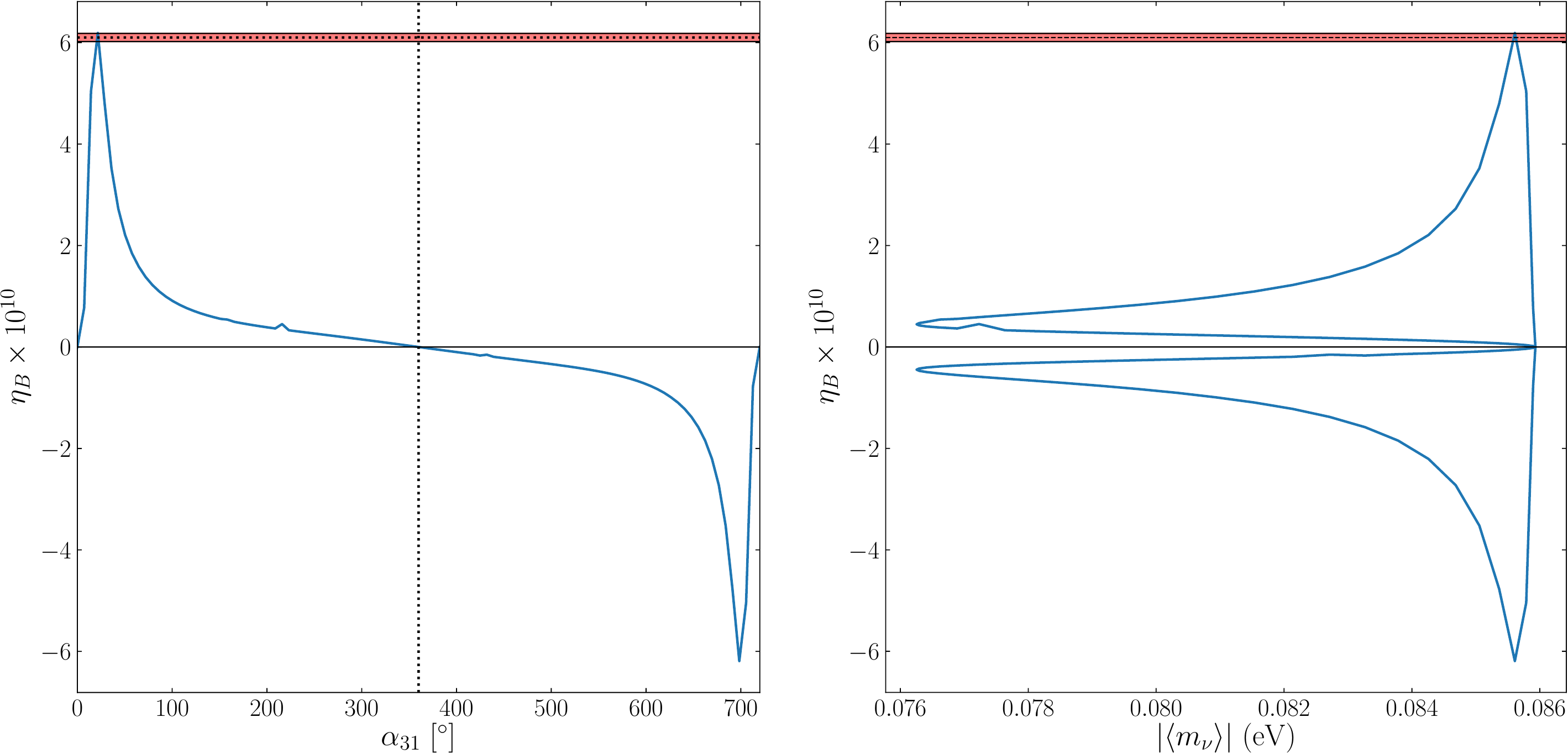}
\caption{Leptogenesis at intermediate scales ($M_1 = 7.00 \times 10^8$ GeV) 
when $CP$ violation is provided solely by $\alpha_{31}$ with 
$\delta = 0^{\circ}$, $\alpha_{21} = 180^{\circ}$. The red band indicates 
the $1 \sigma$ observed values for $\eta_B$ with the best-fit value 
indicated by the horizontal black dotted lines. 
Left: The baryon asymmetry as a function of $\alpha_{31}$. 
Exact $CP$-invariance exists 
 for 
$\alpha_{31} = 180^{\circ}$ and $360^{\circ}$
(vertical black dotted lines). 
Right: A parametric plot of $\eta_B$ against the effective 
neutrino mass $\lvert \langle m_{\nu} \rangle \rvert$ as 
$\alpha_{31}$ is varied with successful leptogenesis at $\lvert \langle m_{\nu} \rangle \rvert = 0.0856$ eV. 
 See the text for further details.}
\label{fig:alpha31_etaB_plot_example}
\end{figure}

Finally, consider the case of $CP$ violation from 
$\alpha_{31}$ where $\delta = 0^{\circ}$, $\alpha_{21} = 180^{\circ}$ 
and for which the baryon asymmetry is plotted in the left plot 
of  \figref{fig:alpha31_etaB_plot_example} and on the right we 
show the parametric dependence of the effective neutrino mass 
$\lvert \langle m_{\nu} \rangle \rvert$ with $\alpha_{31}$ 
against that of $\eta_B$.
The baryon asymmetry $\eta_B$ vanishes 
at the $CP$-conserving values of $\alpha_{31} = 0^\circ$,
$360^{\circ}$ and  $720^{\circ}$.

The efficiency
function $\Delta F$ in this case is qualitatively similar 
to that for the case of $\delta$: $CP$ violation only strongly 
peaks at 
$\alpha_{31}$ close to $0^{\circ}$ and to $720^{\circ}$.
Thus, we do not observe, in the left plot of 
\figref{fig:alpha31_etaB_plot_example}, the simple dependence, 
$\eta_B \propto \sin \left( \alpha_{31}/2 \right)$ as may be expected 
 from the expression for $\epsilon_{\tau \tau}$,
\begin{equation}
\begin{aligned}
\Im\left[ U^{*}_{\tau 3} \left( U_{\tau 1} + i U_{\tau 2} \right) \right] & = c_{13} c_{23} ((c_{12} - s_{12}) s_{23} + (c_{12} + s_{12}) s_{13} c_{23}) \sin \frac{\alpha_{31}}{2} \approx -0.662 \sin \frac{\alpha_{31}}{2}.
\end{aligned}
\end{equation}
Rather, we find 
an enhanced positive peak near $\alpha_{31} = 0^{\circ}$ 
and an enhanced negative peak near $\alpha_{31} = 720^{\circ}$. 

\subsubsection{Summary of Fine-Tuned Solutions with High Energy $CP$-Symmetry}
\label{sec:summary}

The fine-tuned solutions we have discussed in this section share 
the property that they enhance $\eta_B$ through the production of a 
peak in the efficiency factor
$\Delta F$. In each projection coefficient,
\begin{equation}
P^{0(1)}_{\alpha \alpha} 
= \frac{\lvert Y_{\alpha 1} \rvert^2}{\left(Y^{\dagger} Y\right)_{11}},
\end{equation}
the PMNS matrix cancels from the denominator such that 
all phase-dependence comes from that of  $\lvert Y_{\alpha 1} \rvert^2$ 
in the numerator. In this subsection, we may safely use the usual 
Casas-Ibarra parametrisation (obtained by the replacement of 
$f$ with $\sqrt{M^{-1}}$ in \equaref{eq:CIloop}), 
to obtain
\begin{equation}\label{eq:YukawaExpansion}
Y_{\alpha 1} = \sqrt{M_1} \left( \sqrt{m_1} R_{11} U_{\alpha 1} 
+ \sqrt{m_2} R_{12} U_{\alpha 2} + \sqrt{m_3} R_{13} U_{\alpha 3} \right).
\end{equation}
The absolute value $\lvert Y_{\alpha 1}  \rvert$ is extremised when 
each term in the parentheses in \equaref{eq:YukawaExpansion} 
has a common complex phase or when terms may differ in complex phase 
by $\pi$. This occurs at $CP$-conserving values of the PMNS phases and 
so the enhancement expected in the functions $\Delta F$ is 
likely to occur at $CP$-conserving phases also. As an example, 
around the benchmark point of \tabref{tab:BFIntermediateNO}, 
we find that with only $\alpha_{21}$ contributing 
to $CP$ violation ($\delta = \alpha_{31} = 0^\circ$),
\begin{equation}
\label{eq:YukawaTau}
Y_{\tau 1} = \left( 2.16+2.23e^{i \frac{\alpha_{21}+\pi}{2}} \right) \times 10^{-3}.
\end{equation}
The absolute value of this function has extrema when 
$\alpha_{21} = 180^{\circ}$ or $\alpha_{21} = 540^{\circ}$ 
- the $CP$-conserving values. Moreover, the cancellation that 
occurs at $\alpha_{21} = 180^{\circ}$ is strong because of 
the similarity in magnitude of the two terms in \equaref{eq:YukawaTau}. 
As a result of this there are strong peaks in 
$\Delta F$ which enhance $\eta_B$. 
 
 This is a way in which the solutions are found to 
be fine-tuned as there is  no reason to expect these 
two terms to be so similar in size. At these same points, 
the asymmetries $\epsilon^{(1)}_{\alpha \alpha}$ are vanishing 
as $CP$ is a symmetry in the leptonic sector. 
Thus $\eta_B$, being proportional to the product of 
$\epsilon^{(1)}_{\tau \tau}$ and $\Delta F$, is strongly enhanced 
on either side of the $CP$-invariant points 
(for instance, around $\alpha_{21} = 180^{\circ}$ in the left plot of  
\figref{fig:alpha21_etaB_plot_example}). Thus the fine-tuned solutions 
tend to achieve large $\eta_B$ of one-sign on one side of a 
$CP$-invariant point and large $\eta_B$ of the opposite sign 
on the other side. Similarly, this effect persists when all 
phases may contribute together to $CP$ violation 
(\figref{fig:2DIntermediatePlots} and \figref{fig:2DPlot108}). 
Thus, successful leptogenesis tends to occur near 
$\alpha_{21} \sim 180^{\circ}$, $\alpha_{31} \sim 0^{\circ} \text{, } 
720^{\circ}$ when leptogenesis is achieved with fine-tuned 
light neutrino masses, as it is at intermediate scales 
($M_1 \lesssim 10^9$ GeV). Note that although we made these arguments 
based on the two-flavoured Boltzmann equations, very similar conclusions 
are reached based on considerations of $\Delta F_{\tau e}$ and 
$\Delta F_{\mu e}$ for the solutions of the three-flavoured 
Boltzmann equations. For this reason, one expects similar behaviour 
to hold even for lower values $M_1$ such as in \figref{fig:2DIntermediatePlots}.

 Furthermore, one may usually argue that Dirac-phase leptogenesis 
suffers a suppression not present in Majorana phase leptogenesis 
due to the factors of $s_{13}$ that appear in the $CP$-asymmetries 
as shown in \equaref{eq:intdeltacp}. However, for Dirac-phase leptogenesis 
$\eta_B \propto \sin \delta \Delta F (\delta)$, where the maximum 
absolute value of $\Delta F (\delta)$ is $\sim 408$, whereas 
for $\alpha_{21}$ leptogenesis, 
$\eta_B \propto \cos \frac{\alpha_{21}}{2} \Delta F (\alpha_{21})$ 
with maximum absolute value $\sim 77$. Thus, what is more relevant 
when leptogenesis occurs intermediate scales, is the degree of 
enhancement from $\Delta F$ that occurs due to fine-tuning.

 Finally, as we observe  in \figref{fig:2DPlot108}, the contours 
for $\alpha_{21}$, $\alpha_{31}$ show a strong dependence on 
$\alpha_{31}-\alpha_{21}$. A rough explanation of this is given by 
the dependence of $\epsilon^{(1)}_{\tau \tau}$ on the Majorana phases.
With $\delta$ 
fixed at its benchmark value, but $\alpha_{21}$ and $\alpha_{31}$ 
free to vary, this CP-asymmetry is given by
\begin{equation}
\epsilon^{(1)}_{\tau \tau} 
\approx \left(1.46 \cos \frac{\left( \alpha_{31} - \alpha_{21} \right) }{2} 
+  0.869 \sin \frac{\alpha_{31}}{2} \right) \times 10^{-7},
\end{equation}
which exhibits a slightly dominant, $(\alpha_{31}-\alpha_{21})$-dependent 
contribution.
 This contribution
is maximised when $\alpha_{31} = \alpha_{21}$.

\section{Leptogenesis in the regime $\mathbf{M_1 > 10^{12}}$ GeV}\label{sec:HSLG}
In previous studies in which a connection between low-energy 
$CP$ violation ($CP$-conserving $R$) and 
leptogenesis was established~\cite{Pascoli:2006ci}, the scale of leptogenesis was 
 limited to 
$M_1 \leq 5\times 10^{11} \text{ GeV}$.
This allowed for the use of the two-flavour Boltzmann equations as shown in 
\equaref{eq:BE2F} where the $CP$-asymmetries $\epsilon^{(1)}_{\tau \tau}$ 
and $\epsilon^{(1)}_{\tau^{\perp} \tau^{\perp}}$ appear separately. 
The expectation had been that for $M_1 \gg 10^{12} \text{ GeV}$, 
the single-flavour Boltzmann equation \equaref{eq:BE1F} would be 
appropriate. In this equation, the $CP$-asymmetries appear only in 
the factor $\Tr \epsilon^{(1)} = 0$ and hence no baryon asymmetry 
may be produced.  In \secref{sec:DMEtoBE1F} we argue that even 
at high scales $M_1 \gg 10^{12} \text{ GeV}$, if $R$ is $CP$-conserving, 
then flavour effects are significant and that the density matrix 
equations do not reduce to the single flavour Boltzmann equations. 
Hence we conclude that viable leptogenesis may result from low energy 
$CP$ violation. Finally, in \secref{sec:highscaleresults} 
we proceed to numerically analyse this possibility in detail.

\subsection{Flavour Effects with $M_1 \gg 10^{12} \text{ GeV}$ 
and High Energy $CP$-Symmetry}
\label{sec:DMEtoBE1F}
\begin{figure}[t!]
  \includegraphics[width=1.0\textwidth]{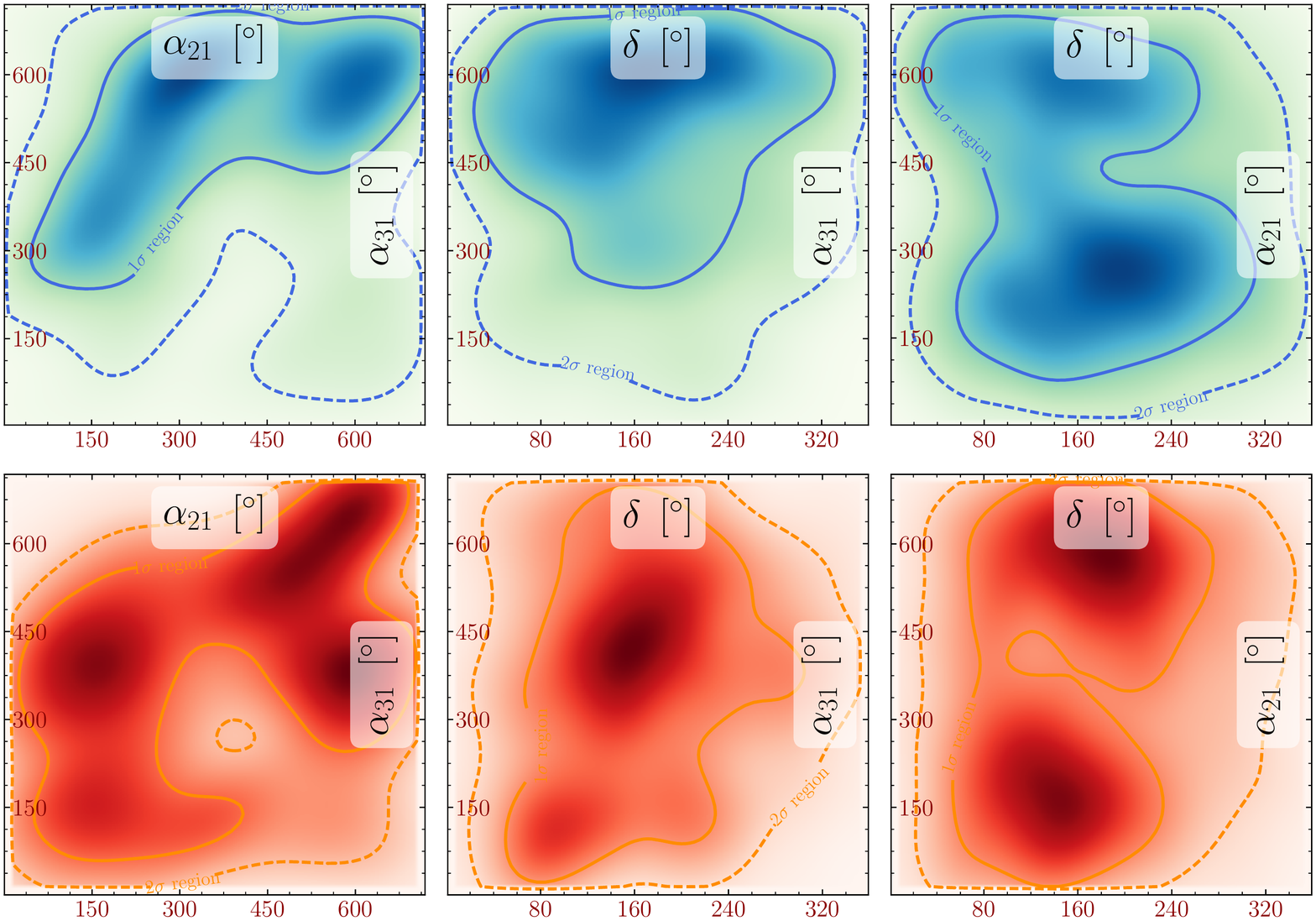}
 \caption{The two-dimensional projections for high-scale leptogenesis with 
$M_1 = 10^{13}$ GeV with $CP$ violation provided only by the phases 
of the PMNS matrix. The 
NO case is coloured blue/green and 
the IO one is orange/red. The contours correspond to $68\%$ and $95\%$ 
confidence levels. This plot was created using 
{\sc SuperPlot}~\cite{Fowlie:2016hew}.}
\label{fig:2DProjections_highscale}
\end{figure}

In \appref{appendix:DMEtoBE1F}, we demonstrate that the complete formal 
solution of the density matrix equations (\equaref{eq:full3}), 
with one decaying 
heavy Majorana neutrino is
\begin{equation}
\begin{aligned}\label{eq:solnBminusL1}
n_{B-L}(z_f) = \int_{z_0}^{z_f} e^{-\int_{z'}^{z_f} W_1(z'') dz''} 
\left(\Tr \epsilon^{(1)} D_1(z') 
(n^{}_{N_1}(z') - n_{N_1}^{\text{eq}}(z')) + W_1(z') \lambda (z') \right) dz',
\end{aligned}
\end{equation}
with
\begin{equation}
\lambda (z) \equiv 
2 \int_{z_0}^z dz' \Re \left[C_{1 \tau^{\perp}} C_{1\tau}^{*} 
\frac{\Im(\Lambda_{\tau})}{Hz'} n_{\tau \tau^{\perp}}(z')\right].
\end{equation}
In a typical leptogenesis scenario, if $M_1 \gg 10^{12} \text{ GeV}$, 
flavour effects are negligible and one obtains the well-known result:
\begin{equation}
\begin{aligned}\label{eq:solnBminusL2}
n_{B-L}(z_f) = \int_{z_0}^{z_f} e^{-\int_{z'}^{z_f} W_1(z'') dz''} 
\Tr \epsilon^{(1)} D_1(z') (n^{}_{N_1}(z') - n_{N_1}^{\text{eq}}(z')) dz',
\end{aligned}
\end{equation}
which may be found by solving the single flavour Boltzmann equation. 
However, with a $CP$-conserving $R$-matrix, such that $CP$ violation 
is provided solely by the PMNS phases, one has $\Tr \epsilon^{(1)} = 0$ and 
so the $\lambda$ term in \equaref{eq:solnBminusL1} becomes the dominant one:
\begin{equation}\label{eq:OnlyFlavSol}
n_{B-L}(z_f) = \int_{z_0}^{z_f} e^{-\int_{z'}^{z_f} W_1(z'') dz''} W_1(z') \lambda(z') dz'.
\end{equation}
If this is the case, then the baryon asymmetry is produced 
purely through flavour-effects from $\Im(\Lambda_{\tau})/Hz$.

 The physical effect of  $\Tr \epsilon^{(1)} = 0$ is that opposite 
asymmetries are produced in the $\tau$ and $\tau^{\perp}$ flavours due to 
the decay of $N_1$: 
$\epsilon^{(1)}_{\tau \tau} = - \epsilon^{(1)}_{\tau^{\perp} \tau^{\perp}}$. 
However, with flavour effects, 
the lepton asymmetries $\epsilon^{(1)}_{\tau \tau}$ 
and $\epsilon^{(1)}_{\tau^{\perp} \tau^{\perp}}$ produced in the decay experience
differing washouts 
such that $n_{\tau \tau} \neq - n_{\tau^{\perp} \tau^{\perp}}$ and 
$n_B = n_{\tau \tau} + n_{\tau^{\perp} \tau^{\perp}} \neq 0$.
It is an asymmetry 
produced by this method that is described in \equaref{eq:OnlyFlavSol}. 
The obvious question at this point is whether this can ever be 
large enough to produce the observed baryon asymmetry when 
$M_1 \gg 10^{12} \text{ GeV}$.
\begin{table}[t!]
\centering
\begin{tabular}{ c  c  c  c  c  c  c  c  c  }
 \toprule
$\delta$ & $\alpha_{21}$ & $\alpha_{31}$
&$M_1$ & $M_2$ & $M_3$ 
&$x_1$ & $x_2$ & $x_3$  \\
(${}^{\circ}$) & (${}^{\circ}$) & (${}^{\circ}$)
& ($\text{GeV})$ & $(\text{GeV})$ & $(\text{GeV})$ 
& (${}^{\circ}$) & (${}^{\circ}$) & (${}^{\circ}$) \\
\specialrule{2.5pt}{1pt}{1pt}
$228$ & $200$ & $175$
&$10^{13}$ & $1.2 \times 10^{15}$ & $10^{16}$
&$-96.55$ & $-105.2$ & $141.4$\\
\bottomrule
\end{tabular}
\caption{The benchmark values for high-scale leptogenesis with normal ordering. Here we have $m_{1} = 0.0159$ eV and $y_{1}=y_{2}=y_{3}=0^\circ$.}
\label{tab:tableBFhighscaleNO}
\end{table}

The density matrix equations may be conveniently expressed 
in terms of the vectors
\begin{align}
\boldsymbol{n} & \equiv (n_{\tau^{\perp} \tau^{\perp}},n_{\tau \tau^{\perp}},n_{\tau^{\perp} \tau},n_{\tau \tau})^T,\\
\boldsymbol{\epsilon^{(1)}} & \equiv (\epsilon^{(1)}_{\tau^{\perp} \tau^{\perp}},\epsilon^{(1)}_{\tau \tau^{\perp}},\epsilon^{(1)}_{\tau^{\perp} \tau},\epsilon^{(1)}_{\tau \tau})^T,
\end{align}
as
\begin{equation}
\frac{d \boldsymbol{n}}{dz} = \boldsymbol{\epsilon^{(1)}} D_1 \left( n^{}_{N_1} - n_{N_1}^{\text{eq}} \right) - \frac{1}{2} \mathcal{W}_1 \boldsymbol{n}-\frac{\Im\left(\Lambda_{\tau}\right)}{Hz} \mathcal{I} \boldsymbol{n},
\end{equation}
where 
\begin{equation}
\mathcal{W}_1 \equiv 
W_1
\left(\begin{array}{cccc}
2 \lvert C_{1\tau^{\perp}} \rvert^2      & C_{1 \tau}^{*} C_{1 \tau^{\perp}} & C_{1 \tau} C_{1 \tau^{\perp}}^{*} & 0\\
C_{1 \tau} C_{1 \tau^{\perp}}^{*}        & 0                                                                & 1                                 & C_{1 \tau} C_{1 \tau^{\perp}}^{*}\\
C_{1 \tau}^{*} C_{1 \tau^{\perp}}        & 0                                                                & 1 & C_{1 \tau}^{*} C_{1 \tau^{\perp}}\\
0 & C_{1 \tau}^{*} C_{1 \tau^{\perp}} & C_{1 \tau} C_{1 \tau^{\perp}}^{*} & 2 \lvert C_{1\tau^{\perp}} \rvert^2
\end{array}\right) \quad\text{ and } \quad \mathcal{I} \equiv
\left(\begin{array}{cccc}
0      & 0 & 0 & 0\\
0      & 1 & 0 & 0\\
0      & 0 & 1 & 0\\
0      & 0 & 0 & 0\\
\end{array}\right).
\end{equation}
In terms of these quantities, the formal solution, with flavour effects 
neglected is:
\begin{equation}
\label{eq:formalvectorsolution}
\boldsymbol{n}\left( z_f \right) 
= \int_{z_0}^{z_f} e^{\int_{z'}^{z_f} \frac{1}{2} 
\mathcal{W}_1 \left( z'' \right) dz''} \boldsymbol{\epsilon^{(1)}} 
D(z') (n_{N_1}(z') - n_{N_1}^{\text{eq}}(z')) dz'.
\end{equation}
Although flavour effects in  high-scale leptogenesis may be negligible, 
this solution may not be accurately applied for finding $\eta_B$ 
in the case that $\Tr \epsilon^{(1)} = 0$.
This is because there is 
a strong cancellation of components of the density matrix when 
computing $\eta_B = n_{\tau^{\perp} \tau^{\perp}} + n_{\tau \tau}$, such that the 
errors made in neglecting flavour effects are dominant. For this reason, 
we make use of it only for finding the approximate behaviour of individual 
components of the density matrix and avoid applying it to situations 
where this cancellation occurs.

 If the heavy Majorana neutrino masses $M_i$ are scaled by a common factor 
$x$, such that $M_i \rightarrow x M_i$, then: $\epsilon^{(1)}$ scales 
in proportion with $x$, $D_1$ and $W_1$ do not scale with $x$ and 
$\Im \left(\Lambda_{\tau}\right)/Hz$ varies inversely with $x$. Consequently, 
according to \equaref{eq:formalvectorsolution}, 
$n_{\alpha \beta} (z)$ scales in proportion to $x$, 
(with increasing precision for larger $x$ since we can better neglect 
the thermal widths). In $\lambda$ the scaling of $\Im(\Lambda_{\tau})/Hz$ 
cancels that of $n_{\tau \tau^{\perp}}$ and so 
$\lambda$ does not scale with $x$ 
if $M_1 \gg 10^{12} \text{ GeV}$.
Thus, at sufficiently large values of $M_1$, $\eta_B$, 
given by \equaref{eq:OnlyFlavSol}, asymptotically approaches a 
non-zero constant. This is shown in \figref{fig:HS_NO_CPValpha21_etaB_M1} (d)
over a range of $M_{1}$ values in which the ratios $M_{1}/M_{2}$ and 
$M_{2}/M_{3}$ are fixed. The curve increases ever more slowly for larger 
$M_1$ as the approximation leading to 
\equaref{eq:formalvectorsolution} becomes ever more precise. 
This may be interpreted as the transition region between the two flavour 
regime and the single flavour having grown infinitely large.\footnote{If, contrary to our scenario of interest, $R$ is $CP$-violating 
($\Tr \epsilon^{(1)} \neq 0$), then the first term in parantheses of 
\equaref{eq:solnBminusL1} eventually dominates the second for 
sufficiently large $x$ and the single flavour regime is entered.}
In each of the plots of \figref{fig:HS_NO_CPValpha21_etaB_M1}, we see a 
dip in the density matrix solution curve near $10^{12}$ GeV. 
This feature is due to the difference in sign of the two-flavour solutions, 
for which  $\eta_B \propto \epsilon^{(1)}_{\tau \tau}/P^{0(1)}_{\tau \tau}
+\epsilon^{(1)}_{\tau^\perp \tau^\perp}/P^{0(1)}_{\tau^\perp \tau^\perp} 
=  3.99 \times 10^{-4}$, compared to that of the single flavour solutions, 
where 
$\eta_B \propto \epsilon^{(1)}_{\tau \tau}+\epsilon^{(1)}_{\tau^\perp \tau^\perp} 
= -7.5 \times 10^{-6}$ (where these numbers are valid for
 plot (a) of \figref{fig:HS_NO_CPValpha21_etaB_M1}  corresponding to $CP$-violating $R$-matrix).
The dip appears as a result of plotting the absolute 
value of $\eta_B$ on a logarithmic scale when $\eta_B$ 
passes through zero during the transition between these regimes.

In \appref{sec:RobustnessOfHighScalePlateau}, we discuss the robustness 
of the plateau that forms for large heavy Majorana neutrino masses when 
the effects of scattering and when a more realistic treatment of the 
right-handed taus are incorporated. In the next section we explore 
the parameter space of the three heavy Majorana neutrino type I seesaw 
with regard to the solutions of \equaref{eq:DMEmodified} 
 with $CP$-conserving
$R$-matrix and $M_1 \gg 10^{12} \text{ GeV}$.

\subsection{Results of Parameter Exploration}
\label{sec:highscaleresults}

At values of $M_1 \gg 10^{12}$ GeV, fine-tuning through large elements of 
the $R$-matrix is not required for successful leptogenesis (if Majorana phases are allowed to play a role, otherwise large fine-tuning is required if only Dirac phases take effect). 
Thus, in this section we analyse the parameter space corresponding to real, 
and therefore $CP$-conserving, $R$-matrices ($y_i=0^\circ$), using the 
same numerical technique as described in \secref{sec:resultsintlepto}. 
In order to perform this analysis we fix $M_1= 10^{13} \text{ GeV}$ and 
again require $M_3 > 3 M_2 > 9 M_1$ in order to avoid the resonant 
regime.
With a much higher value of $M_1$, one would need a correspondingly 
a higher temperature of inflation. For this reason, we choose to 
illustrate the possibility of successful thermal leptogenesis at 
just one order of magnitude beyond the two-flavour to single-flavour 
transition temperature of $10^{12}$ GeV. In  
\figref{fig:2DProjections_highscale}, we display the two-dimensional 
projection plots for both normal ordering and inverted ordering.

 In the NO case, we find that the observed baryon asymmetry may be obtained 
to within $1 \sigma$ ($2 \sigma$) with $\delta$ between 
$[240, 331]^\circ$ ($[0, 360]^\circ$). 
In the IO one, the $1 \sigma$ ($2 \sigma$) range is 
$[50, 304]^\circ$ ($[20,352]^\circ$). In what follows, we analyse 
these results by considering separately the cases of purely Dirac or purely 
Majorana $CP$ violation.
 For brevity we consider only the 
case of NO spectrum.

\subsection{Dependence of $\eta_B$ on the Dirac and Majorana Phases}

\begin{figure}[h!]
\centering
\includegraphics[width=\textwidth]{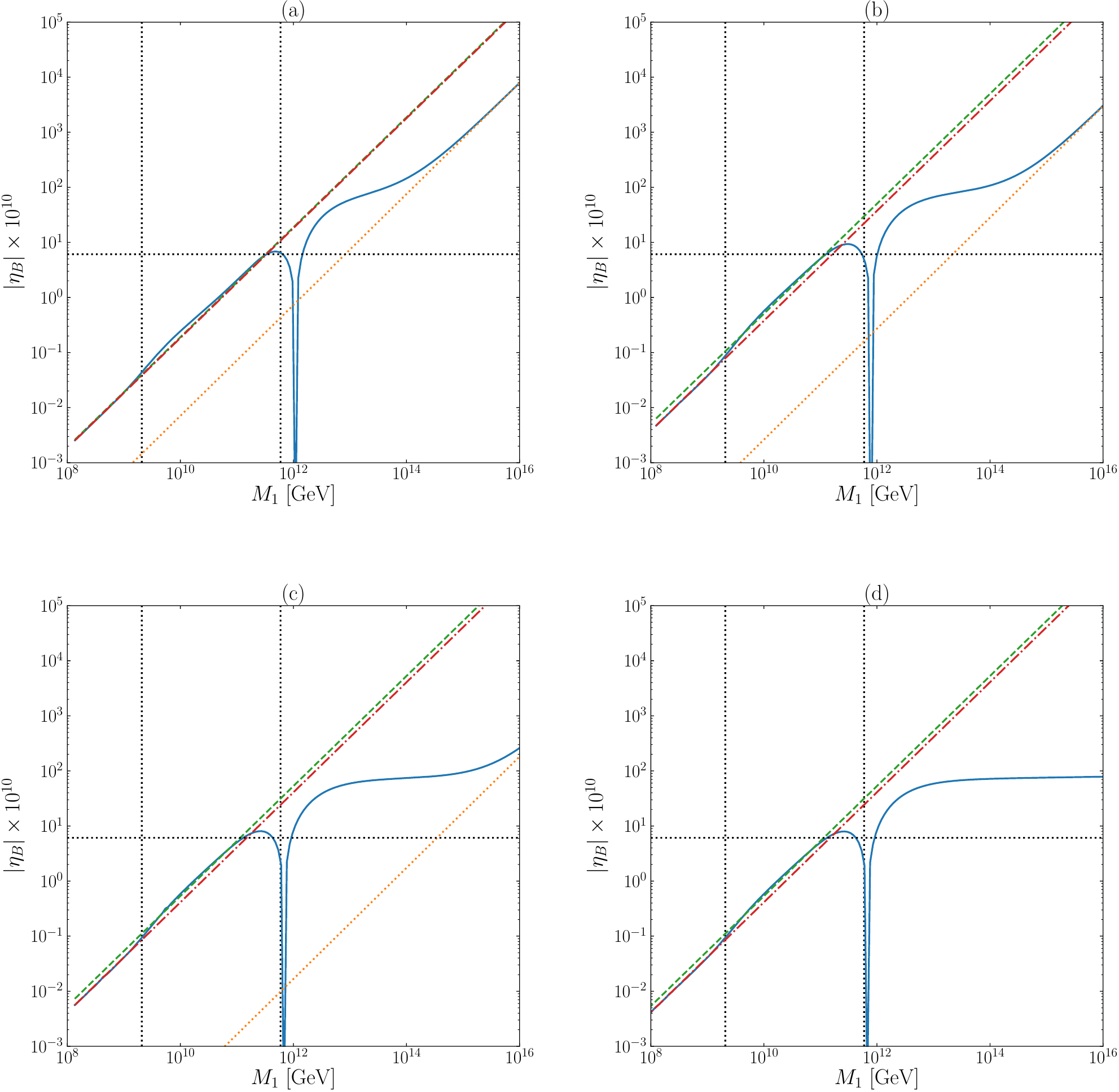}
\caption{The magnitude of the baryon asymmetry as a function of the 
heavy Majorana neutrino masses (assuming the modified Casas-Ibarra 
parametrisation) at a specific point in parameter space. 
The dotted orange line corresponds to solutions of the single-flavour 
Boltzmann equations, the dashed green line to those of the two-flavour 
Boltzmann equation, the red dot-dashed line to those of the 
three-flavour Boltzmann equations and the solid blue line to solutions 
of the density matrix equations. The horizontal black dotted line 
is the observed value of $\eta_{B_{CMB}}$ and the vertical dotted lines 
to the values of the muon and tau thermal widths. We vary $y_3$ such 
that in (a) $y_3 = 30^\circ$, in (b) $y_3 = 5^\circ$, in 
(c) $y_3 = 0.3^\circ$ and in (d) $y_3 = 0^\circ$. As $y_3$ is the only 
complex parameter of the $R$-matrix for this parameter point, 
then plot (d) corresponds to the case of purely low-energy $CP$ violation. 
As the $CP$ violation becomes solely low energy (going from (a) to (d)), 
then the transition of the density matrix equations to the 
single-flavour regime becomes longer. This culminates in an infinite 
transition width in plot (d) --- a plateau in the baryon asymmetry 
for high-scale leptogenesis. The dip in all of the blue lines 
occurs as a consequence of the change in sign of the produced asymmetry.
}
\label{fig:HS_NO_CPValpha21_etaB_M1}
\end{figure}

\begin{figure}[h!]
\centering
\includegraphics[width=1.0\textwidth]{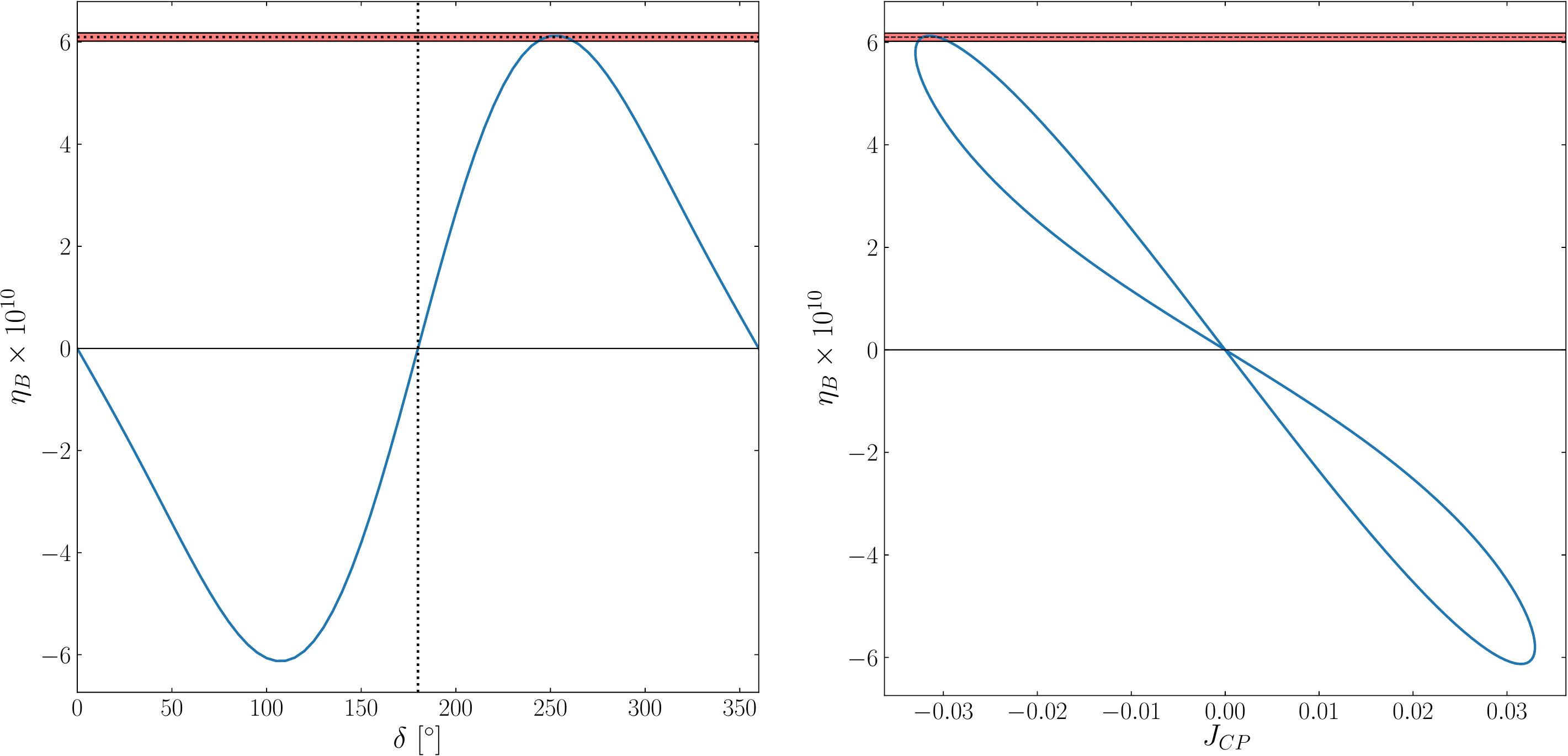}
\caption{Leptogenesis at high scales ($M_1 = 3.16 \times 10^{13}$ GeV) 
with $CP$ violation provided solely by $\delta$, 
with $\alpha_{21} = 0^{\circ}$ and $\alpha_{31} = 0^{\circ}$. The red band 
indicates the $1 \sigma$ observed values for $\eta_B$ with 
the best-fit value indicated by the horizontal black dotted lines. 
Left: The baryon asymmetry as a function of $\delta$ with exact 
$CP$-invariance exists for $\delta = 0^{\circ}$ and $180^{\circ}$ 
(vertical black dotted lines). In order to make the maximum value
touch on the observed baryon asymmetry, an amount of 
fine-tuning $\mathcal{F}=105$ is needed. 
Right: The corresponding variation of $\eta_B$ against 
$J_{CP}$ parametrically plotted with $\delta$.
 See the text for further details.
}
\label{fig:delta_etaB_plot_highscale_NO}
\end{figure}

\begin{figure}[h!]
\centering
\includegraphics[width=1.0\textwidth]{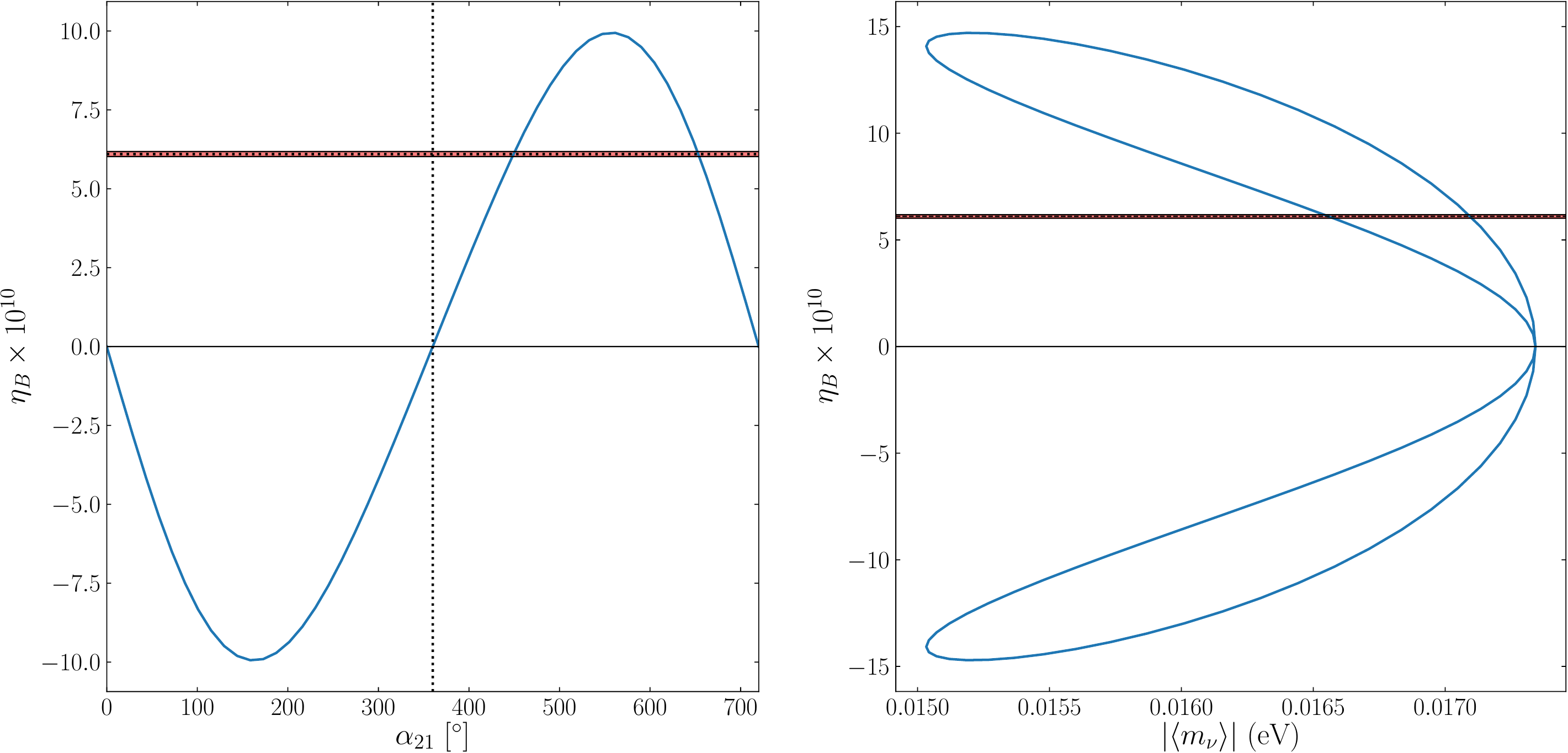}
\caption{Leptogenesis at high scales ($M_1 = 3.16 \times 10^{13}$ GeV) 
when $CP$ violation is provided solely by $\alpha_{21}$, 
with $\delta = 0^{\circ}$, $\alpha_{31} = 0^{\circ}$. The red bands 
indicate the $1 \sigma$ observed values for $\eta_B$ with the best-fit 
value indicated by the horizontal black dotted lines. 
Left: The baryon asymmetry as a function of $\alpha_{21}$ with exact 
$CP$-invariance at $\alpha_{21} = 0^{\circ}$ and $360^{\circ}$ 
(vertical black dotted lines). 
Right: The variation of $\eta_B$ against 
$\lvert \langle m_{\nu} \rangle \rvert$ parametrically plotted 
as a function of $\alpha_{21}$. Successful leptogenesis occurs for $\alpha_{21} \approx 449^\circ$ and $\alpha_{21} \approx 653^\circ$ for which $\lvert \langle m_{\nu} \rangle \rvert = 0.0171$ eV and $\lvert \langle m_{\nu} \rangle \rvert = 0.0166$ eV respectively.
 See text for further details.
}
\label{fig:alpha21_etaB_plot_highscale_NO}
\end{figure}
 As the final value of the baryon asymmetry becomes approximately constant 
for $M_1 \gg 10^{12} \text{ GeV}$ (see \figref{fig:HS_NO_CPValpha21_etaB_M1}) 
with a $CP$-conserving $R$-matrix, then one can can use the value of 
$\eta_B$ that is predicted by the two-flavour Boltzmann equations (2FBE) 
at the start of the transition $M_1 \sim 10^{12} \text{ GeV}$ as a proxy for 
the full solution of the density matrix equations (DME). That is,
\begin{equation}
\eta_B^{\text{DME}} \left( M_1 \gg 10^{12} \text{ GeV} \right) 
\approx \eta_B^{\text{2FBE}}\left( M_1 \sim 10^{12} \text{ GeV} \right),
\end{equation}
provided that the ratios $M_2/M_1$ and $M_3/M_1$ are fixed. This has the 
advantage that we may again make use of the result in \equaref{eq:2FSol}
\begin{equation}
\begin{aligned}
n_{B - L} \approx  n_B^{\text{2FBE}}\left(M_1 \sim 10^{12} \text{ GeV}\right) 
= \frac{\pi^2}{6 z_d K_1} n^{\text{eq}}_{N_1} (z_0) 
\epsilon^{(1)}_{\tau \tau} \Delta F,
\end{aligned}
\end{equation}
in order to gain an analytical understanding of the numerical solutions.

 As in the analysis of \secref{sec:resultsintlepto},  we investigate the 
cases where $CP$ violation comes from precisely one of 
$\delta$, $\alpha_{21}$ or $\alpha_{31}$. Unlike the fine-tuned scenario 
previously considered, $P^{0(1)}_{\tau \tau}$ and consequently $\Delta F$ are 
approximately constant with the PMNS phases as would be expected from our 
discussion of the fine-tuned solutions in \secref{sec:summary}. 
Hence the phase-dependence of the $\eta_B$ can be understood by reference to 
$\epsilon^{(1)}_{\tau \tau}$ alone. This may also be understood by reference 
to the Yukawa couplings when $CP$ violation comes only from 
$\delta$, $\alpha_{21}$ or $\alpha_{31}$ respectively:
\begin{equation}
\begin{aligned}
Y_{\tau 1} & = -0.0476 - 0.000364 e^{i \delta},\\
Y_{\tau 1} & = -0.0541 + 0.00614 e^{i \frac{\alpha_{21}}{2}},\\
Y_{\tau 1} & = 0.00972 - 0.0576 e^{i \frac{\alpha_{31}}{2}}.
\end{aligned}
\end{equation}
 The difference in scale of the two terms means that the cancellation 
is never strong for any value of the phase and so the peaks in 
$\Delta F$ are not large. In this high scale case, $\Delta F$ 
is approximately constant and thus the plots of $\eta_B$ exhibit a 
nearly pure sinusoidal variation given by the $CP$-asymmetries below.

\subsubsection{Dirac Phase $CP$ violation}

In this case, with a real $R$, we have $\alpha_{21} = \alpha_{31} = 0^\circ$ 
such that $\delta$ is the sole provider of all $CP$ violation. 
The asymmetry is given by
\begin{equation}
\epsilon^{(1)}_{\tau \tau} = -1.25 \times 10^{-6} \sin \delta,
\end{equation}
in this scenario. Thus we obtain a sinusoidal dependence 
with $\eta_B = 0$ when $\delta = 0^{\circ}$ or $180^{\circ}$. Fixing all other 
parameters at their benchmark value
with $y_1 = y_2 = y_3 = 0$, no value of $\delta$ can 
produce the observed baryon asymmetry of the Universe. Unlike in the case of 
intermediate scale leptogenesis, a small scaling of the heavy Majorana 
neutrino masses will not much increase the value of $\eta_B$ because of 
the plateau of  \figref{fig:HS_NO_CPValpha21_etaB_M1}. At the best-fit point 
of \tabref{tab:tableBFhighscaleNO}, with $\alpha_{21} = \alpha_{31} = 0^\circ$, 
allowing $CP$ violation only from $\delta$, the largest $\eta_B$ achieved is a 
factor $\sim 9$ smaller than the observed value. This is large enough that 
even enormously larger scales of the heavy masses cannot make $\delta$-only 
leptogenesis a viable option.

 An alternative for producing the observed baryon asymmetry of the Universe 
with $CP$ violation only from $\delta$ is  to work with an $R$-matrix 
containing both 
zero and purely imaginary
components which are $CP$-conserving and may potentially 
be large in magnitude.
If for example, we 
 choose $x_i=0^\circ$ such that all $w_i$ are either 
purely imaginary or zero, and take $y_2 = 0^\circ$ also, then by setting 
$\alpha_{21} = 180^{\circ}$ and $\alpha_{31} = 0^{\circ}$, 
all $CP$ violation will be due to $\delta$. Varying $y_1$ and $y_2$ together 
in this setup, we find that $y_1 = y_2 = 169^{\circ}$ is the smallest value for 
which the observed baryon asymmetry of the Universe is produced. With all 
other parameters equal to the values in \tabref{tab:tableBFhighscaleNO}, 
this corresponds to $\mathcal{F} = 105$. Hence a noticeable degree of 
fine-tuning is required even at high scales to make $\delta$ the sole 
contributor to $CP$ violation with viable leptogenesis. 
In  \figref{fig:delta_etaB_plot_highscale_NO}, we plot the variation 
of $\eta_B$ with pure $\delta$ $CP$ violation for this fine-tuned scenario 
in the left plot, and on the right we parametrically plot $\eta_B$ against $J_{CP}$ as a function of $\delta$.

\subsubsection{$CP$ Violation from the Majorana Phase $\alpha_{21}$}
Similarly, when $\delta =\alpha_{31} = 0^\circ$, the $CP$-asymmetry is
\begin{equation}
\epsilon^{(1)}_{\tau \tau} = 1.98 \times 10^{-5} \sin \frac{\alpha_{21}}{2}.
\end{equation}
 It follows from the preceding expression for 
$\epsilon^{(1)}_{\tau \tau}$ that at the $CP$-conserving values of 
 $\alpha_{21} = 180^\circ,540^\circ$ we have $\epsilon^{(1)}_{\tau \tau}\neq 0$.
This corresponds to the case of $CP$-conserving $R$-matrix 
($y_i = 0$), $CP$-conserving PMNS matrix, 
but $CP$-violating interplay between the 
$R$ and PMNS matrix elements in leptogenesis \cite{Pascoli:2006ie}.

 The corresponding $\eta_B$, plotted in the left plot of  
\figref{fig:alpha21_etaB_plot_highscale_NO} is thus a factor 
of $\mathcal{O}(10)$ higher and of opposite sign than in the previous 
case without fine-tuning. Thus, we obtain the observed baryon asymmetry 
of the Universe (or higher) for values of $\alpha_{21}$ between 
about $450^{\circ}$ and $650^{\circ}$. In the right plot of 
\figref{fig:alpha21_etaB_plot_highscale_NO} is $\eta_B$ for the same 
scenario parametrically plotted against the effective neutrino 
mass with parameter $\alpha_{21}$.

\subsubsection{$CP$ Violation from the Majorana Phase $\alpha_{31}$}

\begin{figure}[h!]
\centering
\includegraphics[width=1.0\textwidth]{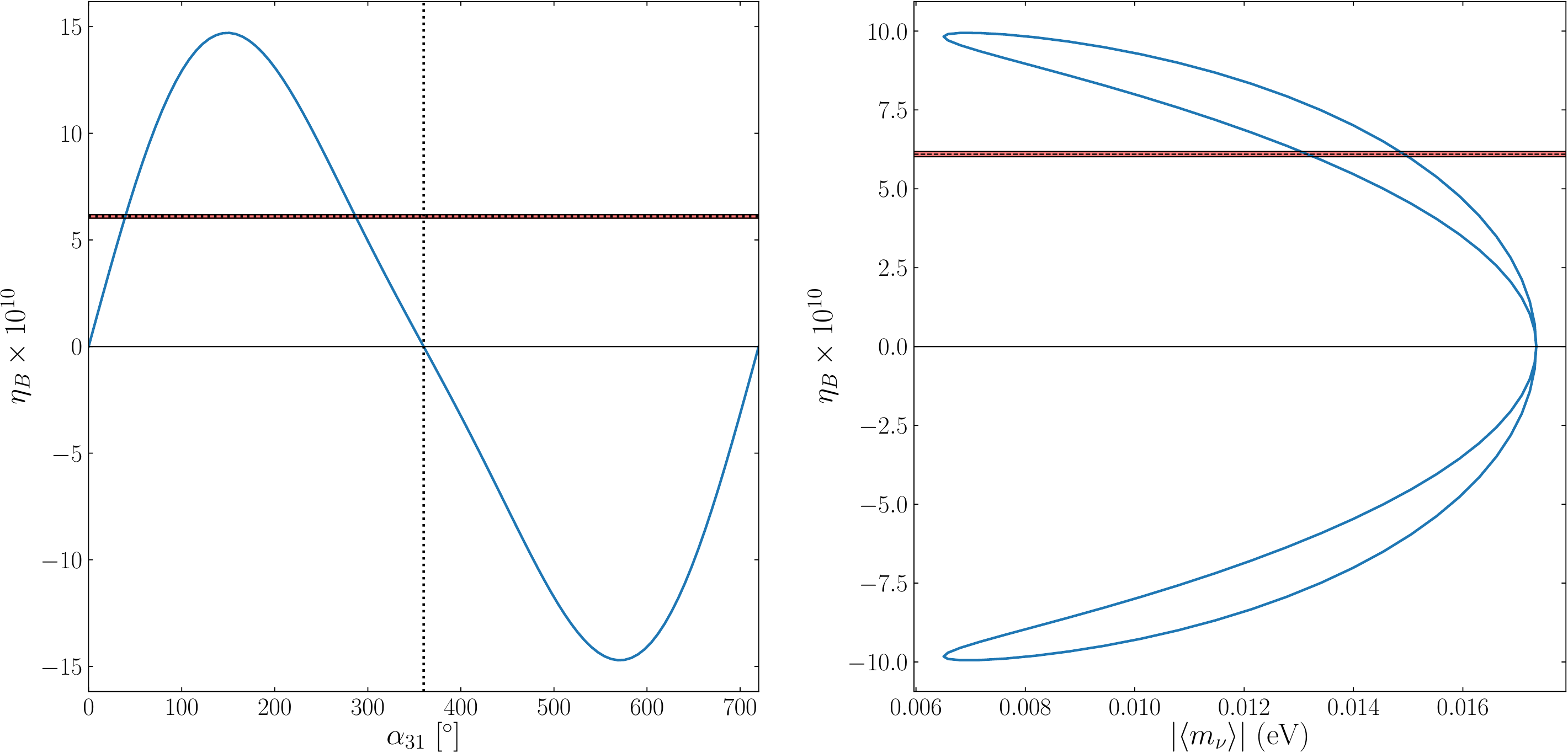}
\caption{High-scale leptogenesis ($M_1 = 3.16 \times 10^{13}$ GeV) with 
$CP$ violation is provided solely by $\alpha_{31}$, with 
$\delta = 0^{\circ}$, $\alpha_{21} = 0^{\circ}$. The red bands indicate 
the $1 \sigma$ observed values for $\eta_B$ with the best-fit value 
indicated by the horizontal black dotted lines. 
Left: The baryon asymmetry as a function of $\alpha_{31}$,  
with exact $CP$-invariance 
when $\alpha_{31} = 0^{\circ}$ and $360^{\circ}$ 
(vertical black dotted lines). 
Right: The parametric plot of $\eta_B$ against the effective neutrino 
Majorana mass $\lvert \langle m_{\nu} \rangle \rvert$ as 
$\alpha_{31}$ is varied.  At the values $\alpha_{31} = 17^{\circ}, 43^{\circ}$, 
$\eta_B$ takes on its observed values corresponding to 
$\lvert \langle m_{\nu} \rangle \rvert =0.0131 \text{ eV, } 
0.0149 \text{ eV}$ respectively. See the text for further details.}
\label{fig:alpha31_etaB_plot_highscale_NO}
\end{figure}

Finally, we turn to the scenario in which $CP$ violation is provided 
entirely by $\alpha_{31}$, plotted on the left in of 
\figref{fig:alpha31_etaB_plot_highscale_NO} for which
\begin{equation}
\epsilon^{(1)}_{\tau \tau} = -3.22\times 10^{-5} \sin \frac{\alpha_{31}}{2}.
\end{equation}
 Similarly to the case discussed in the preceding 
subsection, we see that $\epsilon^{(1)}_{\tau \tau}\neq 0$ 
at the $CP$-conserving values of  $\alpha_{31} = 180^\circ,540^\circ$.
This again corresponds to the case of
$CP$ violation in leptogenesis due 
to the interplay of the $CP$-conserving $R$-matrix
($y_i = 0$) and $CP$-conserving PMNS matrix 
\cite{Pascoli:2006ie}.

  Compared with the previous scenario, there is a sign flip and an 
enhancement by a factor $\sim 1.6$ of the resulting baryon asymmetry 
of the Universe. Thus, the observed BAU is achieved and exceeded for 
smaller values of $\alpha_{31}$, between about $50^{\circ}$ and $300^{\circ}$. 
On the right of \figref{fig:alpha31_etaB_plot_highscale_NO}, we display 
a parametric plot for the same scenario with $\eta_B$ against the effective neutrino Majorana mass with the parameter $\alpha_{31}$.

\section{Conclusions}
In this paper we have investigated the connection between leptogenesis 
and low energy leptonic $CP$ violation over a large range of scales 
$\left( 10^6  < M_1\text{ (GeV)} < 10^{13} \right)$. We summarise our main findings below:

\begin{itemize}
\item Firstly, we revisited the question of the possibility of successful 
thermal leptogenesis at scales $10^9 < M_1 \text{ (GeV)} < 10^{12}$. 
At such scales, tau-Yukawa interactions are in equilibrium, such that there are
 sufficiently frequent
interactions between the leptons and the early Universe plasma causing 
decoherence between the tau flavour from the other flavour components. 
We show that successful leptogenesis is indeed possible in this range of 
scales in the case that the PMNS phases provide all of the $CP$ violation 
in the model. By performing parameter explorations at $M_1 = 10^9$ GeV and 
$M_1 = 10^{10}$ GeV, we found that some degree of fine-tuning, 
$\mathcal{F} \sim 10$, is required for these particular mass scales 
(with the degree of fine-tuning diminishing as one goes to higher 
values of $M_1$).

\item By demanding pure Dirac phase or pure Majorana phase $CP$ violation, 
we found that each phase alone can produce the correct $CP$-asymmetry, 
 with the cases of  Majorana phases 
requiring, in general, a somewhat lower value of $M_1$
than those required for the Dirac phase.

\item If leptogenesis takes place at scales $M_1 \ll 10^9$ GeV, then all 
three of the leptonic flavour components involved in leptogenesis will 
decohere. For masses in this range ($M_1 \sim 10^6$ GeV and 
$M_1 \sim 10^8$ GeV), we determined the regions of parameter space in 
which low energy leptonic $CP$ violation, provided by 
 either the Dirac or the Majorana phases individually,
leads to successful intermediate scale leptogenesis. 
At these scales a large amount of fine-tuning 
($\mathcal{F} \sim \mathcal{O}\left( 100\right)$) is required between 
the tree-level and one-loop neutrino masses. We restricted ourselves to 
fine-tuning such that $\mathcal{F} < 1000$ and in doing so have found 
an approximate lower bound of $M_1 \approx 3 \times 10^6 \text{ GeV}$ 
(consistent with the conclusions of~\cite{Moffat:2018wke}).

\item We studied the possibility of pure Dirac phase $CP$ violation and 
showed that for $\mathcal{F} < 1000$, $M_1 \gtrsim 8 \times 10^{6} \text{ GeV}$ in order to produce 
the observed baryon asymmetry. Similarly for the purely Majorana phase 
$CP$ violation and again $\mathcal{F} < 1000$, for $\alpha_{21}$, we obtain a bound 
$M_1 \approx 4.5 \times 10^6 \text{ GeV}$ whereas for $\alpha_{31}$ 
we obtained $M_1 \approx 3 \times 10^6 \text{ GeV}$. Observables depending 
on the Dirac and Majorana phases, for example 
$J_{CP}$ or $\langle m_{\nu} \rangle$, may be well within experimental 
bounds in the same parts of parameter space in which leptogenesis 
is successful. The Dirac phase $\delta$ is only very weakly constrained, 
with the tightest constraint being $\delta \in [16, 263]^\circ$, 
for $1 \sigma$ agreement with the observed BAU, which comes 
from assuming normal ordering and $M_1 = 1.29 \times 10^8$ GeV.

\item If leptogenesis takes place at high scales, with $M_1 \gg 10^{12}$ GeV, interactions between the leptons and the early Universe plasma only very weakly decohere the tau flavour from the other flavour components. Normally, this leads to the conclusion that the single flavour Boltzmann equations are an appropriate description of the process. However, we have demonstrated that, if $CP$ violation arises only in the low energy leptonic sector, the effects of decoherence cannot be neglected. Therefore, one should not ignore the phenomenology of high-scale leptogenesis with purely low-energy $CP$ violation.

\item We explored the parameter space at $M_1 \sim 10^{13}$ GeV, finding regions in which thermal leptogenesis is a viable explanation of the BAU. We found that the strongest constraint on $\delta$ is for normal ordering, for which we require $\delta \in [240,331]^\circ$ to produce a baryon asymmetry within $1 \sigma$ of the observed value. With only Dirac phase $CP$ violation, we have concluded that it is not possible produce the observed baryon asymmetry of the Universe unless one introduces significant fine-tuning $\left( \mathcal{F} \sim 100 \right)$ in the light neutrino masses. We argued that there is no scale of the heavy Majorana neutrino masses beyond $M_1 \gg 10^{12} \text{ GeV}$ for which Dirac phase leptogenesis may be made to work without this fine-tuning. However, with pure Majorana phase violation, we found that successful leptogenesis is possible with essentially no fine-tuning.
\end{itemize}

The results of this article underscore the significance of understanding leptonic $CP$ violation through experimental searches for Dirac and/or Majorana leptonic $CP$ violation. We have departed from previous literature by concluding that low energy leptonic $CP$-violating phases may always be relevant to the production of the baryon asymmetry in the leptogenesis scenario. It has commonly been thought that their relevance was limited to the window of masses $10^9 \lesssim M_1 \text{ (GeV)} \lesssim 10^{12}$. However, we have shown this window to be significantly wider: Dirac and Majorana phases may be crucial to leptogenesis even at scales as low as $M_1 \sim 10^{6} \text{ GeV}$, or as high as $M_1 \gg 10^{12} \text{ GeV}$.

\acknowledgments
We would like to thank Pasquale Di Bari  for helpful conversations regarding the density matrix equations. K.M. and S.P. acknowledge the (partial) support from the European Research Council under the European Union Seventh Framework Programme (FP/2007-2013) / ERC Grant NuMass agreement n. [617143].  S.P. would like to acknowledge partial support from the Wolfson Foundation and the Royal Society, and also thanks SISSA for support and hospitality during part of this work. S.P. and S.T.P. acknowledge partial support from the European Unions Horizon 2020 research and innovation programme under the Marie Sklodowska Curie grant agreements No 690575 (RISE InvisiblesPlus) and No 674896 (ITN ELUSIVE). The work of S.T.P. was supported in part by the INFN program on Theoretical Astroparticle Physics (TASP) and by the World Premier International Research Center Initiative (WPI Initiative), MEXT, Japan. 
 This manuscript has been authored by Fermi Research Alliance, LLC under Contract No. DE-AC02-07CH11359 with the U.S. Department of Energy, Office of Science, Office of High Energy Physics.  
J.T. would like to express a special thanks to the Mainz Institute for Theoretical Physics (MITP) and
 International School for Advanced Studies (SISSA) for their hospitality and support where part of this work was completed.

\appendix

\section{Classes of $CP$-Conserving $R$-matrix}
\label{sec:RMatrices}
With the parameters $x_2$, $y_1$ and $y_3$ left arbitrary, there are $16$ possible $R$-matrices which lead to the fine-tuned light neutrino masses required for successful leptogenesis (\equaref{eq:RStructure}). For any of these matrices, the absolute values of the elements $\lvert R_{ij} \rvert$ are equal, with the elements themselves differing only by factors $\pm 1$ or $\pm i$. When there is an exact $CP$-symmetry, then each $R$-matrix satisfies the condition in \equaref{eq:UCPRCPconditions}. This allows for a scheme of classification according the phases $\rho^{\nu}$, $\rho^N$ they correspond to. In this section we present a single example of a matrix for each class\footnote{Here we neglect terms involving factors $e^{-y_1}$ or $e^{-y_3}$ such that, as given, these matrices are not strictly orthogonal.}:

$\rho^{\nu} = \pm (-1,+1,+1)^T$, $\rho^{N} = \pm (+1,+1,-1)^T$ and $x_1 = 90^{\circ}$ and $x_3 = 90^{\circ}$:
\[
R   \approx 
\left(
\begin{array}{ccc}
 -\frac{i}{2} e^{y_3} \cos x_2 & \frac{1}{2} e^{y_3} \cos x_2 &
   \sin x_2 \\
 \frac{i}{4} e^{y_1+y_3} \left(\sin x_2+1\right) & -\frac{1}{4} e^{y_1+y_3}
   \left(\sin x_2+1\right) & \frac{1}{2} e^{y_1} \cos x_2 \\
 \frac{1}{4} e^{y_1+y_3} \left(\sin x_2+1\right) & \frac{i}{4} e^{y_1+y_3}
   \left(\sin x_2+1\right) & -\frac{i}{2} e^{y_1} \cos x_2 \\
\end{array}
\right),
\]
in which the second form results from the neglect of terms involving factors $e^{-y_1}$ and $e^{-y_3}$.
$\rho^{\nu} = \pm (+1,-1,+1)^T$, $\rho^{N} = \pm (+1,-1,+1)^T$ and $x_1 = 0^{\circ}$ and $x_3 = 0^{\circ}$:
\[
R \approx
\left(
\begin{array}{ccc}
 \frac{1}{2} e^{y_3} \cos x_2 & \frac{i}{2} e^{y_3} \cos x_2 & \sin
   x_2 \\
 -\frac{i}{4} e^{y_1+y_3} \left(\sin x_2+1\right) & \frac{1}{4} e^{y_1+y_3}
   \left(\sin x_2+1\right) & \frac{i}{2} e^{y_1} \cos x_2 \\
 -\frac{1}{4} e^{y_1+y_3} \left(\sin x_2+1\right) & -\frac{i}{4} e^{y_1+y_3}
   \left(\sin x_2+1\right) & \frac{1}{2} e^{y_1} \cos x_2 \\
\end{array}
\right),
\]

$\rho^{\nu} = \pm (+1,-1,+1)^T$, $\rho^{N} = \pm (+1,+1,-1)^T$ and $x_1 = 90^{\circ}$ and $x_3 = 0^{\circ}$:
\[
R \approx
\left(
\begin{array}{ccc}
 \frac{1}{2} e^{y_3} \cos x_2 & \frac{i}{2} e^{y_3} \cos x_2 & \sin
   x_2 \\
 -\frac{1}{4} e^{y_1+y_3} \left(\sin x_2+1\right) & -\frac{i}{4} e^{y_1+y_3}
   \left(\sin x_2+1\right) & \frac{1}{2} e^{y_1} \cos x_2 \\
 \frac{i}{4} e^{y_1+y_3} \left(\sin x_2+1\right) & -\frac{1}{4} e^{y_1+y_3}
   \left(\sin x_2+1\right) & -\frac{i}{2} e^{y_1} \cos x_2 \\
\end{array}
\right),
\]

$\rho^{\nu} = \pm (-1,+1,+1)^T$, $\rho^{N} = \pm (+1,-1,+1)^T$ and $x_1 = 90^{\circ}$ and $x_3 = 0^{\circ}$:
\[
R \approx
\left(
\begin{array}{ccc}
 -\frac{i}{2} e^{y_3} \cos x_2 & \frac{1}{2} e^{y_3} \cos x_2 &
   \sin x_2 \\
 -\frac{1}{4} e^{y_1+y_3} \left(\sin x_2+1\right) & -\frac{i}{4} e^{y_1+y_3}
   \left(\sin x_2+1\right) & \frac{i}{2} e^{y_1} \cos x_2 \\
 \frac{i}{4} e^{y_1+y_3} \left(\sin x_2+1\right) & -\frac{1}{4} e^{y_1+y_3}
   \left(\sin x_2+1\right) & \frac{1}{2} e^{y_1} \cos x_2 \\
\end{array}
\right).
\]

\section{Further results}
\label{appendix:FurtherResults}
\begin{figure}[h!]
\centering
\includegraphics[width=1.0\textwidth]{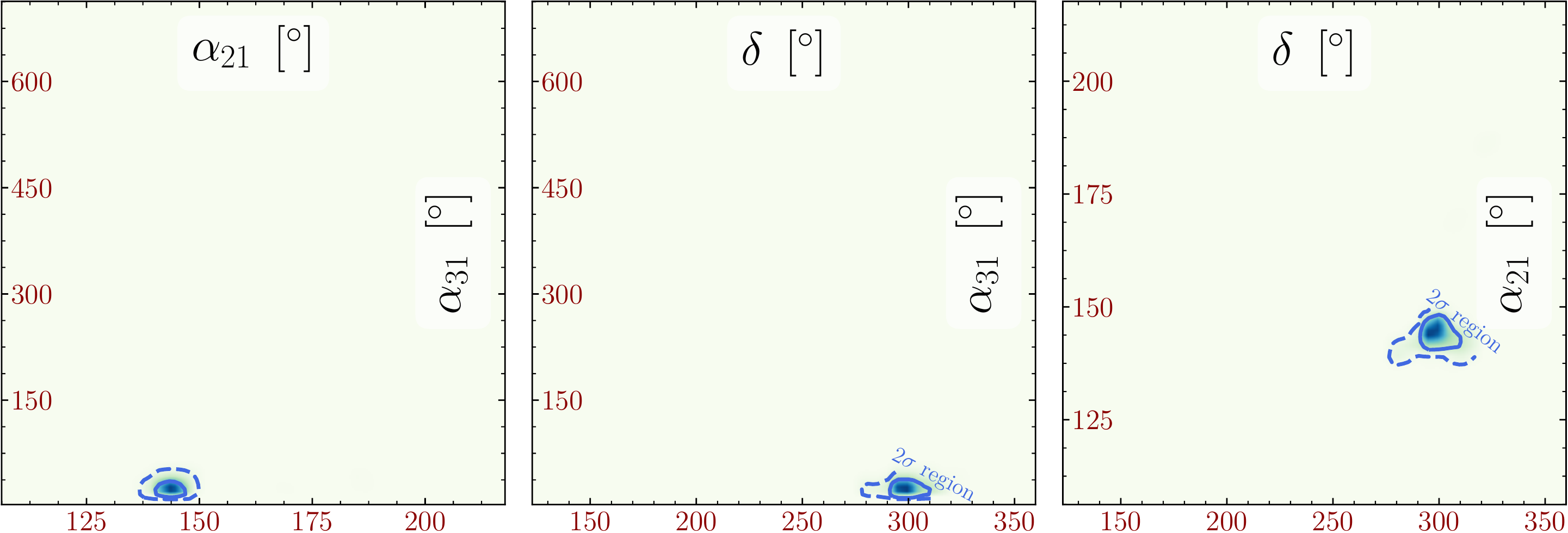}
\caption{The two-dimensional projections for intermediate scale leptogenesis 
with $M_1 = 3.16 \times 10^{6}$ GeV and $m_1 = 0.05$ eV with $CP$ violation provided only 
by the phases of the PMNS matrix. Solid lines correspond to $68\%$ confidence level and dashed to $95\%$ confidence level in agreement with the observed value $\eta_{B_{CMB}}$. This plot was created using {\sc SuperPlot}~\cite{Fowlie:2016hew}.}
\label{fig:lowm1}
\end{figure}

In \figref{fig:lowm1} we demonstrate the possibility of fine-tuned leptogenesis in the case of normal ordering with $M_1 = 3.16 \times 10^6$ GeV and $m_1=0.05$ eV. This is a variant of the case considered in the main body for which the light neutrino masses are significantly reduced below all present cosmological or current generation direct bounds. We note that lowering the light neutrino masses in this way severely constrains the viable parameter space over that in \figref{fig:2DIntermediatePlots} such that $\delta \approx 296^\circ$, $\alpha_{21} \approx 143^\circ$ and $\alpha_{31} \approx 14^\circ$. Typical fine-tuning in the viable regions is $\mathcal{F} \approx 450$.

In the cases of $m_1 =0$ and 
$m_1 = 10^{-3}$ eV with $M_1 = 10^{8}$ GeV, $M_2 = 3M_1$ and $M_3 = 3M_2$ 
we did not find a region in the relevant parameter space 
in which one could have successful leptogenesis.

\label{appendix:M1109}

\begin{figure}[t!]
\centering
\includegraphics[width=1.0\textwidth]{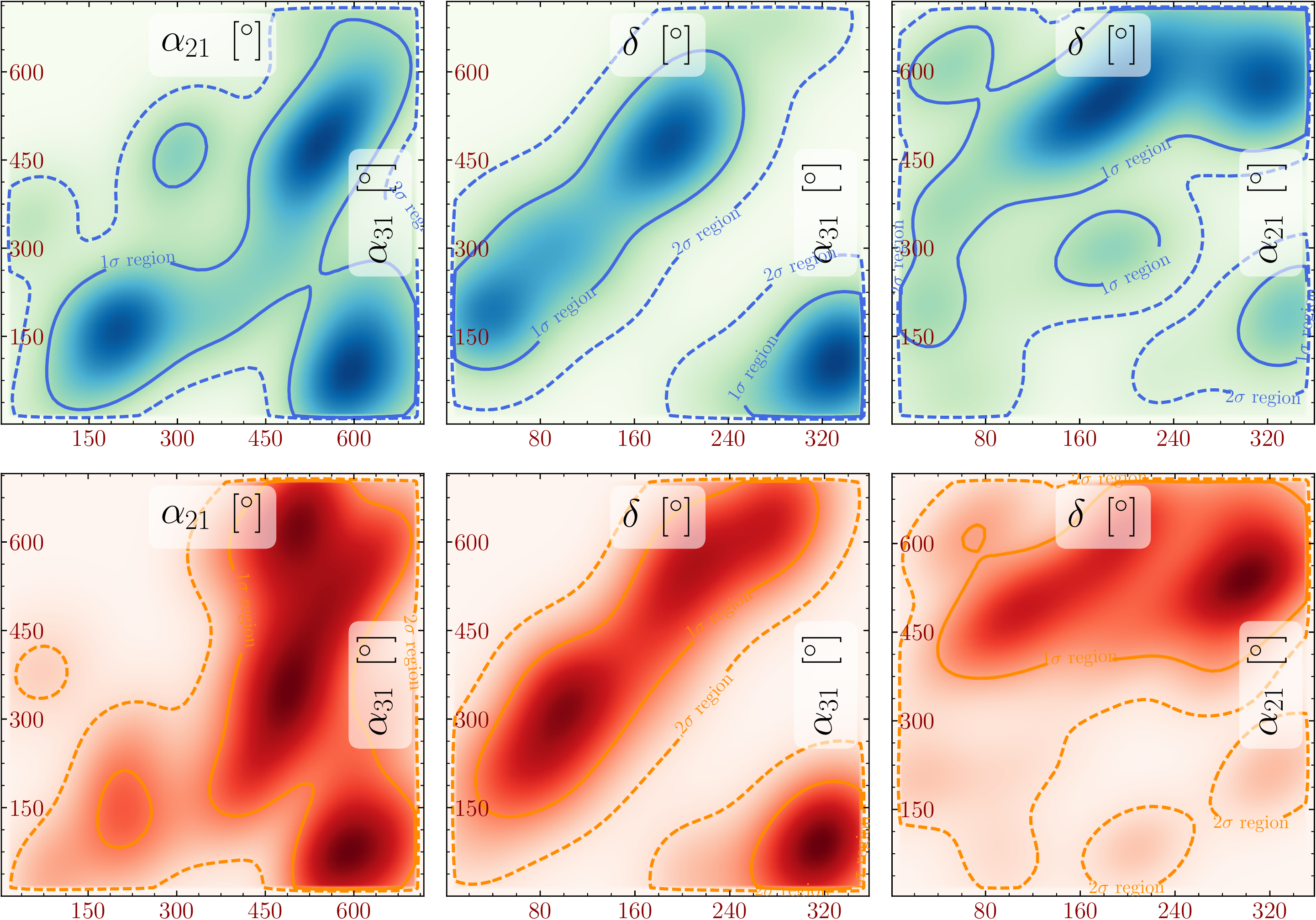}
\caption{The two-dimensional projections for leptogenesis with $M_1 = 1.00 \times 10^{9}$ GeV with $CP$ violation provided only by the phases of the PMNS matrix. The normal ordered case is coloured blue/green and inverted ordering orange/red and contours correspond to $68\%$ and $95\%$ confidence levels. This plot was created using {\sc SuperPlot}~\cite{Fowlie:2016hew}.}
\label{fig:genericscale1}
\end{figure}

In \figref{fig:genericscale1} we present results for $M_1 = 10^{9}$ GeV. We find that a  fine-tuning of the light neutrino masses $\mathcal{F} \approx 14$  at  the best-fit points. In the normal ordered case, we find that the observed baryon asymmetry may be obtained to within $1 \sigma$ ($2 \sigma$) with $\delta$ between $[0,360]^\circ$ ($[0,360]^\circ$). While for inverted ordering, the $1 \sigma$ ($2 \sigma$) range is $[25,360]^\circ$ ($[0,360]^\circ$). This is significantly higher than the case for which $M_1 = 10^{10}$ GeV where the fine-tuning is considerably less at $\mathcal{F} \approx 0.23$. In the normal ordered case, we find that the observed baryon asymmetry may be obtained to within $1 \sigma$ ($2 \sigma$) with $\delta$ between $[95,265]^\circ$ ($[52,282]^\circ$). For inverted ordering, the $1 \sigma$ ($2 \sigma$) range is $[60, 338]^\circ$ ($[8,360]^\circ$).

\section{Single Flavour Boltzmann Equations from Density Matrix Equations}
\label{appendix:DMEtoBE1F}
In this appendix we find the conditions under which the density matrix equations (\equaref{eq:full3}) approximate to the single flavour Boltzmann equations. We begin by analysing the criteria under which the single flavour Boltzmann equation
\begin{equation}
\frac{d n_{B-L}}{d z} = \Tr \epsilon^{(1)} D_1 (n^{}_{N_1} - n_N^{\text{eq}}) - W_1 n_{B-L},
\end{equation}
emerges as an approximation from the density matrix equations, which, written in the ($\tau^{\perp}$, $\tau$)-basis are
\begin{equation}\label{eq:DMEqsAppendix}
\begin{aligned}
\frac{d n_{N_1}}{d z} & = -D_1 \left( n^{}_{N_1} - n_{N_1}^{\text{eq}} \right) \\
\frac{d n_{\tau^{\perp}\tau^{\perp}}}{d z} & = \epsilon^{(1)}_{\tau^{\perp}\tau^{\perp}} D_1 \left( n^{}_{N_1} - n_{N_1}^{\text{eq}} \right) - \frac{1}{2} W_1 \left( 2|C_{1\tau^{\perp}}|^2 n_{\tau^{\perp}\tau^{\perp}}+C^{*}_{1 \tau} C_{1\tau^{\perp}} n_{\tau \tau^{\perp}} + C_{1\tau} C^{*}_{1\tau^{\perp}} n_{\tau^{\perp} \tau} \right) \\
\frac{d n_{\tau \tau}}{d z} & = \epsilon^{(1)}_{\tau \tau} D_1 \left( n^{}_{N_1} - n_{N_1}^{\text{eq}} \right) - \frac{1}{2} W_1 \left( 2|C_{1\tau}|^2 n_{\tau \tau}+C^{*}_{1 \tau} C_{1\tau^{\perp}} n_{\tau \tau^{\perp}} +C_{1\tau} C^{*}_{1\tau^{\perp}} n_{\tau^{\perp} \tau} \right) \\
\frac{dn_{\tau^{\perp} \tau}}{d z} & = \epsilon^{(1)}_{\tau^{\perp} \tau} D_1 \left( n^{}_{N_1} - n_{N_1}^{\text{eq}} \right) - \frac{1}{2} W_1 \left( n_{\tau^{\perp} \tau}+C^{*}_{1 \tau} C_{1 \tau^{\perp}} \left( n_{\tau^{\perp}\tau^{\perp}} + n_{\tau \tau} \right) \right) - \frac{\text{Im} \left( \Lambda_{\tau} \right)}{Hz} n_{\tau^{\perp} \tau}.
\end{aligned}
\end{equation}
As $n_{B-L} = n_{\tau \tau} + n_{\tau^{\perp} \tau^{\perp}}$, we find an equation for the evolution of $n_{B-L}$ by adding the second and third equations together, obtaining
\begin{equation}\label{nBminusL}
\frac{dn_{B-L}}{dz} = D_1 \left(n^{}_{N_1}-n_{N_1}^{\text{eq}}\right) \Tr \epsilon^{(1)} - W_1 \left( \lvert C_{1\tau} \rvert^2 n_{\tau \tau} + \lvert C_{1\tau^\perp} \rvert^2 n_{\tau^{\perp} \tau^{\perp}} + 2 \Re [ C_{1\tau^{\perp}} C_{1\tau}^{*} n_{\tau \tau^{\perp}} ] \right).
\end{equation}
If this were to reproduce the single flavour limit, then we should find that the coefficient of $W_1$:
\begin{equation}
 \lvert C_{1\tau} \rvert^2 n_{\tau \tau} + \lvert C_{1\tau^{\perp}} \rvert^2 n_{\tau^{\perp} \tau^{\perp}} + 2 \Re [ C_{1\tau^{\perp}} C_{1\tau}^{*} n_{\tau \tau^{\perp}} ],
\end{equation}
is equal to $n_{B-L}$ in the limit that $\text{Im} ( \Lambda_{\tau} ) /Hz$ is small. Recalling that $\lvert C_{1\tau^{\perp}} \rvert^2 + \lvert C_{1\tau} \rvert^2 = 1$, then one should expect, that in the limit of small thermal widths,
\begin{equation}
2 \Re \left[ C_{1\tau^{\perp}} C_{1\tau}^{*} n_{\tau \tau^{\perp}} \right] = \lvert C_{1 \tau^{\perp}} \rvert^2 n_{\tau \tau} + \lvert C_{1\tau} \rvert^2 n_{\tau^{\perp} \tau^{\perp}}.
\end{equation}
In order to demonstrate this equality, first we show that the $z$-derivative of $2 \Re [ C_{1\tau^{\perp}} C_{1\tau}^{*} n_{\tau \tau^{\perp}} ]$ equals the $z$-derivative of $\lvert C_{1 \tau^{\perp}} \rvert^2 n_{\tau \tau} + \lvert C_{1\tau} \rvert^2 n_{\tau^{\perp} \tau^{\perp}}$ meaning that the quantities themselves may differ only by a constant. Then we note that, since at $z=z_0$ the quantities are equal, then they must be equal for all $z$.

By multiplication of the relevant equations in \equaref{eq:DMEqsAppendix}, we obtain the $z$-evolution of $\lvert C_{1 \tau^{\perp}} \rvert^2 n_{\tau \tau} + \lvert C_{1\tau} \rvert^2 n_{\tau^{\perp} \tau^{\perp}}$:
\begin{equation}
\begin{aligned}
& \lvert C_{1 \tau^{\perp}} \rvert^2 \frac{d n_{\tau \tau}}{dz} + \lvert C_{1\tau} \rvert^2 \frac{d n_{\tau^{\perp} \tau^{\perp}}}{dz} =\\
& ( \lvert  C_{1 \tau^{\perp}}  \rvert^2 \epsilon^{(1)}_{\tau \tau} + \lvert  C_{1\tau}  \rvert^2 \epsilon^{(1)}_{\tau^{\perp} \tau^{\perp}}) D_1 (n^{}_{N_1} - n_{N_1}^\text{eq}) - W_1( \Re [ C_{1 \tau^{\perp}} C_{1\tau}^{*} n_{\tau \tau^{\perp}} ] +  \lvert C_{1 \tau^{\perp}} \rvert^2 \lvert C_{1\tau} \rvert^2 (n_{\tau \tau}+n_{\tau^{\perp} \tau^{\perp}})).
\end{aligned}
\end{equation}
By similar means we obtain the $z$-evolution of $\Re [ C_{1\tau^{\perp}} C_{1\tau}^{*} n_{\tau \tau^{\perp}} ]$:
\begin{equation}
\begin{aligned}
\Re[C_{1\tau^{\perp}} C_{1\tau}^{*} \frac{d n_{\tau \tau^{\perp}}}{dz}] =
\nonumber &  \Re \left[ C_{1 \tau^{\perp}} C_{1\tau}^{*} \epsilon^{(1)}_{\tau \tau^{\perp}} \right] D_1 (n^{}_{N_1} - n_{N_1}^\text{eq})\\
& - \frac{1}{2} W_1 ( \Re \left[C_{1\tau^{\perp}} C_{1\tau}^{*} n_{\tau \tau^{\perp}} \right] + \lvert C_{1 \tau^{\perp}} \rvert^2 \lvert C_{1\tau} \rvert^2 (n_{\tau \tau} + n_{\tau^{\perp} \tau^{\perp}})) \\ 
\nonumber & - \Re \left[C_{1 \tau^{\perp}} C_{1\tau}^{*} n_{\tau \tau^{\perp}} \right] \frac{\Im(\Lambda_{\tau})}{Hz}.
\end{aligned}
\end{equation}
Neglecting $\Im(\Lambda_{\tau})/Hz$, as we expect this to be small in the single-flavour regime, then we need only show
that 
\begin{equation}\label{eq:RePartMidStep}
2 \Re \left [ C_{1 \tau^{\perp}} C_{1\tau}^{*} \epsilon^{(1)}_{\tau \tau^{\perp}} \right] = \lvert C_{1 \tau^{\perp}} \rvert^2 \epsilon^{(1)}_{\tau \tau} + \lvert C_{1\tau} \rvert^2 \epsilon^{(1)}_{\tau^{\perp} \tau^{\perp}},
\end{equation}
and then it is demonstrated that the coefficient of $W_1$ in \equaref{nBminusL} is approximately equal to $n_{B-L}$ and thus the single flavour equations are recovered.

The relation of \equaref{eq:RePartMidStep} can be put into a more suggestive form if we use $\lvert C_{1\tau^{\perp}} \rvert^2 = 1 - \lvert C_{1\tau} \rvert^2$ to re-express it thus
\begin{equation}
2 \Re [ C_{1 \tau^{\perp}} C_{1\tau}^{*} \epsilon^{(1)}_{\tau \tau^{\perp}} ] + \lvert C_{1 \tau^{\perp}} \rvert^2 \epsilon^{(1)}_{1 \tau^{\perp}} + \lvert C_{1\tau} \rvert^2 \epsilon^{(1)}_{1 \tau} = \epsilon^{(1)}_{\tau \tau} + \epsilon^{(1)}_{\tau^{\perp} \tau^{\perp}}.
\end{equation}
The right-hand side of this equation is merely the trace of the $CP$-asymmetry tensor $\Tr \epsilon^{(1)}$ in the $(\tau^{\perp}, \tau)$-basis. Thus, we suspect that the left-hand side is merely the trace expressed in an unfamiliar basis. This can be confirmed to be the case by construction of the unitary matrix
\begin{equation}
S = \left(\begin{array}{cc}
C_{1\tau} & - C_{1 \tau^{\perp}}^{*}\\
C_{1 \tau^{\perp}} & C_{1\tau}^{*} \\
\end{array}\right),
\end{equation}
then, by explicit calculation it can be seen that the left-hand side is the result of summing the diagonals (evaluating the trace in a particular basis) of
\begin{equation}
S^{\dagger} \epsilon^{(1)} S.
\end{equation}
Thus, we may conclude that, if we set $\Im( \Lambda_{\tau} ) = 0$, we are left with 
\begin{equation}
2 \Re \left[ C_{1 \tau^{\perp}} C_{1\tau}^{*} \frac{d n_{\tau \tau^{\perp}}}{d z} \right] = \lvert C_{1 \tau^{\perp}} \rvert^2 \frac{d n_{\tau \tau}}{d z} +  \lvert C_{1\tau} \rvert^2 \frac{d n_{\tau^{\perp} \tau^{\perp}}}{d z},
\end{equation}
and so
\begin{equation}
\frac{d}{dz} ( \lvert C_{1\tau} \rvert^2 n_{\tau \tau} + \lvert C_{1\tau^{\perp}} \rvert^2 n_{\tau^{\perp} \tau^{\perp}} + 2 \Re [ C_{1 \tau^{\perp}} C_{1\tau}^{*} n_{\tau \tau^{\perp}} ] ) = \frac{d n_{B-L}}{dz}.
\end{equation}
Since $n_{\alpha \beta} = 0$ at the initial $z$, then we may conclude that, if $\Im( \Lambda_{\tau} ) = 0$, then 
\begin{equation}
\frac{d n_{B-L}}{d z} = \Tr \epsilon^{(1)} D (n^{}_{N_1} - n_{N_1}^{\text{eq}}) - W_1 n_{B-L},
\end{equation}
which is the single-flavour limit.

If we don't set  $\Im( \Lambda_{\tau} ) = 0$, then we have
\begin{equation}
\frac{d}{dz} ( \lvert C_{1\tau} \rvert^2 n_{\tau \tau} + \lvert C_{1\tau^{\perp}} \rvert^2 n_{\tau^{\perp} \tau^{\perp}} + 2 \Re [ C_{1 \tau^{\perp}} C_{1\tau}^{*} n_{\tau \tau^{\perp}} ] ) = \frac{d n_{B-L}}{dz}-2 \Re [C_{1 \tau^{\perp}} C_{1\tau}^{*} \frac{\Im(\Lambda_{\tau})}{Hz} n_{\tau \tau^{\perp}}],
\end{equation}
which suggests that we should write the integro-differential equation
\begin{equation}
\frac{d n_{B-L}}{d z} = \Tr \epsilon^{(1)} D_1 (n^{}_{N_1} - n_{N_1}^{\text{eq}}) - W_1 n_{B-L} + 2 W_1 \int_{z_0}^z dz' \Re \left[C_{1 \tau^{\perp}} C_{1\tau}^{*} \frac{\Im(\Lambda_{\tau})}{Hz'} n_{\tau \tau^{\perp}}(z')\right].
\end{equation}
We define
\begin{equation}
\lambda(z) \equiv 2 \int_{z_0}^z dz' \Re \left[C_{1 \tau^{\perp}} C_{1\tau}^{*} \frac{\Im(\Lambda_{\tau})}{Hz'} n_{\tau \tau^{\perp}}(z')\right],
\end{equation}
for brevity, then using the integrating factor method, arrive at a solution
\begin{equation}
\begin{aligned}
n_{B-L}(z_f) & = e^{-\int_{z_0}^{z_f} W_1(z) dz} \int_{z_0}^{z_f} e^{\int_{z_0}^{z'} W_1(z'') dz''} \left(\Tr \epsilon^{(1)} D_1(z') (n^{}_{N_1}(z') - n_{N_1}^{\text{eq}}(z')) + W_1(z') \lambda(z') \right) dz' \\
\nonumber & = \int_{z_0}^{z_f} e^{-\int_{z'}^{z_f} W(z'') dz''} \left(\Tr \epsilon^{(1)} D_1(z') (n^{}_{N_1}(z') - n_{N_1}^{\text{eq}}(z')) + W_1(z') \lambda(z') \right) dz'.
\end{aligned}
\end{equation}
For large $M_1$, the thermal width is very small and so the term in $\lambda$ is usually neglected in comparison with the first.

\section{Robustness of the High-Scale Plateau}
\label{sec:RobustnessOfHighScalePlateau}

In the transition region, the approximation that left-handed $\tau$ leptons are produced and destroyed at the same rate by flavour effects is somewhat inaccurate.  In fact we should consider a slightly more accurate version of the density matrix equations in which the asymmetry density of right-handed $\tau$ leptons, $n_{\tau R}$ is computed. Then, the density matrix equations are
\begin{equation}\label{eq:DMEmodified}
\begin{aligned}
\frac{d n_{N_1}}{d z} & = -D_1 \left( n^{}_{N_1} - n_{N_1}^{\text{eq}} \right) \\
\frac{d n_{\tau^{\perp} \tau^{\perp}}}{d z} & = \epsilon^{(1)}_{\tau^{\perp} \tau^{\perp}} D_1 \left( n^{}_{N_1} - n_{N_1}^{\text{eq}} \right) - \frac{1}{2} W_1 \left( 2|C_{1 \tau^{\perp}}|^2 n_{\tau^{\perp} \tau^{\perp}}+C^{*}_{1 \tau} C_{1 \tau^{\perp}} n_{\tau \tau^{\perp}} + C_{1\tau} C^{*}_{1 \tau^{\perp}} n_{\tau^{\perp} \tau} \right) \\
\frac{d n_{\tau \tau}}{d z} & = \epsilon^{(1)}_{\tau \tau} D_1 \left( n^{}_{N_1} - n_{N_1}^{\text{eq}} \right) - \frac{1}{2} W_1 \left( 2|C_{1\tau}|^2 n_{\tau \tau}+C^{*}_{1 \tau} C_{1 \tau^{\perp}} n_{\tau \tau^{\perp}} +C_{1\tau} C^{*}_{1 \tau^{\perp}} n_{\tau^{\perp} \tau} \right) \\ & - 2 \frac{\Im(\Lambda_{\tau})}{Hz} (n_{\tau \tau} - 2 n_{\tau_R}) \\
\frac{dn_{\tau^{\perp} \tau}}{d z} & = \epsilon_{\tau^{\perp} \tau} D_1 \left( n^{}_{N_1} - n_{N_1}^{\text{eq}} \right) - \frac{1}{2} W_1 \left( n_{\tau^{\perp} \tau}+C^{*}_{1 \tau} C_{1 \tau^{\perp}} \left( n_{\tau^{\perp} \tau^{\perp}} + n_{\tau \tau} \right) \right) - \frac{\text{Im} \left( \Lambda_{\tau} \right)}{Hz} n_{\tau^{\perp} \tau}\\
\frac{d n_{\tau_R}}{d z} & = 2 \frac{\Im(\Lambda_{\tau})}{Hz} (n_{\tau \tau} - 2 n_{\tau_R}).
\end{aligned}
\end{equation}
The simpler set we previously considered result from the assumption that $\Im(\Lambda_{\tau})/Hz$ is large enough to enforce $n_{\tau \tau} = 2 n_{\tau_R}$. Clearly this is inaccurate for the situation under consideration where $M_1 \gg 10^{12} \text{ GeV}$. We should now append to $\lambda(z)$ an extra term such that 
\begin{equation}
 \lambda(z) \rightarrow \lambda'(z) = 2 \int_{z_0}^z dz' \left( \Re \left[C_{1 \tau^{\perp}} C_{1\tau}^{*} \frac{\Im(\Lambda_{\tau})}{Hz'} n_{\tau \tau^{\perp}}(z')\right] - 2 \frac{\Im(\Lambda_{\tau})}{Hz'} (n_{\tau \tau}(z') - 2 n_{\tau R}(z')) \right).
\end{equation}
Now in this solution, there is a term in $n_{\tau \tau} \Im(\Lambda_{\tau})/Hz$ which scales approximately as $x x^{-1} = x^0$ and a term $n_{\tau R} \Im(\Lambda_{\tau})/Hz$ in which, it may be shown $n_{\tau R} \propto x$ and thus $\lambda'(z)$ exhibits a approximate invariance under a scaling $x$ as does $\lambda(z)$. 

It may be added that scattering effects can be incorporated by modifying the decay function $D_1(z) \rightarrow D_1'(z) = D_1(z) + S_1(z)$ and the washout $W_1(z) \rightarrow W_1'(z)=j(z) W_1(z)$~\cite{Buchmuller:2004nz}. The new decay function $D'_1(z)$ which depends on a scattering part $S_1(z)$ is still multiplied by zero in the $\Tr \epsilon = 0$ case and is thus unimportant. The new washout function is multiplied by $j(z)$ which depends on $M_1$ through $\log (M_1/m_H)$. Thus, the plateau demonstrated in \figref{fig:HS_NO_CPValpha21_etaB_M1} picks up some unimportant logarithmic dependence on $M_1$ in addition to the small variation when scattering is neglected. In the numerical calculations of \secref{sec:HSLG}, the effects of scattering are included.

\bibliography{lepbib}{}
\bibliographystyle{JHEP}

\end{document}